\documentclass[journal,a4paper]{IEEEtran}

\usepackage{cite}
\newcommand{\subparagraph}{}

\usepackage{graphicx}
\graphicspath{{./figures}}
\DeclareGraphicsExtensions{.pdf}
\usepackage[cmex10]{amsmath}
\usepackage{amssymb}
\usepackage{amsthm}
\usepackage{hyperref}

\usepackage{multirow}
\usepackage{bm}
\usepackage{algorithmic}
\usepackage{xcolor}
\usepackage{tikz}
\usetikzlibrary{spy,shapes,arrows,positioning,plotmarks,shadows,calc,matrix,fit,patterns}
\usepackage{pgfplots}
\pgfplotsset{compat=1.8}
\usepgfplotslibrary{groupplots}
\usepackage{todonotes}
\usepackage{datetime}
\usepackage{cancel}
\usepackage{hhline}
\usepackage{caption}
\usepackage{subcaption}
\usepackage[overload]{empheq}
\usepackage{cases}
\usepackage{nicefrac}
\usepackage{etoolbox}
\usepackage{multirow}
\usepackage{algorithm}
\usepackage{algorithmic}
\usepackage{cleveref}
\usepackage{balance}

\newcommand{\mybold}[1]{\bm{#1}}
\newcommand{\figref}[1]{Fig.~\ref{#1}}
\newcommand{\tabref}[1]{Tab.~\ref{#1}}
\newcommand{\algref}[1]{Alg.~\ref{#1}}

\newcommand{\expv}{\mathbb{E}}

\newcommand{\cgauss}[1]{\mathcal{CN}\left(#1\right)}

\newcommand{\eye}[1]{\mybold{I}_{#1}}
\newcommand{\zeros}[2]{\mybold{0}_{#1\times{#2}}}
\newcommand{\diag}{\text{diag}}

\newcommand{\tr}{\text{tr}}

\newcommand{\timeVar}{k}
\newcommand{\frameVar}{\kappa}
\newcommand{\firLengthSI}{L}
\newcommand{\firLengthWC}{L}
\newtheoremstyle{remarkmod}
  {\topsep}   % ABOVESPACE
  {\topsep}   % BELOWSPACE
  {\normalfont}  % BODYFONT
  {0pt}       % INDENT (empty value is the same as 0pt)
  {\itshape} % HEADFONT
  {.}         % HEADPUNCT
  {5pt plus 1pt minus 1pt} % HEADSPACE
  {}          % CUSTOM-HEAD-SPEC
\theoremstyle{remarkmod}

\makeatletter
\newcommand*{\textoverline}[1]{$\overline{\hbox{#1}}\m@th$}
\makeatother
\def\subcaptionSpacing{0.75}

\tikzstyle{sum} = [draw, fill=blue!20, circle, node distance=1cm]
\tikzstyle{dot} = [draw, circle, minimum size=0.2pt,scale=0.3,fill=black,black]

\newcommand{\myadd}[2]{% args: label, coordinate
	\node (#1) at (#2) [sum, label={center:\phantom{\large $+$}},scale=1.35] {};
	\draw[black,line width=0.2mm] ($(#1.north)+(0,-0.1)$) to ($(#1.south)+(0,0.1)$);
	\draw[black,line width=0.2mm] ($(#1.west)+(0.1,0)$) to ($(#1.east)+(-0.1,0)$);
}

\newcommand{\mymod}[2]{% args: label, coordinate
\node (#1) at (#2) [sum, label={center:\phantom{\large $+$}},scale=1.35] {};
\draw[black,line width=0.2mm] (#1.north east) to (#1.south west);
\draw[black,line width=0.2mm] (#1.north west) to (#1.south east);
}

\crefname{figure}{Fig.}{Figs.}
\Crefname{figure}{Fig.}{Figs.}
\Crefname{equation}{Eq.}{Eqs.}

\pgfdeclarelayer{background}
\pgfdeclarelayer{foreground}
\pgfsetlayers{background,main,foreground}

\newcounter{hints}
\renewcommand{\thehints}{\alph{hints}}
\newcommand{\hintedrel}[2][]{%
  \stepcounter{hints}%
  \if\relax\detokenize{#1}\relax\else\csxdef{hint@#1}{\thehints}\fi
  \mathrel{\overset{(\thehints)}{\vphantom{\le}{#2}}}%
}

\newcommand{\hintref}[1]{\csuse{hint@#1}}

\makeatletter
\newcommand*{\rom}[1]{\expandafter\@slowromancap\romannumeral #1@}
\makeatother

\makeatletter
\newcommand{\ALC@comblock}[1]{\ifthenelse{\equal{#1}{default}}%
{}{\textbf{#1}}}
\newenvironment{ALC@bl}{\begin{ALC@g}}{\end{ALC@g}}
\newcommand{\BLOCK}[2][default]{
	\ALC@it\ALC@comblock{#1}\ #2\begin{ALC@bl}
}
\newcommand{\ENDBLOCK}{
	\end{ALC@bl}
}
\newcommand{\PREDICT}{
	\BLOCK[Predict]
}
\newcommand{\UPDATE}{
	\BLOCK[Update]
}
\newcommand{\GAIN}{
	\BLOCK[Kalman gain]
}

\makeatother

\newtoggle{showproof}
\toggletrue{showproof}

\def\resRateVsSinr{./resRateVsSinr}
\def\resSrinrConvVsSinr{./resSrinrConvVsSinr}
\def\resSysDistWConvVsSinr{./resSysDistWConvVsSinr}

\def\resSrinrVsSinr{./resSrinrVsSinr}
\def\resSysDistWVsSinr{./resSysDistWVsSinr}
\def\resSysDistAOneVsSinr{./resSysDistA1VsSinr}
\def\resSysDistATwoVsSinr{./resSysDistA2VsSinr}

\pgfmathtruncatemacro{\numSamples}{400}
\pgfmathtruncatemacro{\markRepeat}{\numSamples/10}
\pgfmathtruncatemacro{\markPOne}{\markRepeat/8*0}
\pgfmathtruncatemacro{\markPTwo}{\markRepeat/8*1}
\pgfmathtruncatemacro{\markPThree}{\markRepeat/8*2}
\pgfmathtruncatemacro{\markPFour}{\markRepeat/8*3}
\pgfmathtruncatemacro{\markPFive}{\markRepeat/8*4}
\pgfmathtruncatemacro{\markPSix}{\markRepeat/8*5}
\pgfmathtruncatemacro{\markPSeven}{\markRepeat/8*6}
\pgfmathtruncatemacro{\markPEight}{\markRepeat/8*7}

\newif\ifshowTikzPictures \showTikzPicturestrue
\begin{document}
\title{State-Space Adaptive Nonlinear Self-Interference Cancellation for Full-Duplex Communication\thanks{This work is supported by the Federal Ministry of Education and Research (BMBF) of the Federal Republic of Germany (F\"orderkennzeichen 16KIS0658K, SysKit\_HW).}\vspace{-0ex}}
%This work is supported by the German Research Foundation under Grant S02 of SFB/TRR 196 (MARIE).

\author{Hendrik Vogt, Gerald Enzner, Aydin Sezgin\\
Department of Electrical Engineering and Information Technology\\
Ruhr-Universit\"at Bochum, Germany\\
Email: \{hendrik.vogt, gerald.enzner, aydin.sezgin\}@rub.de }

\maketitle

\IEEEpeerreviewmaketitle

\begin{abstract}
Full-duplex transmission comprises the ability to transmit and receive at the same time on the same frequency band. It allows for more efficient utilization of spectral resources, but raises the challenge of strong self-interference (SI). Cancellation of SI is generally implemented as a multi-stage approach. This work proposes a novel adaptive SI cancellation algorithm in the digital domain and a comprehensive analysis of state-of-the-art adaptive cancellation techniques. Inspired by recent progress in acoustic echo control, we introduce a composite state-space model of the nonlinear SI channel in cascade structure. 
We derive a SI cancellation algorithm that decouples the identification of linear and nonlinear elements of the composite state. They are estimated separately and consecutively in each adaptation cycle by a Kalman filter in DFT domain. We show that this adaptation can be supported by a-priori signal orthogonalization and decoding of the signal-of-interest (SoI). 
In our simulation results, we analyze the performance by evaluating residual interference, system identification accuracy and communication rate. Based on the results, we provide recommendations for system design.  
In case of input orthogonalization, our Kalman filter solution in cascade structure delivers best performance with low computational complexity. In this configuration, the performance lines up with that of the monolithic (parallel) Kalman filter or the recursive-least squares (RLS) algorithms. We show that the Kalman-based algorithm is superior over the RLS under time-variant conditions if the SoI is decoded and in this way the covariance information required by the Kalman filter can be provided to it.
%This allows for better insights into modeling of time-variant transitions of the SI channel and reduces the number of estimated system variables. 
%We apply suitable approximations to the exact algorithm that significantly reduce the computational complexity.
%A subsequent postfilter improves recovery of the desired signal.  
%in time-variant environments, the new cancellation algorithm provides performance gains in the low signal-to-interference-and-noise ratio (SINR) regime compared to other proposed algorithms. Furthermore, we provide a comprehensive overview on the impact of algorithmic design choices, for instance, we show that orthogonalization of the nonlinear input components is blurring the performance differences between the algorithms in static environments, but it is not helpful for time-variant SI channels. In addition, it is highly recommended to apply decoding of the SoI before the update step in order to improve estimation accuracy.
\end{abstract}

\section{Introduction}
The sheer amount of data conveyed over the wireless medium requires novel concepts to be integrated into 5G (and beyond) transmission standards. A promising technique for more efficient spectrum utilization is RF (radio frequency) in-band full-duplex communication, where transmission and reception of signals are run simultaneously on the same frequency band~\cite{src:sexton20175g}. Theoretically such an approach supports an improved capacity of the wireless link~\cite{src:sabharwal2014band,src:aijaz2017simultaneous}. However, some considerable signal power from the transmitter inevitably leaks as self-interference (SI) into the receiver chain, and therefore any signal-of-interest (SoI) from a distant communication node cannot be reliably recovered at the receiver. Cancellation of the strong SI is a challenging task. Several landmark papers have successfully demonstrated the feasibility of such an approach~\cite{src:jain2011practical,src:riihonen2011hybrid,src:duarte2012experiment}. Common state-of-the-art SI cancellation is generally implemented as a multi-stage operation. Full-duplex prototypes have combined passive SI suppression, active analog cancellation, and digital cancellation~\cite{src:khandani2013two,src:duarte2014design,src:debaillie2014analog,src:ahmed2015all,src:askar2016agile}.
% The option of a fully-digital design has been shown~\cite{}. Compact solutions of analog RF cancellation~\cite{} enable the integration of full-duplex capability into existing wireless standards~.

The availability of SI cancellation allows for a wide range of full-duplex applications. It can increase the communication rate in a wireless network by using full-duplex relays~\cite{src:avestimehr2008approximate,src:nunn2017antenna}, reduce end-to-end delay~\cite{src:kariminezhad2017fullduplex}, provide enhanced security at the physical layer by friendly jamming~\cite{src:zheng2013improving} or key agreement~\cite{src:vogt2016practical}, and it implies the potential of energy-harvesting~\cite{src:zeng2015full, src:bi2016accumulate}.

To remove the impact of SI in the digital domain, a variety of digital SI channel estimation metrics have been studied such as least-squares (LS)~\cite{src:bharadia2013full}, minimum-mean squared error (MMSE)~\cite{src:day2012fullduplex} or maximum-likelihood (ML) estimation~\cite{src:masmoudi2016maximum}. Non-uniform sampling can avoid undesired SI in Orthogonal Frequency-Division Multiplexing (OFDM) systems~\cite{src:bernhardt2018self}. Under time-variant conditions, adaptive algorithms generally can cover the task of adjusting an SI estimate to gradual changes in the SI channel~\cite{src:heino2015recent}. The linear SI channel model has been extended to nonlinear forms, where the Hammerstein polynomial model is the most popular~\cite{src:bharadia2014full}. It takes the form of a parallel, multi-channel representation with monolithic structure. Alternatively, the nonlinear SI channel can be characterized by an artificial neural network~\cite{src:stimming2018nonlinear}. Beyond the nonlinear SI model, previous work established enhancements like signal orthogonalization, least-mean square (LMS) adaptation~\cite{src:korpi2015adaptive,src:ferrand2017multi} or recursive-least square (RLS) algorithms~\cite{src:lemos2015fullduplex,src:emara2017nonlinear}. Adaptive approaches can be used in hybrid digital/analog cancellation designs~\cite{src:kiayani2018adaptive}. Many prototypes use an analog cancellation stage to remove the line-of-sight component, which is the dominant and, in most cases, the time-invariant part of the SI. However, moving objects close to the antenna can significantly change the multipath characteristics of the SI channel due to power backscattering~\cite{src:everett2014passive} and therefore introduce a significant time-variant contribution to the SI. The general problem of adaption to time-variant SI channels requires further research in the digital domain. 

The SI is a challenge not only in wireless communication. Consider hands-free voice communication, where microphone and loudspeaker signals require full-duplex operations. Here, similar to the SI in wireless full-duplex applications, the unacceptable acoustic echo signal needs to be effectively removed. It was, however, found %Haensler2004,
that the eventual echo cancellation system involves a number of challenges, such as the fast adaptivity to time-variant loudspeaker-room-microphone systems~\cite{src:benesty2001advances,Haensler97,Haensler2006}. %and the simultaneous robustness against near-end/far-end double talk scenarios. The latter implies a large dynamic range of the signal-to-observation noise ratio~\cite{Haensler97,Haensler2006}. 
The standard for acoustic echo control is the LMS and normalized least means-square (NLMS) type adaptive-filter algorithm in single- and multi-channel configurations~\cite{Haykin2002,Enzner2014}. Similarly, RLS-based approaches of the multi-channel type have been introduced~\cite{src:benesty2001advances}.  As a compromise between NLMS and RLS in terms of fast adaptation and computational complexity, approaches based on affine projection algorithms (APA)~\cite{src:gay1995the} have been proposed and applied to nonlinear echo cancellation~\cite{src:glicacho2012nonlinear}. State-space modeling of the time- and frequency-domain echo channel~\cite{Enzner06,Enzner2010} has been introduced. 
%These models imply the respective time- or frequency-domain adaptive Kalman filter algorithm to handle the fast and robust adaptation of echo cancellation filters~\cite{Enzner06}. 
Variants of the Kalman filter algorithm were provided in multichannel~\cite{Malik2011a} and nonlinear configurations \cite{MalikEnzner2012b,src:malik2013variational}.

 %In addition, the theoretical performance bounds of adaptation need to be established in order to obtain deeper insights into the problem.  
%such as nonlinear Bayesian algorithms~\cite{src:malik2013variational}.

Both the domains of acoustic echo control and wireless SI cancellation share many aspects of the system modeling. Thus, we can transfer concepts and insights from the acoustic echo cancellation. Most prominently, the nonlinear echo/SI models of both the acoustic loudspeaker and the RF power amplifiers are very similar. In the acoustic echo channel, however, impulse responses can have up to a few thousand of significant taps. In the wireless domain, the delay spread is much shorter, but the accurate depiction of fractional signal delays still requires a significant amount of taps~\cite{src:laakso96splitting}. There is a notable difference between acoustic echo control and wireless SI cancellation: in the handling of the SoI. In acoustics, the SoI is usually modeled as a random process and therefore contributes to the observation noise during adaptation. In wireless communications, however, the SoI is deliberately encoded and thus can be removed before adaptation.

%of the adaptive filter and has been ideally covered with the famous statement ``From Algorithms to Systems - It's a Rocky Road''  that went ahead of current developments.

In this work, we set our focus on the digital part of SI cancellation by an adaptive algorithm. Consider the system overview of~\figref{fig:overview}. 
\def\antenna{%
   -- +(0mm,4.0mm) -- +(2.625mm,7.5mm) -- +(-2.625mm,7.5mm) -- +(0mm,4.0mm)
}	
\begin{figure}
	\centering
	\begin{tikzpicture}[
	sysBlock/.style={draw, fill=white, rectangle, minimum height=2em, minimum width=3.5em},
	nlBlock/.style={draw, fill=white, rectangle, minimum height=2em, minimum width=4em},
	shadowBlock/.style={draw, drop shadow, fill=white, rectangle},
	triangle/.style = {draw, regular polygon, regular polygon sides=3 },
	%node rotated/.style = {rotate=90},
	border rotated/.style = {shape border rotate=90}
	]

	\node (inTx) {$x_{\timeVar}$};
	\node (outRx) at ($(inTx)+(0,-3)$) {};
	
	\node[shadowBlock, minimum height=4em](dec) at ($(inTx)!0.5!(outRx)+(0.75,0.25)$) {Decoder};
	\myadd{add2}{$(dec)+(1.5,0)$}
	\node[shadowBlock, minimum height=4em, minimum width=4em, align=center, fill=red!10] at ($(add2)+(1.5,0)$) (adapt) {Adaptive\\algorithm};	
	\myadd{add1}{$(adapt|-outRx)+(0,0)$}	
	\node[dot] (dot1) at ($(adapt|-inTx)$) {};	
	\node[dot] (dot2) at ($(add2|-outRx)+(0,0)$) {};
	
	\node  at ($(add1.east)+(0.1,0.2)$) {$-$};
	\node  at ($(add2.west)+(-0.1,0.2)$) {$-$};
	
	\node[shadowBlock, align=center, minimum height=10em] (analog) at ($(adapt)+(1.6,-0.25)$) {Tx/Rx\\path};
	\node (ant) at ($(analog)+(1.1,0)$) {};
	\draw[very thick] (ant.center) \antenna;
	\draw[very thick] (analog.east) -- (ant.center);
	
	\node[shadowBlock, align=center, minimum height=2em] (dist) at ($(analog)+(2.3,-1)$) {Distant\\node};
	\node (antDist) at ($(dist)+(0,0.5)$) {};
	\draw[very thick] (antDist.center) \antenna;
	\draw[very thick] (dist.north) -- (antDist.center);
	
	\draw[thick,->] (inTx) -- (dot1);
	\draw[thick,->] (dot1) -- (analog.west|-dot1);
	\draw[thick,->] (dot1) -- (adapt);
		
	\draw[thick,->] (analog.west|-add1) -- node[below] {$y_{\timeVar}$} (add1);
	\draw[thick,->] (adapt.south-|add1) -- node[right] {$\hat{x}_{\text{si},\timeVar}$} (add1);
	\draw[thick,->] (dot2) -- (add2);
	\draw[thick,->] (add1) -- node[above] {$e_{\timeVar}$} (dot2);
	\draw[thick,->] (dot2) -| (dec);
	\draw[thick,->] (dec) -- node[below] {$\hat{d}_{\timeVar}^h$} (add2);
	\draw[thick,->] (add2) -- node[above] {$\tilde{e}_{\timeVar}$} (adapt);
	
	\draw[dashed,->] ($(antDist)+(-0.2,0.2)$) -- node[below,pos=0.75] {$d_{\timeVar}^h$} ($(ant)+(0.2,0.2)$);
	
	\end{tikzpicture}
\caption{Proposed full-duplex SI estimation and cancellation.}
\label{fig:overview}
\end{figure}
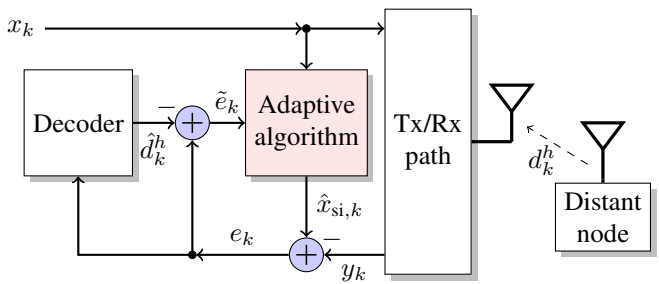
A distant node sends a SoI~$d^h_k$ at time $k$ to the receiver, while the local transmitter conveys $x_k$. At the local receiver, an adaptive algorithm reconstructs an estimated SI~$\hat{x}_{\text{si},\timeVar}$ and subtracts its contribution from the received signal $y_k$, which leaves the error signal $e_k$. Next, this signal $e_k$ is fed into a decoder which retrieves the decoded SoI~$\hat{d}^h_k$. By removing $\hat{d}^h_k$ from the error signal $e_k$, the residual signal $\tilde{e}_k$ is given to the algorithm for adaptation. Inspired by recent progress in the area of acoustic echo control, we intend to tailor acoustic echo cancellation methods to wireless SI cancellation by employing a nonlinear, composite state-space system model in cascade structure.
%, where the SI path is the state. 
%In that sense, we explicitly describe the inner transition of the nonlinear SI channel by specifying a coherence time. 
The state-space representation allows more dedicated treatment of time-variant SI channels, since its a-priori knowledge is embedded directly into the system model. 
%Thus, certain   of temporal variations can be purposefully utilized. 
%The requirements of a time-variant application are then more carefully matched, 
This leads to a better control over the adaptation process and an improved performance of the solution. 
%In our proposed model, linear and nonlinear elements of the SI path appear in product form inside the composite state-space model. 
Unlike in the parallel nonlinear SI channel model, the cascade structure is reducing the number of variables to be estimated and simultaneously the susceptibility to over-fitting~\cite{src:malik2013variational}. 
%The algorithm linearizes the model by decoupling the estimations. As a consequence, the linear SI channel component and nonlinear coefficients are estimated separately and iteratively by linear algorithms in each adaptation cycle. To reduce computational complexity, we approximately assume during the estimation of the nonlinear coefficients that the nonlinear input components are orthogonal. 
We linearize the model by decoupling linear SI path and nonlinear coefficients, which are then estimated separately and iteratively by linear algorithms in each adaptation cycle.
The natural algorithmic solution to the estimations based on MMSE criterion turns out to be the Kalman filter. This Kalman filter algorithm is derived in DFT domain. Frequency domain approaches, while often explicitly tailored to OFDM systems~\cite{src:sohaib2017alow}, have the potential of reducing computational complexity while maintaining estimation accuracy~\cite{src:komatsu2017freq}. 
%To further reduce the complexity, we employ approximations by neglecting intra-channel correlation of the linear SI channel and by removing the overlap-save constraint from the fast convolution in DFT domain.
In previous work, adaptive algorithms with input orthogonalization have been proposed~\cite{src:korpi2015adaptive,src:emara2017nonlinear} to meet the aforementioned assumptions. We maintain this feature to decouple the nonlinear basis function for improved speed of convergence. We pursue a systematic treatise of that design feature and explore its impact on the performance.
Furthermore, we integrate the decoding of the signal-of-interest (SoI), which has been conveyed by a distant node. This approach is beneficial, since in wireless full-duplex communication, the SoI contains structure which is a-priori known and therefore can be exploited. In the development of adaptive SI cancellation algorithms, the a-priori knowledge of SoI signal statistics have been rather neglected so far.
%Furthermore, we provide additional SI suppression by employing a postfilter after the cancellation step.  
%The simulation results indicate that within time-variant environments, it is significantly important to match the SI cancellation algorithm to the scale of state changes. However, the case of non-stationary random processes, which might lead to further mismatches of modeling the SI, is beyond the scope of this work.\todo{Discussion on results?}

We study the performance by considering certain metrics like the signal-to-residual-interference-and-noise ratio, system identification accuracy and the communication rate. Both the time convergence behavior and the global performance with respect to input signal-to-interference-and-noise ratio are considered. We provide comparisons of the proposed algorithm in exact and approximated form to other approaches like Kalman-based, NLMS and RLS algorithms from the literature. This leads to a comprehensive analysis of certain design options  (such as input orthogonalization, complexity or model structure) on the performance of the novel and state-of-the-art adaptive algorithms. 
%If the input signals are orthogonalized, our Kalman filter solution in cascade structure is sufficient for best performance with low computational complexity. Without orthogonalization, the Kalman algorithm in parallel implementation represents the best trade-off between complexity and performance due to its monolithic structure. 
We show that the temporal variations impose a fundamental performance limitation, regardless of whether input orthogonalization is applied or not. The simulation results indicate that the decoding of the SoI is generally beneficial to both speed of convergence and cancellation performance, especially under time-variant conditions.

%Recent progress on acoustic echo cancellation for hands-free devices motivates state-space models for full-duplex SI cancellation. 
%\subsection{Notation}
Throughout the paper, we print column vectors and matrices as bold lower-case and upper-case letters, respectively. The operators $\expv\left[\cdot\right]$, $\tr\left[\cdot\right]$, $\left|\cdot\right|$, $\left\lVert\cdot\right\rVert_2$, $\left(\cdot\right)^T$, $\left(\cdot\right)^H$ denote expectation, trace of a matrix, absolute value, Euclidean norm, matrix transpose and Hermitian transpose, respectively. The operator $\diag\left[ \mybold{x}\right]$ represents a diagonal matrix with the elements of vector $\mybold{x}$ on its main diagonal. Signal vectors in DFT domain are marked by underline $\underline{\mybold{x}}$. The term $\mybold{x}\circ\mybold{y}$ denotes the Hadamard product, i.e., the element-wise multiplication of $\mybold{x}$ and $\mybold{y}$. The identity matrix of size $M\times M$ is given by $\eye{M}$.

%\subsection{Organization}
The paper is organized as follows. Section~\ref{sec:systemmodel} introduces the system model with more details on the Tx/Rx path of~\figref{fig:overview}, its DFT domain representation and the state-space model. We derive the nonlinear adaptive algorithm in Section~\ref{sec:algorithm} and discuss useful approximations in Section~\ref{sec:approximations}. In Section~\ref{sec:results}, we define suitable metrics and evaluate the performance by considering computational complexity, time convergence behavior, global performance and various use cases. Finally, Section~\ref{sec:conclusion} concludes the paper. 

\section{System model}
\label{sec:systemmodel}
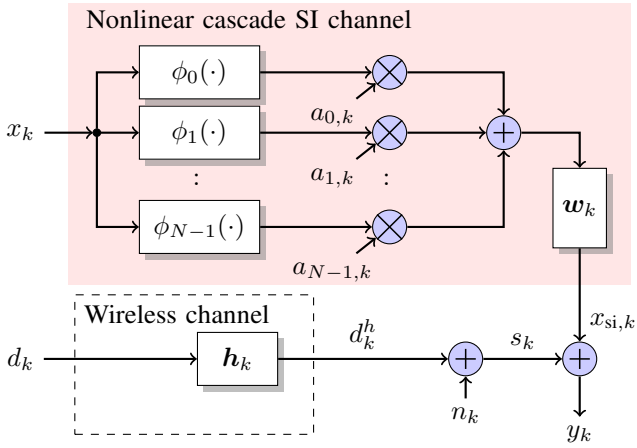
\begin{figure}
	\centering
	\begin{tikzpicture}[
	sysBlock/.style={draw, drop shadow, fill=white, rectangle, minimum height=2em, minimum width=3em},
	nlBlock/.style={draw, drop shadow, fill=white, rectangle, minimum height=2em, minimum width=4.5em},
	chanBlock/.style={draw, drop shadow, fill=white, rectangle, minimum height=3em, minimum width=2em}
	]
	
	\node (in1) {$x_{\timeVar}$};
	\node[dot] (dot2) at ($(in1)+(1,0)$) {};
	
	\node[nlBlock] at ($(dot2)+(1.35,0.8)$) (nlfunc0) {$\phi_0(\cdot)$};
	\node[nlBlock] at ($(dot2)+(1.35,0)$) (nlfunc1) {$\phi_1(\cdot)$};
	\node at ($(dot2)+(1.35,-0.6125)+(0,0)$) (colon1) {$\colon$};
	\node[nlBlock] at ($(dot2)+(1.35,-1.25)$) (nlfuncN) {$\phi_{N-1}(\cdot)$};	
	
	\mymod{mult1}{$(nlfunc0)+(2.5,0)$}
	\mymod{mult2}{$(nlfunc1)+(2.5,0)$}
	\node at ($(colon1)+(2.5,0)+(0,0)$) (colon1) {$\colon$};
	\mymod{mult3}{$(nlfuncN)+(2.5,0)$}
	
	\node at ($(mult1)+(-0.75,-0.6)$) (a0) {$a_{0,\timeVar}$};
	\node at ($(mult2)+(-0.75,-0.6)$) (a1) {$a_{1,\timeVar}$};
	\node at ($(mult3)+(-0.75,-0.6)$) (aN) {$a_{N-1,\timeVar}$};
	
	\myadd{add1}{$(mult2)+(1.5,0)$}	
	
	\node[chanBlock] at ($(add1)+(1,-1)$) (sichan) {$\mybold{w}_{\timeVar}$};
	%\node (out) at ($(sichan)+(1.5,0)$) {$x_{\text{si},\timeVar}$};
	
	\draw[thick,->] (in1) to (dot2);
	
	\draw[thick,->] (dot2) |- (nlfunc0);
	\draw[thick,->] (dot2) -- (nlfunc1);
	\draw[thick,->] (dot2) |- (nlfuncN);	
	\draw[thick,->] (a0) -- (mult1);
	\draw[thick,->] (a1) -- (mult2);
	\draw[thick,->] (aN) -- (mult3);
	\draw[thick,->] (nlfunc0) -- (mult1);
	\draw[thick,->] (nlfunc1) -- (mult2);
	\draw[thick,->] (nlfuncN) -- (mult3);
	\draw[thick,->] (mult1) -| (add1);
	\draw[thick,->] (mult2) -- (add1);
	\draw[thick,->] (mult3) -| (add1);
	
	%\draw[thick,->] (add1) to (sichan);
	%\draw[thick,->] (sichan) to (out);

	\myadd{add3}{$(sichan)+(0,-2)$}
	\myadd{add4}{$(add3)+(-1.5,0)$}
	\node (noise) at ($(add4)+(0,-0.75)$) {$n_{\timeVar}$};
	\node[sysBlock] at ($(add4)+(-3,0)$) (comchan) {$\mybold{h}_{\timeVar}$};
	\node (distant) at (in1 |- add4) {$d_{\timeVar}$};	
	\node (out) at ($(add3)+(0,-1)$) {$y_{\timeVar}$};
	
	\node (labelSi) at ($(dot2) + (2,1.5)$) {Nonlinear cascade SI channel};
	\begin{pgfonlayer}{background}
		\node [fill=red!10,fit={($(labelSi.west) + (0,0.1)$) ($(sichan) + (0.55,-0.9)$)}] {};
	\end{pgfonlayer}	
	
	\node (labelWc) at ($(comchan) + (-0.8,0.6)$) {Wireless channel};
%	\begin{pgfonlayer}{background}
%		\node [fill=green!10,fit={($(labelWc.west) + (0,0.1)$) ($(comchan) + (1.5,-0.75)$)}] {};
%	\end{pgfonlayer}
	
	%\draw[dashed] ($(dot1) + (-0.3,0.5)$) rectangle ($(add1) + (0.3,-1.7)$);
	%\draw[dashed] ($(dot2) + (-0.3,1.3)$) rectangle ($(sichan) + (0.5,-1.1)$);
	\draw[dashed] ($(labelWc.west) + (0,0.25)$) rectangle ($(comchan) + (1,-1)$);
	
	%	\draw[thick,->] (sichanN) -| (add1);
	\draw[thick,->] (add1) -| (sichan);
	\draw[thick,->] (sichan) to node[right,pos=0.8] {$x_{\text{si},\timeVar}$} (add3);
	
	\draw[thick,->] (noise) to (add4);
	
	\draw[thick,->] (add4) to node[above] {$s_{\timeVar}$} (add3);
	\draw[thick,->] (distant) to (comchan);
	\draw[thick,->] (comchan) to node[above] {$d^{h}_{\timeVar}$} (add4);
	\draw[thick,->] (add3) to (out); 
	\end{tikzpicture}
	\caption{Cascade model of the nonlinear SI channel with input $x_{\timeVar}$ and the signal-of-interest $d_{\timeVar}$.}
	\label{fig:systemmodel}
\end{figure}

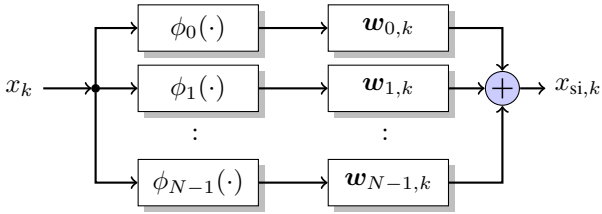
\begin{figure}
	\centering
	\begin{tikzpicture}[
	nlBlock/.style={draw, drop shadow, fill=white, rectangle, minimum height=1.75em, minimum width=4.5em},
	chanBlock/.style={draw, drop shadow, fill=white, rectangle, minimum height=1.75em, minimum width=4.5em}
	]
	
	\node (in1) {$x_{\timeVar}$};
	\node[dot] (dot2) at ($(in1)+(1,0)$) {};
	
	\node[nlBlock] at ($(dot2)+(1.35,0.8)$) (nlfunc0) {$\phi_0(\cdot)$};
	\node[nlBlock] at ($(dot2)+(1.35,0)$) (nlfunc1) {$\phi_1(\cdot)$};
	\node at ($(dot2)+(1.35,-0.6125)+(0,0)$) (colon1) {$\colon$};
	\node[nlBlock] at ($(dot2)+(1.35,-1.25)$) (nlfuncN) {$\phi_{N-1}(\cdot)$};	
		
	\node[chanBlock] at ($(nlfunc0)+(2.5,0)$) (sichan0) {$\mybold{w}_{0,\timeVar}$};
	\node[chanBlock] at ($(nlfunc1)+(2.5,0)$) (sichan1) {$\mybold{w}_{1,\timeVar}$};
	\node at ($(colon1)+(2.5,0)+(0,0)$) (colon1) {$\colon$};
	\node[chanBlock] at ($(nlfuncN)+(2.5,0)$) (sichanN) {$\mybold{w}_{N-1,\timeVar}$};
	
	\myadd{add1}{$(sichan1)+(1.5,0)$}	
	
	%\node (out) at ($(sichan)+(1.5,0)$) {$x_{\text{si},\timeVar}$};
	
	\draw[thick,->] (in1) to (dot2);
	
	\draw[thick,->] (dot2) |- (nlfunc0);
	\draw[thick,->] (dot2) -- (nlfunc1);
	\draw[thick,->] (dot2) |- (nlfuncN);	
	\draw[thick,->] (nlfunc0) -- (sichan0);
	\draw[thick,->] (nlfunc1) -- (sichan1);
	\draw[thick,->] (nlfuncN) -- (sichanN);
	\draw[thick,->] (sichan0) -| (add1);
	\draw[thick,->] (sichan1) -- (add1);
	\draw[thick,->] (sichanN) -| (add1);
		
	\node (out) at ($(add1)+(1,0)$) {$x_{\text{si},\timeVar}$};
	\draw[thick,->] (add1) to (out);
	\end{tikzpicture}
	\caption{Parallel model of the nonlinear SI channel.}
	\label{fig:parallelmodel}
\end{figure}

Consider the system model of~\figref{fig:systemmodel}, which depicts a nonlinear cascaded SI channel.
The local transmitter creates an input signal $x_{\timeVar}$ to be conveyed to a distant receiver. Due to hardware impairments like I/Q imbalance in the quadrature mixer or high-order harmonics of the power amplifier, the signal is distorted by nonlinear components. This effect is especially severe at high transmitter output powers~\cite{src:korpi2014full}. We model the nonlinear contribution as $N^{\text{th}}$-order memoryless expansion, where the $i^{\text{th}}$ component consists of a nonlinear basis function $\phi_i\left(\cdot\right)$ and the coefficient $a_{i,\timeVar}$. Now, the linear SI channel is modeled by a linear, time-variant finite-impulse response (FIR) filter.  Let the aforementioned FIR filter have $\firLengthSI$ leading coefficients given by
\begin{align}
\label{eq:siChanDef}
\mybold{w}_{\timeVar}=\left[ w_{\timeVar,0}, w_{\timeVar,1}, \ldots, w_{\timeVar,\firLengthSI-1}\right]^T.
\end{align}
Alternatively, the nonlinear SI channel can be modeled in parallel structure, which is shown in~\figref{fig:parallelmodel}. 
In that case, each of the nonlinear basis functions is followed by a different FIR filter with coefficients $\mybold{w}_{i,\timeVar}$ for $0\leq i\leq N-1$. However, the parallel approach significantly increases the degrees of freedom to be estimated. Conventional SI channel models require a large number of filter taps, which has been identified as an obstacle for practical systems~\cite{src:nadh2017ataylor}. Specifically, such an architecture might lead to ambiguous solutions. Thus, we focus on the cascade modeling of~\figref{fig:systemmodel} in our work. Then, the SI signal of~\figref{fig:systemmodel} can be written in a generalized linear model~\cite{src:bishop2006pattern}
\begin{align}
x_{\text{si},\timeVar}
&= \sum^{\firLengthSI-1}_{l=0} w_{\timeVar,l} \sum_{i=0}^{N-1}a_{i,\timeVar}\phi_i\left(x_{\timeVar-l}\right), \nonumber\\
\label{eq:nldef}
&= \sum_{i=0}^{N-1}a_{i,\timeVar}\sum^{\firLengthSI-1}_{l=0} w_{\timeVar,l} \phi_i\left(x_{\timeVar-l}\right), 
\end{align}
where $w_{\timeVar,l}$ is taken from~\eqref{eq:siChanDef}.
%, $\phi_i\left(\cdot\right)$ is the $i^{\text{th}}$ nonlinear basis function, and the $a_{i,\timeVar}$ are the nonlinear coefficients. 

At a distant node, the transmitter broadcasts a SoI, denoted by $d_{\timeVar}$ with zero mean and power $\expv\left[|d_{\timeVar}|^2\right]=P_d$ over a wireless channel. Similar to the SI channel, the wireless channel is modeled as linear FIR filter with $\firLengthWC$ coefficients
\begin{align}
\label{eq:wChanDef}
\mybold{h}_{\timeVar}=\left[ h_{\timeVar,0}, h_{\timeVar,1}, \ldots, h_{\timeVar,\firLengthWC-1}\right]^T
\end{align}
with the same number of coefficients $L$ as in the case of the linear SI channel for convenience. Hence, at the receiver, the SoI is designated as
\begin{align}
\label{eq:rxSoi}
d^{h}_{\timeVar} = \sum_{l=0}^{\firLengthWC-1}h_{\timeVar,l}d_{\timeVar-l}.
\end{align}
The signal $s_{\timeVar}$ contains all elements at the receiver that are not related to the SI, such as the SoI~\eqref{eq:rxSoi} and the additive noise $n_{\timeVar}$, i.e.,
\begin{align}
s_{\timeVar} &= d^{h}_{\timeVar} + n_{\timeVar} \notag \\
\label{eq:notSIrelatedTime}
&= \sum_{l=0}^{\firLengthWC-1}h_{\timeVar,l}d_{\timeVar-l} + n_{\timeVar}
\end{align} 
and hence we have the overall observed signal at the receiver
\begin{align}
y_{\timeVar} = x_{\text{si},\timeVar} + s_{\timeVar}.
\end{align}
%\subsection{Inpu\timeVar-output model}

\begin{figure}
	\centering
	\newcommand*{\timeBlockLength}{1.75}%
	\begin{tikzpicture}
	[
	timeBlock/.style={draw, drop shadow, fill=white, rectangle, minimum height=2em, minimum width=\timeBlockLength cm},
	dimen/.style={<->,>=latex,thin,every rectangle node/.style={fill=white,midway}},
	dimenright/.style={->,>=latex,thin,every rectangle node/.style={fill=white,right}}
	]
	\pgfmathsetmacro{\dotsGap}{1.25}
	\node[timeBlock] (t1) at (0,0) {\scriptsize $\frameVar R-M+1$};
%	\node[timeBlock] (t2) at ($(t1)+(\timeBlockLength,0)$) {\scriptsize $\frameVar R-M+2$};
	\node[] (c1) at ($(t1)+(\dotsGap,0)$) {$\dots$};
	\node[timeBlock] (t2) at ($(c1)+(\dotsGap,0)$) {\scriptsize $(\frameVar-1)R$};
	\node[timeBlock] (t3) at ($(t2)+(\timeBlockLength,0)$) {\scriptsize $(\frameVar-1)R+1$};
	\node[] (c2) at ($(t3)+(\dotsGap,0)$) {$\dots$};
	\node[timeBlock] (t4) at ($(c2)+(\dotsGap,0)$) {\scriptsize $\frameVar R$};
	
	\draw (t3.south west) -- ++(0,-0.5) coordinate (D1) -- +(0,-5pt);
	\draw (t4.south east) -- ++(0,-0.5) coordinate (D2) -- +(0,-5pt);
	\draw [dimen] (D1) -- (D2) node {$R$};
	\draw (t1.south west) -- ++(0,-1) coordinate (D3) -- +(0,-5pt);
	\draw (t4.south east) -- ++(0,-1) coordinate (D4) -- +(0,-5pt);
	\draw [dimen] (D3) -- (D4) node {$M$};
	
	\draw (t1.north west) -- ++(0,0.25) coordinate (time1) -- +(0,5pt);
	\draw [dimenright] (time1) -- ($(time1)+(1,0)$) node {$\timeVar$};
	
	\end{tikzpicture}
	\caption{Frame of $M$ samples with frame shift $R$ at index $\frameVar$.}
	\label{fig:frames}
\end{figure}
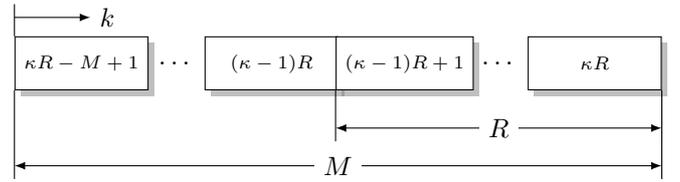

Next, we introduce a vector notation for all signals in time domain. We group consecutive samples in time into frames. This is depicted in \figref{fig:frames}. Suppose that there are frames consisting of $M$ samples, where each frame is indexed by $\frameVar$. We form a new frame on every $R^{\text{th}}$ sample, such that we have 
%a shifting of $R$ samples with $R\leq M$ and 
an overlap of neighboring frames by $\firLengthSI=M-R$ samples.
On this basis, we can express the convolution of the transmitted signal and the SI channel by the overlap-save method. Let
%This requires the condition $M=R+\firLengthSI$ to be fulfilled.
%\begin{align}
%\label{eq:fvec}
%\mybold{f}_\frameVar = \left[ f\left(x'_{\text{iq},\frameVar R-M+1}\right),\ldots,f\left(x'_{\text{iq},\frameVar R}\right) \right]^T
%\end{align}
%and
\begin{align}
\label{eq:nlBasisVec}
\mybold{\phi}_{i,\frameVar} = \left[ \phi_{i}\left(x_{\frameVar R-M+1}\right),\phi_{i}\left(x_{\frameVar R-M+2}\right),\ldots,\phi_{i}\left(x_{\frameVar R}\right) \right]^T
\end{align} 
be the $\frameVar^{\text{th}}$ frame vector for the $i^{\text{th}}$ nonlinear basis function, applied to the input $x_{\timeVar}$. 
Furthermore, let
\begin{align}
\label{eq:nlBasisMatrix}
\mybold{\Phi}_{\frameVar}=\left[ \mybold{\phi}_{0,\frameVar}, \mybold{\phi}_{1,\frameVar},\ldots, \mybold{\phi}_{N-1,\frameVar} \right]
\end{align}
be a $M\times N$ matrix with the signal of all nonlinear basis functions from~\eqref{eq:nlBasisVec}, stacked as column vectors.

% By combining~\eqref{eq:fvec} and~\eqref{eq:phivec}, we have
%\begin{align}
%\mybold{f}_\frameVar = \sum^{N-1}_{i=0} a_{i,\frameVar} \mybold{\phi}''_{i,\frameVar}, 
%\end{align} 
%where we assume that the nonlinear coefficients remain approximately constant within one block such that
%\begin{align*}
%a_{i,\frameVar} := a''_{i,\frameVar R-M+1} \approx \ldots \approx a''_{i,\frameVar R}.
%\end{align*} 
%Furthermore, the SI channel consists of $L=M-R$ non-zero FIR coefficients
%\begin{align}
%\mybold{w}'_{\timeVar}=\left[ w'_{0,\timeVar}, w'_{1,\timeVar}, \ldots, w'_{L-1,\timeVar}\right]^T.
%\end{align}

%Now consider the FIR filter coefficients $\mybold{w}_{\timeVar}$ of the SI channel from~\eqref{eq:siChanDef}. 
We assume the SI channel to be slowly varying and therefore constant within a single frame, similar to block fading. Then, we form a FIR channel vector of size $M\times 1$ by appending $R$ zeros to~\eqref{eq:siChanDef}, and get
\begin{align}
\label{eq:siChanVec}
\mybold{w}_{\frameVar} = \left[ \mybold{w}^T_{\frameVar R}, \mybold{0}^T_{R\times 1} \right]^T.
\end{align}  
%Similarly, we define a $M\times 1$ wireless channel vector % for each frame
%\begin{align}
%\mybold{h}_\frameVar = \left[ \mybold{h}_{\frameVar R}^T, \mybold{0}^T_{R\times 1} \right]^T.
%\end{align}
Similarly, we define a $M\times 1$ frame vector for the desired $d_{\timeVar}$
\begin{align}
\mybold{d}_{\frameVar} &= \left[ d_{\frameVar R-M+1},d_{\frameVar R-M+2},\ldots,d_{\frameVar R} \right]^T
%\mybold{n}'_{\frameVar} &= \left[ n'_{(\frameVar-1)R+1},n'_{(\frameVar-1)R+2},\ldots,n'_{\frameVar R} \right], \\
%\mybold{s}'_{\frameVar} &= \left[ s'_{(\frameVar-1)R+1},s'_{(\frameVar-1)R+2},\ldots,s'_{\frameVar R} \right], \\
%\mybold{y}'_{\frameVar} &= \left[ y'_{(\frameVar-1)R+1},y'_{(\frameVar-1)R+2},\ldots,y'_{\frameVar R} \right].
\end{align}
and a $R\times 1$ frame vector for the receiver noise
\begin{align}
\mybold{n}_{\frameVar}=\left[n_{(\frameVar-1)R+1},n_{(\frameVar-1)R+2},\ldots,n_{\frameVar R} \right]
\end{align}
and we define $\mybold{y}_{\frameVar}$, $\mybold{s}_{\frameVar}$ and $\mybold{x}_{\text{si},\frameVar}$ similarly as $\mybold{n}_{\frameVar}$. 
%Similarly, we define $R\times 1$ frame vectors $\mybold{n}_{\frameVar}$, $\mybold{x}_{\text{si},\frameVar}$ and $\mybold{y}_{\frameVar}$ for the receiver noise $n_{\timeVar}$, the SI signal $x_{\text{si},\timeVar}$, and the overall received signal $y_{\timeVar}$ including SI, respectively.

\subsection{DFT-domain representation}
\label{sec:dft}
Next, we transform selected signals into DFT domain and express the convolutions of~\eqref{eq:nldef} and~\eqref{eq:rxSoi} as multiplications.
Let
\begin{align}
\underline{\mybold{\Phi}}_{\frameVar} &=\mybold{F}_M\mybold{\Phi}_{\frameVar}
% \nonumber\\&
 = \left[ \underline{\mybold{\phi}}_{0,\frameVar},
\label{eq:basisFrameMatrix}
\underline{\mybold{\phi}}_{1,\frameVar},\ldots, \underline{\mybold{\phi}}_{N-1,\frameVar} \right]
\end{align}
denote the $M\times N$ DFT-domain representations of the basis frame, where $\mybold{F}_M$ is the $M\times M$ DFT matrix. Similarly, let
\begin{align}
\label{eq:siChanVecDft}
\underline{\mybold{w}}_{\frameVar} &=\mybold{F}_M\mybold{w}_{\frameVar} 
\end{align}
be the DFT-domain transform of the linear SI channel coefficients from~\eqref{eq:siChanVec}. It is likewise assumed that the nonlinear coefficients $a_{i,\timeVar}$ only vary slowly over time, thus they are approximately constant within a frame. We write
\begin{align}
	\label{eq:nlCoeffsVecEqual}
	\underline{a}_{i,\frameVar} = a_{i,(\frameVar-1)R} = a_{i,(\frameVar-1)R+1} = \ldots = a_{i,\frameVar R}
\end{align}
and group the nonlinear coefficients~\eqref{eq:nlCoeffsVecEqual} in vector form
\begin{align}
	\label{eq:nlCoeffsVec}
\underline{\mybold{a}}_{\frameVar}= \left[\underline{a}_{0,\frameVar},\underline{a}_{1,\frameVar},\ldots,\underline{a}_{N-1,\frameVar}\right]^T.
\end{align}
Now we are equipped to express the SI signal of~\eqref{eq:nldef} in DFT domain. The nonlinear basis frames are given by $\underline{\mybold{\Phi}}_{\frameVar}\underline{\mybold{a}}_{\frameVar}$ based on~\Cref{eq:basisFrameMatrix,eq:nlCoeffsVec}. They are multiplied with the linear SI channel coefficients~\eqref{eq:siChanVecDft}, since the convolution is transformed into a multiplication in DFT domain. Thus, we construct
\begin{align}
\label{eq:obsCompact}
\underline{\mybold{x}}_{\text{si},\frameVar}
&=  \diag\left[\underline{\mybold{\Phi}}_{\frameVar}\underline{\mybold{a}}_{\frameVar}\right] \underline{\mybold{w}}_{\frameVar} \\
\label{eq:obsDetail}
&\hintedrel[eq:xsiDFT1]{=} \left(\sum_{i=0}^{N-1}\underline{a}_{i,\frameVar}\diag\left[ \underline{\mybold{\phi}}_{i,\frameVar}\right]\right) \underline{\mybold{w}}_{\frameVar},
\end{align} 
where we introduce~$(\hintref{eq:xsiDFT1})$ to prepare a form that later on allows to efficiently infer the coefficients $\underline{a}_{i,\frameVar}$ by a linear algorithm. 
The corresponding $R\times 1$ time-domain signal of~\eqref{eq:obsDetail}, in line with~\eqref{eq:nldef}, is
\begin{align}
\mybold{x}_{\text{si},\frameVar} &= \mybold{\Upsilon}^T\mybold{F}^{-1}_M\underline{\mybold{x}}_{\text{si},\frameVar} \\ 
\label{eq:siTime}
&=  \left(\sum_{i=0}^{N-1}\underline{a}_{i,\frameVar}\mybold{\Upsilon}^T\mybold{F}^{-1}_M\diag\left[ \underline{\mybold{\phi}}_{i,\frameVar}\right]\right) \underline{\mybold{w}}_{\frameVar},
\end{align}
where the matrix
\begin{align}
\mybold{\Upsilon}^T = 
\begin{bmatrix}
    \zeros{R}{\firLengthSI} & \eye{R}
\end{bmatrix}
\end{align}
removes the first $\firLengthSI=M-R$ entries from a vector since these entries are contaminated by aliasing. Next, we address the SoI. In DFT domain, we have
\begin{align}
\label{eq:soiDFT}
\underline{\mybold{d}}_{\frameVar}&=\mybold{F}_M\mybold{d}_{\frameVar} \\
\label{eq:wChanDFT}
\underline{\mybold{h}}_{\frameVar}&=\mybold{F}_M\mybold{h}_{\frameVar}
\end{align}
for the SoI and the coefficients of the wireless channel. 
%Next, we formulate the time-domain vector frame for the signal $s_{\timeVar}$ of~\eqref{eq:notSIrelatedTime}. 
Similar to~\eqref{eq:siTime}, we express the convolution of SoI and the wireless channel as multiplication of~\eqref{eq:soiDFT} and~\eqref{eq:wChanDFT} in DFT domain, and then transform it back into time domain
\begin{align}
\label{eq:soiPlusNoiseTime}
\mybold{s}_{\frameVar} &=  \mybold{\Upsilon}^T\mybold{F}^{-1}_M\diag\left[ \underline{\mybold{d}}_{\frameVar}\right] \underline{\mybold{h}}_{\frameVar}+\mybold{n}_{\frameVar}.
\end{align}
% can be expressed as
%\begin{align}
%\label{eq:recvFrame}
%\mybold{y}_{\frameVar}.
%\end{align}

Next, we transform the overall received signal $\mybold{y}_{\frameVar} = \mybold{x}_{\text{si},\frameVar} + \mybold{s}_{\frameVar}$ back into DFT domain by first prepending $\firLengthSI$~zeros and then applying the DFT matrix. Thus we have
\begin{align}
\underline{\mybold{y}}_{\frameVar}&= \mybold{F}_M\mybold{\Upsilon}\mybold{y}_{\frameVar} \nonumber\\
&= \mybold{F}_M\mybold{\Upsilon}\mybold{x}_{\text{si},\frameVar} + \mybold{F}_M\mybold{\Upsilon}\mybold{s}_{\frameVar} \\
\label{eq:obsDFT}
&\hintedrel[eq:obsDFT1]{=} \left(\sum_{i=0}^{N-1} \underline{a}_{i,\frameVar}\underline{\mybold{C}}_{i,\frameVar}\right) \underline{\mybold{w}}_{\frameVar} + \underline{\mybold{s}}_{\frameVar},
\end{align} 
where~$(\hintref{eq:obsDFT1})$ uses~\eqref{eq:siTime} and we employed the definition
\begin{align}
\label{eq:CDef}
 \underline{\mybold{C}}_{i,\frameVar} = \mybold{F}_M\mybold{\Upsilon}\mybold{\Upsilon}^T\mybold{F}^{-1}_M\diag\left[ \underline{\mybold{\phi}}_{i,\frameVar}\right]
\end{align}
and the DFT domain representation
\begin{align}
\label{eq:soiPlusNoiseDft}
\underline{\mybold{s}}_{\frameVar}&=\mybold{F}_M\mybold{\Upsilon}\mybold{s}_{\frameVar}.
\end{align}
The received signal without SI contribution~\eqref{eq:soiPlusNoiseDft} can be alternatively expressed by using~\eqref{eq:soiPlusNoiseTime} by
\begin{align}
\underline{\mybold{s}}_{\frameVar}&= 
%\mybold{F}_M\mybold{\Upsilon}
%\left(
%\mybold{\Upsilon}^T\mybold{F}^{-1}_M\diag\left[ \underline{\mybold{d}}_{\frameVar}\right] \underline{\mybold{h}}_{\frameVar}+\mybold{n}_{\frameVar}
%\right) \notag\\
%\label{eq:soiPlusNoiseDftAlt}
%&\hintedrel[eq:soiPlusNoiseDftAlt1]{=}
\underline{\mybold{d}}^{h}_{\frameVar} + \underline{\mybold{n}}_{\frameVar},
\end{align}
where the term $\underline{\mybold{n}}_{\frameVar}$ denotes the DFT domain representation of the additive receiver noise and
%\begin{align}
%\label{eq:desiredWirelessDef}
%\underline{\mybold{d}}^{h}_{\frameVar}= 
%\mybold{F}_M\mybold{\Upsilon}
%\mybold{\Upsilon}^T\mybold{F}^{-1}_M\diag\left[ \underline{\mybold{d}}_{\frameVar}\right] \underline{\mybold{h}}_{\frameVar}
%\end{align}
$\underline{\mybold{d}}^{h}_{\frameVar}$ is the DFT domain equivalent of the SoI at the receiver.

\subsection{State-space model}
We propose state-space models for both the linear SI channel~\eqref{eq:siChanVecDft} and the nonlinear coefficients~\eqref{eq:nlCoeffsVec}. 
%and define its second-order moment as $N\times N$ matrix by
%\begin{align}
%\underline{\mybold{P}}^{a}_{\frameVar} &= \expv\left[ \underline{\mybold{a}}_{\frameVar}\underline{\mybold{a}}^{H}_{\frameVar} \right] \nonumber\\
%&= 
%\begin{pmatrix}
%	p^a_{0,0,\frameVar} & p^a_{0,1,\frameVar} & \cdots & p^a_{0,N-1,\frameVar} \\
%	p^a_{1,0,\frameVar} & p^a_{1,1,\frameVar} & \cdots & p^a_{1,N-1,\frameVar} \\
%	\vdots  & \vdots  & \ddots & \vdots  \\
%	p^a_{N-1,0,\frameVar} & p^a_{N-1,1,\frameVar} & \cdots & p^a_{N-1,N-1,\frameVar} 
%\end{pmatrix}.
%\end{align}
The time-variant nature of the nonlinear SI channel is specifically modeled by first-order Markov models of the linear SI channel and the nonlinear coefficients, i.e.,
\begin{align}
\label{eq:siChanMarkov}
\underline{\mybold{w}}_{\frameVar} &=
\underline{A}^{w}\underline{\mybold{w}}_{\frameVar-1}+\underline{\mybold{\Delta}}^{w}_{\frameVar},
 \\
\label{eq:nlCoeffsMarkov}
\underline{\mybold{a}}_{\frameVar} &=
\underline{\mybold{A}}^{a}\underline{\mybold{a}}_{\frameVar-1}+\underline{\mybold{\Delta}}^{a}_{\frameVar},
\end{align}
where the parameter $\underline{A}^{w}$ with $0\leq|\underline{A}^{w}|\leq 1$ and the diagonal matrix $\underline{\mybold{A}}^{a}$ denote the transition factor between consecutive channel realizations and nonlinear coefficients over time, respectively. We assume a scalar $\underline{A}^{w}$, since the linear SI channel coefficients are assumed to be similar in transient behavior. On the other hand, the matrix $\underline{\mybold{A}}^{a}$ reflects different degrees of temporal variations among the nonlinear coefficients. The Gaussian system noise variables $\underline{\mybold{\Delta}}^{w}_{\frameVar}\sim\cgauss{\mybold{0},\underline{\psi}^{\Delta w}_{\frameVar}\eye{M}}$ and $\underline{\mybold{\Delta}}^{a}_{\frameVar}\sim\cgauss{\mybold{0},\underline{\mybold{\Psi}}^{\Delta a}_{\frameVar}}$ are independent with variance $\underline{\psi}^{\Delta w}_{\frameVar}$ and covariance matrix $\underline{\mybold{\Psi}}^{\Delta a}_{\frameVar}$. By definition, we assume the top path from~\figref{fig:systemmodel} with index $i=0$ represents the linear component of the SI, thus, without loss of generality, we have $\phi_{0}(x_{\timeVar})=x_{\timeVar}$ and fix $\underline{a}_{0,\frameVar}=1$, $\underline{\Delta}^{a}_{0,\frameVar}=0$ in the following.
%, where the $M\times M$ covariance matrix $\underline{\mybold{\Psi}}^{\Delta w}_{\frameVar}$ and the $N\times N$ covariance matrix $\underline{\mybold{\Psi}}^{\Delta a}_{\frameVar}$ are assumed to be diagonal. 

Let $\underline{\mybold{R}}^{w}_{\frameVar}=\expv\left[ \underline{\mybold{w}}_{\frameVar}\underline{\mybold{w}}^{H}_{\frameVar} \right]$ be the autocorrelation matrix of the random SI channel coefficients,
% and $\underline{\mybold{R}}^{a}_{\frameVar}=\expv\left[ \underline{\mybold{a}}_{\frameVar}\underline{\mybold{a}}^{H}_{\frameVar} \right]$, 
thus, from~\eqref{eq:siChanMarkov}, we have
\begin{align}
\label{eq:siChanCorr}
\tr\left[\underline{\mybold{R}}^{w}_{\frameVar}\right]=\left|\underline{A}^{w}\right|^{2}\tr\left[\underline{\mybold{R}}^{w}_{\frameVar-1}\right]+M\underline{\psi}^{\Delta w}_{\frameVar}.
\end{align}
%By considering the initial condition $\tr\left[\underline{\mybold{R}}^{w}_{0}\right]$ and $\underline{\psi}^{\Delta w}_{0}=0$, we can solve for the recursion of~\eqref{eq:siChanCorr} by
%\begin{align}
%\label{eq:siChanCorrTr}
%\tr\left[\underline{\mybold{P}}^{w}_{\frameVar}\right] =
%\tr\left[\underline{\mybold{R}}^{w}_{0}\right]\left|\underline{A}^{w}\right|^{2\frameVar}+M\frac{\underline{\psi}^{\Delta w}_{\frameVar}}{1-\left|\underline{A}^{w}\right|^{2}},\quad\frameVar\geq 0.
%\end{align}
Apparently, the system model is characterized by a significant number of parameters. Since these parameters are difficult to determine, we reduce their number by using certain relations between them. For instance, from a practical perspective, it is reasonable to assume that the statistical properties of the linear SI channel remain similar over time. Thus, we assume
\begin{align}
\label{eq:siChanCorrTrEqual}
\tr\left[\underline{\mybold{R}}^{w}_{\frameVar}\right] &= \tr\left[\underline{\mybold{R}}^{w}_{\frameVar+1}\right] = \ldots = \tr\left[\underline{\mybold{R}}^{w}_{\infty}\right].
% \\
%\underline{\mybold{R}}^{a}_{\frameVar} &=
%\underline{\mybold{R}}^{a}_{\frameVar-1}, \quad\forall\frameVar.
\end{align} 
%By using~\eqref{eq:siChanCorrTrEqual}, we find the covariance of the system noise from~\eqref{eq:siChanCorrTr} by
%\begin{align}
%\label{eq:siChanSystemNoiseCov}
%\underline{\psi}^{\Delta w}_{\frameVar}=\frac{1}{M}\tr\left[\underline{\mybold{R}}^{w}_{0}\right]
%\left(1-\left|\underline{A}^{w}\right|^{2}\right)\left(1-\left|\underline{A}^{w}\right|^{2\frameVar}\right).
%\end{align} 
and thus we find the covariance of the system noise as
 \begin{align}
 \label{eq:siChanSystemNoiseCovLongTerm}
\underline{\psi}^{\Delta w} = %\lim\limits_{\frameVar\to\infty}\underline{\psi}^{\Delta w}_{\frameVar}=
\frac{1}{M}\tr\left[\underline{\mybold{R}}^{w}_{\infty}\right]
 \left(1-\left|\underline{A}^{w}\right|^{2}\right).
 \end{align} 
The steady-state parameter $\tr\left[\underline{\mybold{R}}^{w}_{\infty}\right]$ determines the power of the linear SI channel components. In practice, this can be obtained from a-priori knowledge, since the relation of SI power level to the SoI is at least approximately known. 

Furthermore, the \textit{coherence time} reflects the variability of the communication channel in time, and therefore is regarded as main characteristic of temporal variations. To best of our knowledge, the coherence time has not been systematically studied for SI channels so far. However, we believe it to be on the same order as the wireless channel, as indicated in some previous work~\cite{src:jain2011practical}. To integrate the notion of a channel coherence time into our framework, we introduce a coherence frame index $\frameVar^{w}_{\text{coh}}$, which denotes the time when the correlation of subsequent channel variables has dropped by half, thus
\begin{align}
\label{eq:channelCoherence}
\left| A^{w} \right|^{\frameVar^{w}_{\text{coh}}} = \frac{1}{2}.
\end{align}

For the zero-mean nonlinear coefficients $\underline{\mybold{a}}_{\frameVar}$, we define the diagonal covariance matrix $\underline{\mybold{R}}^{a}_{\frameVar}= \expv\left[ \underline{\mybold{a}}_{\frameVar}\underline{\mybold{a}}^{H}_{\frameVar} \right]$ with the $i^{\text{th}}$ diagonal element $\underline{p}^{a}_{i,\frameVar}=\expv\left[ \left|\underline{a}_{i,\frameVar}\right|^2 \right]$. From~\eqref{eq:nlCoeffsMarkov}, we can derive
\begin{align}
\label{eq:nlCorr}
\underline{\mybold{R}}^{a}_{\frameVar}=\underline{\mybold{A}}^{a}\underline{\mybold{R}}^{a}_{\frameVar-1}\underline{\mybold{A}}^{a^H}+\underline{\mybold{\Psi}}^{\Delta a}_{\frameVar}.
\end{align}
%since $\underline{\mybold{A}}^{a}$ is diagonal. 
With the elements of $\underline{\mybold{R}}^{a}_{\frameVar}$ approximately constant over time,
\begin{align}
\underline{\mybold{R}}^{a}_{\frameVar}=\underline{\mybold{R}}^{a}_{\frameVar+1}=\ldots=\underline{\mybold{R}}^{a}_{\infty},
\end{align}
the system noise covariance matrix is expressed by
\begin{align}
%\underline{\mybold{\Psi}}^{\Delta a}_{\frameVar} = \left(\eye{N}- \underline{\mybold{A}}^{a}\underline{\mybold{A}}^{a^H}\right)\underline{\mybold{P}}^{a}_{\infty}.
\underline{\mybold{\Psi}}^{\Delta a}_{\frameVar} =
\underline{\mybold{R}}^{a}_{\infty}
 \left(\mybold{I}- \underline{\mybold{A}}^{a}\underline{\mybold{A}}^{a^H}\right),
\end{align}
since $\underline{\mybold{A}}^{a}$ is diagonal.
Similar to~\eqref{eq:channelCoherence}, we define a coherence time $\frameVar^{a}_{i,\text{coh}}$ for the $i^{\text{th}}$ nonlinear coefficient by
\begin{align}
\label{eq:nlCoherence}
\left| A_i^{a} \right|^{\frameVar^{a}_{i,\text{coh}}} = \frac{1}{2},
\end{align}
where $A_i^{a}$ denotes the $i^{\text{th}}$ element on the main diagonal of $\underline{\mybold{A}}^{a}$.
%, one can follow similar steps as in~\cref{eq:siChanCorr,eq:siChanCorrTrEqual,eq:siChanSystemNoiseCovLongTerm}\todo{Individual equation for each $a$}.
%eq:siChanCorrTr,eq:siChanCorrTrEqual,eq:siChanSystemNoiseCov,eq:siChanSystemNoiseCovLongTerm,eq:channelCoherence}.

\section{State-space nonlinear adaptive algorithm}
\label{sec:algorithm}
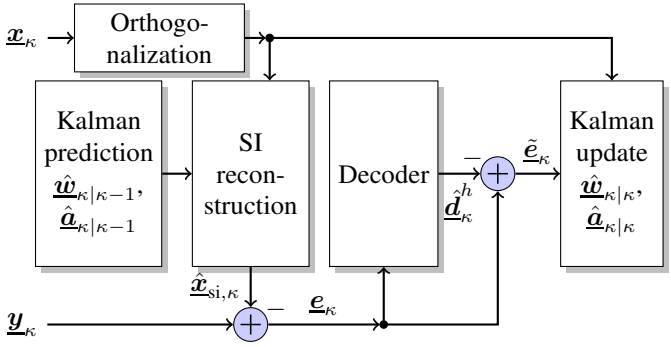
\begin{figure}
	\begin{tikzpicture}[
	sysBlock/.style={draw, fill=white, rectangle, minimum height=2em, minimum width=3.5em},
	nlBlock/.style={draw, fill=white, rectangle, minimum height=2em, minimum width=4em},
	shadowBlock/.style={draw, drop shadow, fill=white, rectangle},
	triangle/.style = {draw, regular polygon, regular polygon sides=3 },
	%node rotated/.style = {rotate=90},
	border rotated/.style = {shape border rotate=90}
	]
	
	\node (inTx) {$\underline{\mybold{x}}_{\frameVar}$};
			
	\node[shadowBlock, align=center, minimum height=7em] (kalpred) at ($(inTx) +(1,-1.8)$) {Kalman\\prediction\\$\underline{\hat{\mybold{w}}}_{\frameVar|\frameVar-1}$,\\$\underline{\hat{\mybold{a}}}_{\frameVar|\frameVar-1}$};
	\node (inRx) at ($(kalpred)+(-1,-2)$) {$\underline{\mybold{y}}_{\frameVar}$};
		
	\node[shadowBlock, align=center, text width=1.3cm, minimum height=7em] (sirec) at ($(kalpred) + (2,0)$) {SI\\recon\-struc\-tion};
	
	\node[shadowBlock, align=center, text width=2cm] (ortho) at ($(inTx)+(1.8,0)$) {Ortho\-go\-n\-ali\-za\-tion};
	\myadd{add1}{$(inRx-|sirec)$}
	\node  at ($(add1.east)+(0.1,0.2)$) {$-$};
	
	\node[shadowBlock, align=center, minimum height=7em] (dec) at ($(sirec)+(1.75,0)$) {Decoder};
	\myadd{add2}{$(dec)+(1.5,0)$}
	\node  at ($(add2.west)+(-0.1,0.2)$) {$-$};
	
	\node[shadowBlock, align=center, minimum height=7em] (kalupd) at ($(add2)+(1.5,0)$) {Kalman\\update\\$\underline{\hat{\mybold{w}}}_{\frameVar|\frameVar}$,\\$\underline{\hat{\mybold{a}}}_{\frameVar|\frameVar}$};
	
	\node[dot] (dot1) at ($(sirec|-ortho)+(0.25,0)$) {};
	\node[dot] (dot2) at ($(dec|-inRx)$) {};
	
	\draw[thick,->] (inTx) to (ortho);
	\draw[thick,->] (ortho) -- (dot1);
	\draw[thick,->] (dot1) -- (sirec.north-|dot1);
	\draw[thick,->] (dot1) -| (kalupd);
	
	\draw[thick,->] (kalpred) -- (sirec);
	\draw[thick,->] (inRx) -- (add1);
	\draw[thick,->] (sirec) -- node[left] {$\underline{\hat{\mybold{x}}}_{\text{si},\frameVar}$} (add1);
	\draw[thick,->] (add1) -- node[above] {$\underline{\mybold{e}}_{\frameVar}$} (dot2);
	\draw[thick,->] (dot2) -- (dec);
	\draw[thick,->] (dot2) -| (add2);
	
	\draw[thick,->] (dec) -- node[below] {$\underline{\hat{\mybold{d}}}^h_{\frameVar}$} (add2);
	\draw[thick,->] (add2) -- node[above] {$\underline{\tilde{\mybold{e}}}_{\frameVar}$} (kalupd);
\end{tikzpicture}
\caption{Proposed nonlinear SI estimation and cancellation.}
\label{fig:algorithm}
\end{figure}

Consider the proposed SI estimation and cancellation algorithm depicted in~\figref{fig:algorithm}.
All signals are represented here in DFT domain. The adaptive algorithm has access to the transmitted signal $\underline{\mybold{x}}_{\frameVar}$. Initially, the input is subject to an orthogonalization process, where the nonlinear contributions are transformed to uncorrelated equivalents. Then, the adaptation is performed in an iterative process over time. At each step $\frameVar$, we first acquire a predicted Kalman estimate of the linear SI channel and the nonlinear coefficients.
%(to be defined later).  
We intend to cancel as much SI as possible, thus the algorithm produces a reconstruction $\underline{\hat{\mybold{x}}}_{\text{si},\frameVar}$ of the SI signal. After subtracting~$\underline{\hat{\mybold{x}}}_{\text{si},\frameVar}$  from the received signal $\underline{\mybold{y}}_{\frameVar}$, the algorithm outputs an error signal $\underline{\mybold{e}}_{\frameVar}$. This error signal is then provided to the decoder which produces a reconstructed SoI $\underline{\hat{\mybold{d}}}^h_{\frameVar}$. Finally, after $\underline{\hat{\mybold{d}}}^h_{\frameVar}$ is subtracted from the error signal $\underline{\mybold{e}}_{\frameVar}$, the residual error $\underline{\tilde{\mybold{e}}}_{\frameVar}$ supports the Kalman update estimations for both the linear SI channel and the nonlinear coefficients. 
% is fed into a postfilter. Since the residual signal contains the SI estimation error, a postfilter can improve the recovery of the desired signal from the residual SI, and outputs the signal~$\underline{\tilde{\mybold{s}}}_{\frameVar}$. 

%An approach based on the full state-space model of Subsection~\ref{sec:fullss} has been proposed in the literature already.
\subsection{Orthogonalization}
\label{sec:ortho}
The nonlinear basis functions $\phi_i(\cdot)$ do not necessarily represent an orthogonal basis for the SI signal. Therefore, we intend to change the basis by a matrix transformation. There are many approaches available that achieve this orthogonalization, such as eigenvalue decomposition~\cite{src:korpi2015adaptive}, Cholesky factorization~\cite{src:emara2017nonlinear} or Gram-Schmidt method~\cite{src:proakis2007digital}.
%~\cite{src:emara2017nonlinear}. 
%To simplify the analysis, we assume the input process $x_{\timeVar}$ to approximately behave like white noise. 
We describe the statistical connection between different basis functions by a time-invariant $N\times N$ autocorrelation matrix. Using~\eqref{eq:basisFrameMatrix}, we have
%\begin{align*}
%	\mybold{R}_{\phi} &= \expv\Big[ \left[\phi_0(\cdot),\ldots,\phi_{N-1}(\cdot)\right] \left[\phi_0(\cdot),\ldots,\phi_{N-1}(\cdot)\right]^H\Big]. 
%\end{align*}
\begin{align*}
	\mybold{R}^{\Phi} &= \expv\Big[ \underline{\mybold{\Phi}}^T_{\frameVar} \underline{\mybold{\Phi}}^*_{\frameVar}\Big]. 
\end{align*}
%~\todo{Covariance notation}
%We apply the Cholesky decomposition to $\mybold{R}_\phi$ and get
%\begin{align}
%	\label{eq:cholesky}
%	\mybold{R}_{\phi} &= \mybold{L}^H\mybold{L},
%\end{align}
%where $\mybold{L}$ denotes the $N\times N$ upper-triangular factorization matrix. Thus, we can apply the Cholesky decomposition to~\eqref{eq:basisFrameMatrix} and obtain
Next, we choose a $N\times N$ transform matrix $\mybold{G}$, which is not necessarily unitary, but full-rank, and such that $\mybold{G}\mybold{R}^{\Phi}\mybold{G}^H$ is diagonal.
%\begin{align}
%\label{eq:condDiag}
%\mybold{G}\mybold{R}^{\Phi}\mybold{G}^H = \mybold{D}
%\end{align}
%with a diagonal matrix $\mybold{D}$. 
We apply the transformation to each basis frame by
\begin{align}
	\underline{\tilde{\mybold{\Phi}}}_{\frameVar} &= \underline{\mybold{\Phi}}_{\frameVar} \mybold{G}^T \nonumber\\
	\label{eq:orthoPhi}
	&= \left[ \underline{\tilde{\mybold{\phi}}}_{0,\frameVar}, \underline{\tilde{\mybold{\phi}}}_{1,\frameVar},\ldots, \underline{\tilde{\mybold{\phi}}}_{N-1,\frameVar} \right].
\end{align}
Next, starting from~\eqref{eq:obsCompact}, we have
\begin{align}
\underline{\mybold{x}}_{\text{si},\frameVar}
&=  \diag\left[\underline{\mybold{\Phi}}_{\frameVar}\mybold{G}^{T}\left(\mybold{G}^T\right)^{-1}\underline{\mybold{a}}_{\frameVar}\right] \underline{\mybold{w}}_{\frameVar} \nonumber\\
&\hintedrel[eq:obsOrtho1]{=} \diag\left[\underline{\tilde{\mybold{\Phi}}}_{\frameVar}\underline{\tilde{\mybold{a}}}_{\frameVar}\right]\underline{\mybold{w}}_{\frameVar} \nonumber\\
&= \diag\left[\underline{\tilde{\mybold{\Phi}}}_{\frameVar}\underline{\tilde{\mybold{a}}}_{\frameVar}\frac{1}{\underline{\tilde{a}}_{0,\frameVar}}\right]\underline{\tilde{a}}_{0,\frameVar}\underline{\mybold{w}}_{\frameVar} \nonumber\\
&\hintedrel[eq:obsOrtho2]{=}  \left(\sum_{i=0}^{N-1}\underline{\check{a}}_{i,\frameVar}\diag\left[ \underline{\tilde{\mybold{\phi}}}_{i,\frameVar}\right]\right) \underline{\tilde{\mybold{w}}}_{\frameVar},
\end{align}
where~$(\hintref{eq:obsOrtho1})$ uses~\eqref{eq:orthoPhi} and $\underline{\tilde{\mybold{a}}}_{\frameVar}=\left(\mybold{G}^T\right)^{-1}\underline{\mybold{a}}_{\frameVar}$, and~$(\hintref{eq:obsOrtho2})$ applies normalization with $\underline{\check{a}}_{i,\frameVar}=\underline{\tilde{a}}_{i,\frameVar}/\underline{\tilde{a}}_{0,\frameVar}$ and $\underline{\tilde{\mybold{w}}}_{\frameVar}=\underline{\tilde{a}}_{0,\frameVar}\underline{\mybold{w}}_{\frameVar}$. Apparently, the change of basis has an effect on the outcome of both the estimated linear SI channel and the nonlinear coefficients as they represent the inner state of the system. However, the overall SI reconstruction is left unchanged. In the following, for the sake of simplicity, we drop the ``tilde'', but orthogonalization is always applied unless noted otherwise.   
%Accordingly, the FIR channel coefficients from~\eqref{eq:W} are changing due to the orthogonalization of the SI signal, thus we write 
%\begin{align}
%	\label{eq:wOrthoMatrix}
%	\tilde{\mybold{W}}_{\frameVar} = \left[ \mybold{w}_{0,\frameVar}, \mybold{w}_{1,\frameVar},\ldots, \mybold{w}_{N-1,\frameVar} \right]
%\end{align}
%instead of $\mybold{W}_{\frameVar}$ in the following.  

\subsection{Predictions}
\label{sec:predictions}
Consider the state equations for the linear SI channel~\eqref{eq:siChanMarkov} and the nonlinear coefficients~\eqref{eq:nlCoeffsMarkov}. We intend to derive estimators for $\underline{\mybold{w}}_{\frameVar}$ and $\underline{\mybold{a}}_{\frameVar}$. Let 
\begin{align}
\label{eq:siChanEstimatorPred}
\underline{\mybold{w}}_{\frameVar} &= \hat{\underline{\mybold{w}}}_{\frameVar|\frameVar-1} + \underline{\mybold{n}}^w_{\frameVar|\frameVar-1}\\
\label{eq:nlCoeffsEstimatorPred}
\underline{\mybold{a}}_{\frameVar} &= \hat{\underline{\mybold{a}}}_{\frameVar|\frameVar-1} + \underline{\mybold{n}}^a_{\frameVar|\frameVar-1},
\end{align}
where $\hat{\underline{\mybold{w}}}_{\frameVar|\frameVar-1}$ and $\hat{\underline{\mybold{a}}}_{\frameVar|\frameVar-1}$ are estimators for the linear SI channel and the nonlinear coefficients given all past observations $\left(\underline{\mybold{y}}_{0},\ldots,  \underline{\mybold{y}}_{\frameVar-2}, \underline{\mybold{y}}_{\frameVar-1}\right)$, respectively. The zero-mean a-priori errors $\underline{\mybold{n}}^w_{\frameVar|\frameVar-1}$ and $\underline{\mybold{n}}^a_{\frameVar|\frameVar-1}$ have the covariances $\underline{\mybold{P}}^{w}_{\frameVar|\frameVar-1}$ and $\underline{\mybold{P}}^{a}_{\frameVar|\frameVar-1}$, respectively. 
%Similar to~\Cref{eq:siChanEstimatorPred,eq:nlCoeffsEstimatorPred}, let 
%\begin{align}
%\underline{\mybold{w}}_{\frameVar} &= \hat{\underline{\mybold{w}}}_{\frameVar|\frameVar} + \underline{\mybold{n}}^w_{\frameVar|\frameVar} \\
%\underline{\mybold{a}}_{\frameVar} &= \hat{\underline{\mybold{a}}}_{\frameVar|\frameVar} + \underline{\mybold{n}}^a_{\frameVar|\frameVar},
%\end{align}
Similarly, we define $\hat{\underline{\mybold{w}}}_{\frameVar|\frameVar}$ and $\hat{\underline{\mybold{a}}}_{\frameVar|\frameVar}$ to be estimators for the linear SI channel and the nonlinear coefficients given the present and all past observations $\left(\underline{\mybold{y}}_{0},\ldots,  \underline{\mybold{y}}_{\frameVar-1}, \underline{\mybold{y}}_{\frameVar}\right)$, respectively. The zero-mean a-posteriori errors %$\underline{\mybold{n}}^w_{\frameVar|\frameVar}$ and $\underline{\mybold{n}}^a_{\frameVar|\frameVar}$ 
have the covariances $\underline{\mybold{P}}^{w}_{\frameVar|\frameVar}$ and $\underline{\mybold{P}}^{a}_{\frameVar|\frameVar}$, respectively. Furthermore, consider the following assumptions:
\begin{enumerate}
\item The linear SI channel $\underline{\mybold{w}}_{\frameVar}$ and the nonlinear coefficients $\underline{\mybold{a}}_{\frameVar}$ are independent.
\item The joint probability density functions of both the estimators 
$\hat{\underline{\mybold{w}}}_{\frameVar|\frameVar-1}$,
$\hat{\underline{\mybold{a}}}_{\frameVar|\frameVar-1}$,
$\hat{\underline{\mybold{w}}}_{\frameVar|\frameVar}$,
$\hat{\underline{\mybold{a}}}_{\frameVar|\frameVar}$ and the a-priori, a-posterori errors
%$\underline{\mybold{n}}^w_{\frameVar|\frameVar-1}$,
%$\underline{\mybold{n}}^a_{\frameVar|\frameVar-1}$, 
%$\underline{\mybold{n}}^w_{\frameVar|\frameVar}$,
%$\underline{\mybold{n}}^a_{\frameVar|\frameVar}$ 
are circular-symmetric, complex Gaussian distributions. 
\item The a-priori, a-posteriori errors 
%$\underline{\mybold{n}}^w_{\frameVar|\frameVar-1}$,
%$\underline{\mybold{n}}^a_{\frameVar|\frameVar-1}$,
%$\underline{\mybold{n}}^w_{\frameVar|\frameVar}$, 
%$\underline{\mybold{n}}^a_{\frameVar|\frameVar}$ 
are independent.
\end{enumerate}
The adaptive algorithms for the estimation of the linear SI channel and of the nonlinear coefficients are derived in the following. We show the overall adaptive algorithm in~\algref{alg:overall}. Following the lines of~\cite{src:scharf1991statistical}, the predictions of the estimates $\hat{\underline{\mybold{w}}}_{\frameVar|\frameVar-1}$, $\hat{\underline{\mybold{a}}}_{\frameVar|\frameVar-1}$ and the estimation error covariances $\underline{\mybold{P}}^{w}_{\frameVar|\frameVar-1}$, $\underline{\mybold{P}}^{a}_{\frameVar|\frameVar-1}$ are computed first. Next, the information encoded in the SoI is retrieved and its contribution is removed from the received signal. This is explained in more detail in Subsection~\ref{sec:decoding}. Finally, the update steps of the linear SI channel and the nonlinear coefficients are computed. This is presented in Subsections~\ref{sec:estLin} and~\ref{sec:estNl}, respectively.
\begin{algorithm}
\caption{Iterative estimation of linear SI channel~$\underline{{\mybold{w}}}_{\frameVar}$ and nonlinear coefficients~$\underline{\mybold{a}}_{\frameVar}$}
\label{alg:overall}
\begin{algorithmic}
\REQUIRE Initialization of $\underline{\hat{\mybold{w}}}_{0|0}$, $\underline{\mybold{P}}^{w}_{0|0}$, $\underline{\hat{\mybold{a}}}_{0|0}$ and $\underline{\mybold{P}}^{a}_{0|0}$
\REPEAT
%$\underline{\mybold{P}}^{w}_{0|-1} \neq \mybold{0}_{M\times M}$
\STATE $\frameVar \leftarrow \frameVar + 1$
\PREDICT{linear SI channel estimate}
\STATE $\underline{\hat{\mybold{w}}}_{\frameVar|\frameVar-1}=\underline{A}^{w} \underline{\hat{\mybold{w}}}_{\frameVar-1|\frameVar-1}$
\ENDBLOCK
\PREDICT{linear SI channel estimation error covariance}
\STATE
$\underline{\mybold{P}}^{w}_{\frameVar|\frameVar-1} = \left|\underline{A}^{w}\right|^{2}\underline{\mybold{P}}^{w}_{\frameVar-1|\frameVar-1}
%\underline{\mybold{A}}^{w^H}_{\frameVar}
+\underline{\mybold{\Psi}}^{\Delta w}_{\frameVar}$
\ENDBLOCK
\PREDICT{coefficients estimate}
\STATE $\underline{\hat{\mybold{a}}}_{\frameVar|\frameVar-1}=\underline{\mybold{A}}^{a} \underline{\hat{\mybold{a}}}_{\frameVar-1|\frameVar-1}$
\ENDBLOCK
\PREDICT{coefficients estimation error covariance}
\STATE
$\underline{\mybold{P}}^{a}_{\frameVar|\frameVar-1} = \underline{\mybold{A}}^{a}\underline{\mybold{P}}^{a}_{\frameVar-1|\frameVar-1}\underline{\mybold{A}}^{a^H}
+\underline{\mybold{\Psi}}^{\Delta a}_{\frameVar}$
\ENDBLOCK
\STATE{\textbf{Decode}~the information from the SoI and subtract its contribution from the received signal, see Section~\ref{sec:decoding}}
\UPDATE{linear SI channel estimate from prediction $\underline{\hat{\mybold{w}}}_{\frameVar|\frameVar-1}$, see~\algref{alg:linearChannel} in Subsection~\ref{sec:estLin}}
\STATE Requires all predictions and provides $\underline{\hat{\mybold{w}}}_{\frameVar|\frameVar}$ and $\underline{\mybold{P}}^{w}_{\frameVar|\frameVar}$  
\ENDBLOCK
\UPDATE{nonlinear coefficients estimate from prediction $\underline{\mybold{P}}^{a}_{\frameVar|\frameVar-1}$, see~\algref{alg:nonlinearCoeffs} in Subsection~\ref{sec:estNl}}
\STATE Requires all predictions, the updated $\underline{\hat{\mybold{w}}}_{\frameVar|\frameVar}$ and $\underline{\mybold{P}}^{w}_{\frameVar|\frameVar}$, and provides $\underline{\hat{\mybold{a}}}_{\frameVar|\frameVar}$ and $\underline{\mybold{P}}^{a}_{\frameVar|\frameVar}$ 
\ENDBLOCK
\UNTIL{forever}
\end{algorithmic}
\end{algorithm}

\subsection{Decoding}
\label{sec:decoding}
After predicting the estimates of both the linear SI channel and the nonlinear coefficients, the SI contribution can be reconstructed based on~\eqref{eq:obsDFT} for the current time frame $\frameVar$ by
\begin{align}
\label{eq:reconstructedSI}
\underline{\hat{\mybold{x}}}_{\text{si},\frameVar}= \sum_{i=0}^{N-1}\underline{\hat{a}}_{i,\frameVar|\frameVar-1}\underline{\mybold{C}}_{i,\frameVar}\underline{\hat{\mybold{w}}}_{\frameVar|\frameVar-1}.
\end{align}
We define the error signal
\begin{align}
\label{eq:errorSignal}
\underline{\mybold{e}}_{\frameVar} &= \underline{\mybold{y}}_{\frameVar} - \underline{\hat{\mybold{x}}}_{\text{si},\frameVar}.
\end{align}  
At this point the error signal~\eqref{eq:errorSignal} comprises many sources of observation noise, such as the residual SI, the SoI and independent additive noise. Large observation noise generally reduces the convergence speed or increases the misalignment of adaptive algorithms~\cite{Haykin2002}, but here we can reduce its contribution. Unlike in acoustic echo cancellation, where the acoustic signal is modeled as a random process, the SoI in wireless communication contains structure from encoded information. By decoding the information from the SoI and subtracting the corresponding SoI signal from~\eqref{eq:errorSignal}, we have better conditioning for subsequent updates of the SI path. 

Let $\underline{\hat{\mybold{d}}}^{h}_{\frameVar}$ be the SoI generated from decoded information in DFT domain. Then, 
%using the error signal~\eqref{eq:errorSignal}, 
we have the residual error
\begin{align}
\label{eq:residualError}
\underline{\tilde{\mybold{e}}}_{\frameVar} = \underline{\mybold{y}}_{\frameVar} - \underline{\hat{\mybold{x}}}_{\text{si},\frameVar} - \underline{\hat{\mybold{d}}}^{h}_{\frameVar}.
\end{align} 
%The residual error~\eqref{eq:residualError} 
It comprises a potential decoding error, the residual SI and independent noise. 
%If the decoding step is reliable without any errors, then $\underline{\tilde{\mybold{e}}}_{\frameVar}$ contains the residual SI only.  
 
\subsection{Update estimation of linear SI channel}
\label{sec:estLin}
During the $\frameVar^{\text{th}}$~step of adaption, we need to know the estimate of the nonlinear coefficients from the previous step. Therefore, in the following, we assume that the estimates $\underline{\hat{a}}_{i,\frameVar|\frameVar-1}$ are given. Recall the observation equation~\eqref{eq:obsDFT}. After the decoding, we subtract the reconstructed SoI and have 
\begin{align}
\underline{\tilde{\mybold{y}}}_{\frameVar} &= \underline{\mybold{y}}_{\frameVar} - \underline{\hat{\mybold{d}}}^{h}_{\frameVar} \notag \\
\label{eq:obsDec}
&\hintedrel[eq:y1]{=} \left(\sum_{i=0}^{N-1} \underline{a}_{i,\frameVar}\underline{\mybold{C}}_{i,\frameVar}\right) \underline{\mybold{w}}_{\frameVar} + \underline{\tilde{\mybold{s}}}_{\frameVar} \\
&\hintedrel[eq:y2]{=}
\left(\sum_{i=0}^{N-1} \underline{\hat{a}}_{i,\frameVar|\frameVar-1}\underline{\mybold{C}}_{i,\frameVar}\right) \underline{\mybold{w}}_{\frameVar} + \underline{\tilde{\mybold{s}}}^{w1}_{\frameVar},
\label{eq:obsLinear}
\end{align} 
where~$(\hintref{eq:y1})$ introduces the noise term $\underline{\tilde{\mybold{s}}}_{\frameVar} = \underline{\mybold{s}}_{\frameVar} - \underline{\hat{\mybold{d}}}^{h}_{\frameVar}$
%\begin{align*}
%\underline{\tilde{\mybold{s}}}_{\frameVar} = \underline{\mybold{s}}_{\frameVar} - \underline{\hat{\mybold{d}}}^{h}_{\frameVar}
%\end{align*}
comprising the residual SI and additive noise with covariance matrix $\underline{\mybold{\Psi}}^{\tilde{s}}_{\frameVar}$. We assume that $\underline{\mybold{\Psi}}^{\tilde{s}}_{\frameVar}$ can be characterized from a-priori information on the noise level. The term~$(\hintref{eq:y2})$ is due to both~\eqref{eq:nlCoeffsEstimatorPred} and the zero-mean augmented noise term
\begin{align}
\label{eq:augNoiseLinear}
\underline{\tilde{\mybold{s}}}^{w1}_{\frameVar} = \left(\sum_{i=0}^{N-1} \underline{n}^{a}_{i,\frameVar|\frameVar-1}\underline{\mybold{C}}_{i,\frameVar}\right) \underline{\mybold{w}}_{\frameVar} + \underline{\tilde{\mybold{s}}}_{\frameVar}.
\end{align}
The covariance matrix of the augmented noise $\underline{\tilde{\mybold{s}}}^{w1}_{\frameVar}$ is 
\begin{align}
\underline{\mybold{\Psi}}^{\tilde{s}^{w1}}_{\frameVar} 
&= \expv\left[ \underline{\tilde{\mybold{s}}}^{w1}_{\frameVar}\underline{\tilde{\mybold{s}}}^{w1^H}_{\frameVar} \right] \nonumber \\
%\sum_{i=1}^{N-1}\sum_{j=1}^{N-1}
%\expv\left[ \underline{a}_{i,\frameVar}\underline{a}^*_{j,\frameVar} \right]
%\underline{\mybold{C}}_{i,\frameVar}
%\expv\left[ \underline{\mybold{w}}_{\frameVar}\underline{\mybold{w}}^H_{\frameVar} \right]
%\underline{\mybold{C}}^{H}_{j,\frameVar}
%+\underline{\mybold{\Psi}}^s_{\frameVar} \nonumber\\
&= 
\sum_{i=0}^{N-1}\underline{p}^{a}_{i,\frameVar|\frameVar-1}
\underline{\mybold{C}}_{i,\frameVar}
%\expv\left[ \underline{\mybold{w}}_{\frameVar}\underline{\mybold{w}}^H_{\frameVar} \right]
\underline{\mybold{R}}^{w}_{\frameVar}
\underline{\mybold{C}}^{H}_{i,\frameVar}
+\underline{\mybold{\Psi}}^{\tilde{s}}_{\frameVar}, 
\label{eq:covAugNoiseLinear}
\end{align}
%and~$(\hintref{eq:psiSPrime2})$ is given by 
%\begin{align}
%\underline{\mybold{C}}^{\hat{w}}_{\frameVar|\frameVar-1} = \left[
%\underline{\mybold{C}}_{0,\frameVar}\underline{\hat{\mybold{w}}}_{\frameVar|\frameVar-1},
%\ldots,
%\underline{\mybold{C}}_{N-1,\frameVar}\underline{\hat{\mybold{w}}}_{\frameVar|\frameVar-1}
%\right].
%\label{eq:cwTauTauM1}
%\end{align}
where we have $\underline{p}^{a}_{i,\frameVar|\frameVar-1}=\expv\left[ |\underline{n}^{a}_{i,\frameVar|\frameVar-1}|^2 \right]$ and $\underline{\mybold{R}}^{w}_{\frameVar}=\expv\left[\underline{\mybold{w}}_{\frameVar}\underline{\mybold{w}}^H_{\frameVar}\right]$. Due to the product term of the nonlinear estimation error and the channel vector, the overall term~\eqref{eq:augNoiseLinear} is non-Gaussian. As a consequence, the derivation of exact Kalman filter equations is not possible, since it requires jointly Gaussian distributions within the state-space model. Therefore, in the following, we approximate~\eqref{eq:augNoiseLinear} by an independent Gaussian random vector $\underline{\tilde{\mybold{s}}}^{w2}_{\frameVar}$, which has the same second-order moment as $\underline{\tilde{\mybold{s}}}^{w1}_{\frameVar}$. 
%This is deduced from a concept in information theory, where the capacity of a channel with additive noise can be always lower bounded by that of an equivalent Gaussian channel, i.e., the Gaussian noise serves as the worst-case additive noise~\cite{src:shomorony2012Is}. 
Thus, the modified observation vector is
\begin{align}
\underline{\tilde{\mybold{y}}}_{\frameVar}
&\approx 
\left(\sum_{i=0}^{N-1} \underline{\hat{a}}_{i,\frameVar|\frameVar-1}\underline{\mybold{C}}_{i,\frameVar}\right) \underline{\mybold{w}}_{\frameVar} + \underline{\tilde{\mybold{s}}}^{w2}_{\frameVar} \nonumber\\
&=
\underline{\mybold{C}}^{\hat{a}}_{\frameVar|\frameVar-1} \underline{\mybold{w}}_{\frameVar} + \underline{\tilde{\mybold{s}}}^{w2}_{\frameVar}
\label{eq:obsLinearGauss}
\end{align}
with 
\begin{align}
\underline{\mybold{C}}^{\hat{a}}_{\frameVar|\frameVar-1} = \sum_{i=0}^{N-1} \underline{\hat{a}}_{i,\frameVar|\frameVar-1}\underline{\mybold{C}}_{i,\frameVar}.
\label{eq:bwTau}
\end{align}
We are now prepared to provide the Kalman gain and update equations~\cite{src:scharf1991statistical}, based on the observation equation~\eqref{eq:obsLinearGauss}, and by using~\eqref{eq:covAugNoiseLinear} and~\eqref{eq:bwTau}, and the definitions of error covariance matrices from Section~\ref{sec:predictions}. The computations for the $\frameVar^{\text{th}}$ step are depicted in~\algref{alg:linearChannel}. 
\begin{algorithm}
\caption{Obtain linear SI channel estimate at step $\frameVar$}
\label{alg:linearChannel}
\begin{algorithmic}
\REQUIRE known $\underline{\hat{\mybold{w}}}_{\frameVar|\frameVar-1}$, $\underline{\mybold{P}}^{w}_{\frameVar|\frameVar-1}$, $\underline{\hat{\mybold{a}}}_{\frameVar|\frameVar-1}$ and $\underline{\mybold{P}}^{a}_{\frameVar|\frameVar-1}$
%$\underline{\mybold{P}}^{w}_{0|-1} \neq \mybold{0}_{M\times M}$
\GAIN{}
\STATE
%\begin{flalign*}
$\underline{\mybold{K}}^{w}_{\frameVar} =
\underline{\mybold{P}}^{w}_{\frameVar|\frameVar-1}
\underline{\mybold{C}}^{\hat{a}^H}_{\frameVar|\frameVar-1}\cdot$\newline
$\phantom{\underline{\mybold{K}}^{w}_{\frameVar} =
\underline{\mybold{P}}^{w}_{\frameVar|\frameVar-1}
}
\left(
\underline{\mybold{C}}^{\hat{a}}_{\frameVar|\frameVar-1}
\underline{\mybold{P}}^{w}_{\frameVar|\frameVar-1}
\underline{\mybold{C}}^{\hat{a}^H}_{\frameVar|\frameVar-1}+
\underline{\mybold{\Psi}}^{\tilde{s}^{w2}}_{\frameVar}
\right)^{-1} 
$
\ENDBLOCK
\UPDATE{linear SI channel estimation}
\STATE $\underline{\hat{\mybold{w}}}_{\frameVar|\frameVar}=
\underline{\hat{\mybold{w}}}_{\frameVar|\frameVar-1}+
\underline{\mybold{K}}^{w}_{\frameVar}
\left(
\underline{\tilde{\mybold{y}}}_{\frameVar} - \underline{\mybold{C}}^{\hat{a}}_{\frameVar|\frameVar-1}\underline{\hat{\mybold{w}}}_{\frameVar|\frameVar-1}
\right)$
\ENDBLOCK
\UPDATE{linear SI channel estimation error covariance}
\STATE
$\underline{\mybold{P}}^{w}_{\frameVar|\frameVar} = 
\left(
\eye{M}-\underline{\mybold{K}}^{w}_{\frameVar}\underline{\mybold{C}}^{\hat{a}}_{\frameVar|\frameVar-1}
\right)
\underline{\mybold{P}}^{w}_{\frameVar|\frameVar-1}$
\ENDBLOCK
\end{algorithmic}
\end{algorithm}

\subsection{Update estimation of nonlinear coefficients}
\label{sec:estNl}
We first provide an alternative expression for the observation~\eqref{eq:obsDec} at the receiver. The adaptation with respect to the nonlinear coefficients~$\underline{\mybold{a}}_{\frameVar}$ requires another linear form of the observation. Thus, we express~\eqref{eq:obsDec} alternatively by using~\eqref{eq:nlCoeffsVec}
\begin{align}
\label{eq:obsAlt}
%\underline{\tilde{\mybold{y}}}_{\frameVar}&= \left(\sum_{i=0}^{N-1} \underline{a}_{i,\frameVar}\underline{\mybold{C}}_{i,\frameVar}\right) \underline{\mybold{w}}_{\frameVar} + \underline{\tilde{\mybold{s}}}_{\frameVar} \\
%&= \sum_{i=0}^{N-1} \underline{\mybold{c}}^{w}_{i,\frameVar} \underline{a}_{i,\frameVar} + \underline{\tilde{\mybold{s}}}_{\frameVar} \\
\underline{\tilde{\mybold{y}}}_{\frameVar}&= \underline{\mybold{C}}^{w}_{\frameVar} \underline{\mybold{a}}_{\frameVar} + \underline{\tilde{\mybold{s}}}_{\frameVar},
\end{align} 
where we define
%\begin{align}
%\underline{\mybold{c}}^{w}_{i,\frameVar} = 
%\underline{\mybold{C}}_{i,\frameVar}\underline{\mybold{w}}_{\frameVar}.
%\end{align}
\begin{align}
\underline{\mybold{C}}^{w}_{\frameVar} = 
\left[
\underline{\mybold{C}}_{0,\frameVar}\underline{\mybold{w}}_{\frameVar},
\ldots,
\underline{\mybold{C}}_{N-1,\frameVar}\underline{\mybold{w}}_{\frameVar}
\right].
\end{align}
Compared to the original form~\eqref{eq:obsDec}, we have essentially swapped the roles of~$\underline{\mybold{a}}_{\frameVar}$ and the linear SI channel~$\underline{\mybold{w}}_{\frameVar}$. Let
\begin{align}
\label{eq:y2dft}
\underline{\tilde{\mybold{y}}}_{\frameVar}=
%&= \underline{\mybold{C}}^{w}_{\frameVar} \underline{\mybold{a}}_{\frameVar} + \underline{\tilde{\mybold{s}}}_{\frameVar} %\\
%&\hintedrel[eq:y2dft1]{=} 
\underline{\mybold{C}}^{\hat{w}}_{\frameVar|\frameVar} \underline{\mybold{a}}_{\frameVar} + \underline{\tilde{\mybold{s}}}^{a1}_{\frameVar} 
\end{align} 
where we use~\eqref{eq:obsAlt} and the definition
\begin{align}
\label{eq:cwtau}
\underline{\mybold{C}}^{\hat{w}}_{\frameVar|\frameVar} &=
\left[
\underline{\mybold{C}}_{0,\frameVar}\underline{\hat{\mybold{w}}}_{\frameVar|\frameVar},
\underline{\mybold{C}}_{1,\frameVar}\underline{\hat{\mybold{w}}}_{\frameVar|\frameVar},
\ldots,
\underline{\mybold{C}}_{N-1,\frameVar}\underline{\hat{\mybold{w}}}_{\frameVar|\frameVar}
\right] \\
&=
\begin{bmatrix}
\underline{\mybold{c}}^{\hat{w}}_{0,\frameVar|\frameVar} &
\underline{\mybold{c}}^{\hat{w}}_{1,\frameVar|\frameVar} &
\ldots &
\underline{\mybold{c}}^{\hat{w}}_{N-1,\frameVar|\frameVar}
\end{bmatrix}.
\label{eq:frameColsDef}
\end{align} 
%\todo{The exact algorithm needs to be updated}
Furthermore, the zero-mean augmented noise here is 
\begin{align}
\label{eq:SignalNoiseA}
\underline{\tilde{\mybold{s}}}^{a1}_{\frameVar} &= 
\left[
\begin{matrix}
\underline{\mybold{C}}_{0,\frameVar}\underline{\mybold{n}}^{w}_{\frameVar|\frameVar}
& \underline{\mybold{C}}_{1,\frameVar}\underline{\mybold{n}}^{w}_{\frameVar|\frameVar}
& \ldots
\end{matrix} \right.\nonumber\\
&\qquad\left.
\begin{matrix}
& \underline{\mybold{C}}_{N-1,\frameVar}\underline{\mybold{n}}^{w}_{\frameVar|\frameVar}
\end{matrix}
\right]
\underline{\mybold{a}}_{\frameVar}
%\left[
%\underline{\mybold{C}}_{0,\frameVar}\underline{\mybold{n}}^{w}_{0,\frameVar|\frameVar-1},
%%& \underline{\mybold{C}}_{1,\frameVar}\underline{\mybold{n}}^{w}_{i,\frameVar|\frameVar-1}
%\ldots,
%\underline{\mybold{C}}_{N-1,\frameVar}\underline{\mybold{n}}^{w}_{N-1,\frameVar|\frameVar-1}
%\right]\underline{\mybold{a}}_{\frameVar}
 + \underline{\tilde{\mybold{s}}}_{\frameVar}.
\end{align}
The covariance matrix of the augmented noise $\underline{\tilde{\mybold{s}}}^{a1}_{\frameVar}$ is 
\begin{align}
\underline{\mybold{\Psi}}^{\tilde{s}^{a1}}_{\frameVar} 
&= 
\sum_{i=0}^{N-1}
%\expv\left[ \underline{a}_{i,\frameVar}\underline{a}^{*}_{j,\frameVar} \right]
\underline{p}^{a}_{i,\frameVar}
\underline{\mybold{C}}_{i,\frameVar}
\underline{\mybold{P}}^{w}_{\frameVar|\frameVar}
\underline{\mybold{C}}^{H}_{i,\frameVar}
+\underline{\mybold{\Psi}}^{\tilde{s}}_{\frameVar}. %\nonumber\\
%&\hintedrel[eq:psiSPrime1]{\approx} 
%\sum_{i=0}^{N-1}\sum_{j=0}^{N-1}
%\left(
%\underline{\hat{a}}_{i,\frameVar|\frameVar-1}\underline{\hat{a}}^{*}_{j,\frameVar|\frameVar-1}+
%\underline{p}^{a}_{i,j,\frameVar|\frameVar-1}
%\right)\nonumber\\
%&\hphantom{\sum_{i=0}^{N-1}\sum_{j=0}^{N-1}\underline{\hat{a}}_{i,\frameVar|\frameVar-1}}
%\cdot\underline{\mybold{C}}_{i,\frameVar}
%\underline{\mybold{P}}^{w}_{\frameVar|\frameVar}
%\underline{\mybold{C}}^{H}_{j,\frameVar}
%+\underline{\mybold{\Psi}}^{\tilde{s}}_{\frameVar}\nonumber\\
%&\hintedrel[eq:psiSPrime2]{=} 
%\underline{\mybold{C}}^{\hat{a}}_{\frameVar|\frameVar-1}\underline{\mybold{P}}^{w}_{\frameVar|\frameVar}\underline{\mybold{C}}^{\hat{a}^H}_{\frameVar|\frameVar-1} \nonumber\\
%&\qquad+\sum_{i=0}^{N-1}\sum_{j=0}^{N-1}
%\underline{p}^{a}_{i,j,\frameVar|\frameVar-1}
%\underline{\mybold{C}}_{i,\frameVar}
%\underline{\mybold{P}}^{w}_{\frameVar|\frameVar}
%\underline{\mybold{C}}^{H}_{j,\frameVar}
\label{eq:SignalNoiseACov}
\end{align}
%where~$(\hintref{eq:psiSPrime1})$ approximates the channel covariance by considering the instantaneous channel representation~\eqref{eq:nlCoeffsEstimatorPred} instead of the average, and~$(\hintref{eq:psiSPrime2})$ is using the definition~\eqref{eq:bwTau}. 
with $\underline{p}^{a}_{i,\frameVar}=\expv\left[ |\underline{a}_{i,\frameVar}|^2 \right]$ and $\underline{\mybold{P}}^{w}_{\frameVar|\frameVar}=\expv\left[ \underline{\mybold{n}}^{w}_{\frameVar|\frameVar}\underline{\mybold{n}}^{w^{H}}_{\frameVar|\frameVar} \right]$.

As a starting point,  it is noteworthy to say that the estimation of the linear SI channel provided in Section~\ref{sec:estLin} depends on the predictions of $\underline{\mybold{w}}_{\frameVar}$ and $\underline{\mybold{a}}_{\frameVar}$ only, while here the updated $\underline{\hat{\mybold{w}}}_{\frameVar|\frameVar}$ is available.
%we assume that the  are given and fully known from the estimation of the linear SI channel provided in Section~\ref{sec:estLin}. The interaction between the two parts is thus feed-forward only.
Similar to~\eqref{eq:augNoiseLinear}, we approximate~\eqref{eq:SignalNoiseA} by an independent Gaussian random vector $\underline{\tilde{\mybold{s}}}^{a2}_{\frameVar}$ with the same second-order moment as $\underline{\tilde{\mybold{s}}}^{a1}_{\frameVar}$. Thus, we have
\begin{align}
\label{eq:yaDef}
\underline{\tilde{\mybold{y}}}_{\frameVar}
&\approx
\underline{\mybold{C}}^{\hat{w}}_{\frameVar|\frameVar} \underline{\mybold{a}}_{\frameVar}
 + \underline{\tilde{\mybold{s}}}^{a2}_{\frameVar}. 
\end{align}
On this basis, we provide the Kalman filter equations, based on the system equation~\eqref{eq:nlCoeffsMarkov} and the observation equation~\eqref{eq:yaDef}, and by considering~\eqref{eq:SignalNoiseACov}. The computations for the $\frameVar^{\text{th}}$ step are depicted in~\algref{alg:nonlinearCoeffs}.
\begin{algorithm}
\caption{Obtain nonlinear coefficients estimate at step $\frameVar$}
\label{alg:nonlinearCoeffs}
\begin{algorithmic}
\REQUIRE known $\underline{\hat{\mybold{w}}}_{\frameVar|\frameVar}$, $\underline{\mybold{P}}^{w}_{\frameVar|\frameVar}$, $\underline{\hat{\mybold{a}}}_{\frameVar|\frameVar-1}$ and $\underline{\mybold{P}}^{a}_{\frameVar|\frameVar-1}$
%$\underline{\mybold{P}}^{w}_{0|-1} \neq \mybold{0}_{M\times M}$
%\PREDICT{coefficients estimate}
%\STATE $\underline{\hat{\mybold{a}}}_{\frameVar|\frameVar-1}=\underline{\mybold{A}}^{a} \underline{\hat{\mybold{a}}}_{\frameVar-1|\frameVar-1}$
%\ENDBLOCK
%\PREDICT{coefficients estimation error covariance}
%\STATE
%$\underline{\mybold{P}}^{a}_{\frameVar|\frameVar-1} = \underline{\mybold{A}}^{a}_{\frameVar}\underline{\mybold{P}}^{a}_{\frameVar-1|\frameVar-1}
%\underline{\mybold{A}}^{a^H}_{\frameVar}
%%\left|\underline{A}^{a}\right|^{2}
%+\underline{\mybold{\Psi}}^{\Delta a}_{\frameVar}$
%\ENDBLOCK
\GAIN{}
\STATE
$\underline{\mybold{K}}^{a}_{\frameVar} =
\underline{\mybold{P}}^{a}_{\frameVar|\frameVar-1}
\underline{\mybold{C}}^{\hat{w}^H}_{\frameVar|\frameVar}
\left(
\underline{\mybold{C}}^{\hat{w}}_{\frameVar|\frameVar}
\underline{\mybold{P}}^{a}_{\frameVar|\frameVar-1}
\underline{\mybold{C}}^{\hat{w}^H}_{\frameVar|\frameVar}+
\underline{\mybold{\Psi}}^{\tilde{s}^{a2}}_{\frameVar}
\right)^{-1}$
\ENDBLOCK
\UPDATE{coefficients estimate}
\STATE $\underline{\hat{\mybold{a}}}_{\frameVar|\frameVar}=
\underline{\hat{\mybold{a}}}_{\frameVar|\frameVar-1}+
\underline{\mybold{K}}^{a}_{\frameVar}
\left(
\underline{\tilde{\mybold{y}}}_{\frameVar} - \underline{\mybold{C}}^{\hat{w}}_{\frameVar|\frameVar}\underline{\hat{\mybold{a}}}_{\frameVar|\frameVar-1}
\right)$
\ENDBLOCK
\UPDATE{coefficients estimation error covariance}
\STATE
$\underline{\mybold{P}}^{a}_{\frameVar|\frameVar} = 
\left(
\eye{M}-\underline{\mybold{K}}^{a}_{\frameVar}\underline{\mybold{C}}^{\hat{w}}_{\frameVar|\frameVar}
\right)
\underline{\mybold{P}}^{a}_{\frameVar|\frameVar-1}$
\ENDBLOCK
\end{algorithmic}
\end{algorithm}

\section{Approximations}
\label{sec:approximations}
In this section, we discuss three approximations of the algorithm derived in Section~\ref{sec:algorithm} such as an intra-channel approximation (Section~\ref{sec:intraChannelApprox}), an overlap-save diagonalization (Section~\ref{sec:overlapSaveDiag}) and an nonlinear diagonalization (Section~\ref{sec:nonlinearDiag}).

\subsection{Intra-channel approximation}
\label{sec:intraChannelApprox}
The computation of noise covariances~\eqref{eq:covAugNoiseLinear} and~\eqref{eq:SignalNoiseACov} during the Kalman update step requires a-priori knowledge of the second moments $\underline{p}^{a}_{i,\frameVar}$, $\underline{\mybold{R}}^{w}_{\frameVar}$ of the linear channel and nonlinear coefficients, respectively. However, this information might not be available, and therefore we approximate the second moments by their estimated counterparts:
\begin{align}
\underline{\mybold{R}}^{w}_{\frameVar} &=
\expv\left[ \underline{\mybold{w}}_{\frameVar}\underline{\mybold{w}}^H_{\frameVar} \right] \nonumber \\
&\approx
\diag\left[
\underline{\hat{\mybold{w}}}_{\frameVar|\frameVar-1}
\circ
\underline{\hat{\mybold{w}}}^{*}_{\frameVar|\frameVar-1} 
\right]
+
\underline{\mybold{P}}^{w}_{\frameVar|\frameVar-1}. \\
\underline{p}^{a}_{i,\frameVar} &=
\expv\left[
|\underline{a}_{i,\frameVar}|^2
\right] \nonumber \\
&\approx 
|\underline{\hat{a}}_{i,\frameVar|\frameVar-1}|^2+
\underline{p}^{a}_{i,\frameVar|\frameVar-1},
\end{align}   
where we approximate the linear channel and nonlinear coefficient covariance by considering the instantaneous representations of the variables~\eqref{eq:siChanEstimatorPred} and~\eqref{eq:nlCoeffsEstimatorPred} instead of the average. 
%This is similar to the assumption used in the derivation of the LMS algorithm~\cite{Haykin2002}.

Furthermore, we assume the error covariance matrices $\underline{\mybold{P}}^{w}_{\frameVar|\frameVar-1}$, $\underline{\mybold{P}}^{w}_{\frameVar|\frameVar}$ and the augmented noise covariance matrix of~\eqref{eq:covAugNoiseLinear} to be diagonal.
%\begin{align}
%\expv\left[ \underline{\mybold{w}}_{\frameVar}\underline{\mybold{w}}^H_{\frameVar} \right]
%\approx
%\underline{\hat{\mybold{w}}}_{\frameVar|\frameVar-1}
%\underline{\hat{\mybold{w}}}^{H}_{\frameVar|\frameVar-1} 
%+
%\underline{\mybold{P}}^{w}_{\frameVar|\frameVar-1}.
%\end{align}

\subsection{Overlap-save diagonalization}
\label{sec:overlapSaveDiag}
The computation of the Kalman gain in \algref{alg:linearChannel} requires the inversion of an $M\times M$ matrix, which is both computationally intensive and numerically unstable. To simplify~\eqref{eq:CDef}, we apply the intra-channel diagonalization from Section~\ref{sec:intraChannelApprox} and use the approximations~\cite{src:benesty2001advances,Enzner06} for each nonlinear contribution
\begin{align}
\label{eq:approx1}
\underline{\mybold{C}}_{i,\frameVar}
&\approx
\frac{R}{M}\diag\left[ \underline{\mybold{\phi}}_{i,\frameVar}\right] \\
\label{eq:approx2}
\underline{\mybold{C}}_{i,\frameVar}\underline{\mybold{P}}^{w}_{\frameVar|\frameVar-1}\underline{\mybold{C}}^{H}_{j,\frameVar}
&\approx
\frac{R}{M}\diag\left[ \underline{\mybold{\phi}}_{i,\frameVar}\right]\underline{\mybold{P}}^{w}_{\frameVar|\frameVar-1}\diag\left[ \underline{\mybold{\phi}}^{*}_{j,\frameVar}\right].
\end{align}
The approximation errors of~\eqref{eq:approx1},~\eqref{eq:approx2} vanish for $R\to M$~\cite{src:malik2013variational}. The diagonalization step essentially ignores the overlap-save constraint during fast convolution, however, this effect is small if the difference $M-R$ is rather low. Considering the aforementioned changes, the prediction, Kalman gain and update steps from~\algref{alg:linearChannel} are diagonalized. 

\subsection{Nonlinear diagonalization}
\label{sec:nonlinearDiag}
The Kalman gain of \algref{alg:nonlinearCoeffs} uses the inversion of an $M\times M$ matrix. We intend to reduce computational complexity of that operation by diagonalizing that matrix. Since $\underline{\mybold{P}}^{a}_{\frameVar|\frameVar-1}$ and $\underline{\mybold{\Psi}}^{s^{a2}}_{\frameVar}$ are already diagonal by the approximations of Sections~\ref{sec:intraChannelApprox} and~\ref{sec:overlapSaveDiag}, only the $M\times N$ frame matrix $\underline{\mybold{C}}^{\hat{w}}_{\frameVar|\frameVar}$ has to be considered. Recall the definition~\eqref{eq:frameColsDef}. Each of the columns represents a nonlinear basis frame after convolution with the linear SI channel. Thus, we can write
\begin{align}
\label{eq:nldiagInv}
&\underline{\mybold{C}}^{\hat{w}}_{\frameVar|\frameVar}
\underline{\mybold{P}}^{a}_{\frameVar|\frameVar-1}
\underline{\mybold{C}}^{\hat{w}^H}_{\frameVar|\frameVar}+
\underline{\mybold{\Psi}}^{\tilde{s}^{a2}}_{\frameVar} = \\
&\qquad=
\sum_{i=0}^{N-1}
\underline{p}^{a}_{i,\frameVar|\frameVar-1}
\underline{\mybold{c}}^{\hat{w}}_{i,\frameVar|\frameVar}
\underline{\mybold{c}}^{\hat{w}^H}_{i,\frameVar|\frameVar}+
\underline{\mybold{\Psi}}^{\tilde{s}^{a2}}_{\frameVar}, \notag \\
\label{eq:nlDiagInvApprox}
&\qquad
\hintedrel[eq:nldiagInv1]{\approx}
\sum_{i=0}^{N-1}
\underline{p}^{a}_{i,\frameVar|\frameVar-1}
\underline{\mybold{c}}^{\hat{w}}_{i,\frameVar|\frameVar}
\underline{\mybold{c}}^{\hat{w}^H}_{i,\frameVar|\frameVar}+
\underline{\sigma}^{2}_{\frameVar}\eye{M}, 
\\
&\qquad
\hintedrel[eq:nldiagInv2]{=}
\mybold{U} \diag\left[ \lambda_0, \lambda_1, \ldots, \lambda_{M-1} \right] \mybold{U}^H 
\end{align}
where in~$(\hintref{eq:nldiagInv1})$, $\underline{\mybold{\Psi}}^{\tilde{s}^{a2}}_{\frameVar}$ is approximated by $\underline{\sigma}^{2}_{\frameVar}\eye{M}$ and $\underline{\sigma}^{2}_{\frameVar}$ denotes the largest element of $\underline{\mybold{\Psi}}^{\tilde{s}^{a2}}_{\frameVar}$. Furthermore, in~$(\hintref{eq:nldiagInv2})$, $\mybold{U}$ denotes the matrix of eigenvectors and $\lambda_i$ are the eigenvalues. The basis frames $\underline{\mybold{c}}^{\hat{w}}_{i,\frameVar|\frameVar}$ are not necessarily orthogonal. However, if we assume they are approximately orthogonal, then the basis frames are eigenvectors of 
$
\underline{\mybold{C}}^{\hat{w}}_{\frameVar|\frameVar}
\underline{\mybold{P}}^{a}_{\frameVar|\frameVar-1}
\underline{\mybold{C}}^{\hat{w}^H}_{\frameVar|\frameVar}
$, and, since $\underline{\sigma}^{2}_{\frameVar}\eye{M}$ is just a scaled identity matrix, they are also eigenvectors of~\eqref{eq:nlDiagInvApprox} with eigenvalues
\begin{align*}
\lambda_i = 
\begin{cases}
\underline{p}^{a}_{i,\frameVar|\frameVar-1}
\underline{\mybold{c}}^{\hat{w}^H}_{i,\frameVar|\frameVar}
\underline{\mybold{c}}^{\hat{w}}_{i,\frameVar|\frameVar}+
\underline{\sigma}^{2}_{\frameVar} &
0\leq i\leq N-1 \\
\underline{\sigma}^{2}_{\frameVar} &
N\leq i\leq M-1.
\end{cases}
\end{align*}
Furthermore, the first $N$ columns of $\mybold{U}$ are normalized versions of the $\underline{\mybold{c}}^{\hat{w}}_{i,\frameVar|\frameVar}$. Next, for the Kalman gain, we have
\begin{align*}
\underline{\mybold{K}}^{a}_{\frameVar} &=
\underline{\mybold{P}}^{a}_{\frameVar|\frameVar-1}
\underline{\mybold{C}}^{\hat{w}^H}_{\frameVar|\frameVar}
\left(
\sum_{i=0}^{N-1}
\underline{p}^{a}_{\frameVar|\frameVar-1}
\underline{\mybold{c}}^{\hat{w}}_{i,\frameVar|\frameVar}
\underline{\mybold{c}}^{\hat{w}^H}_{i,\frameVar|\frameVar}+
\underline{\sigma}^{2}_{\frameVar}\eye{M}
\right)^{-1} \\
&=
\underline{\mybold{P}}^{a}_{\frameVar|\frameVar-1}
\underline{\mybold{C}}^{\hat{w}^H}_{\frameVar|\frameVar}
\left(
\mybold{U} \diag\left[ \frac{1}{\lambda_0}, \frac{1}{\lambda_1}, \ldots, \frac{1}{\lambda_{M-1}} \right] \mybold{U}^H 
\right).
\end{align*}
Let $\underline{\mybold{k}}^{a^{T}}_{i,\frameVar}$ be the $i$th row of  $\underline{\mybold{K}}^{a}_{\frameVar}$, then we have
\begin{align}
\label{eq:kalmanApprox}
\underline{\mybold{k}}^{a^{T}}_{i,\frameVar}=
\frac{
\underline{p}^{a}_{i,\frameVar|\frameVar-1}
}{
\underline{p}^{a}_{i,\frameVar|\frameVar-1}
\underline{\mybold{c}}^{\hat{w}^H}_{i,\frameVar|\frameVar}
\underline{\mybold{c}}^{\hat{w}}_{i,\frameVar|\frameVar}+
\underline{\sigma}^{2}_{\frameVar}.
}
\underline{\mybold{c}}^{\hat{w}^H}_{i,\frameVar|\frameVar},
\end{align}
since the $i$th row of $\underline{\mybold{C}}^{\hat{w}^H}_{\frameVar|\frameVar}$ is orthogonal to all columns of $\mybold{U}$ except the $i$th one. The computation of~\eqref{eq:kalmanApprox} is much less expensive in terms of computational complexity. Finally, we apply the overlap-save diagonalization of~\eqref{eq:approx1} to~\eqref{eq:kalmanApprox}, and get
\begin{align*}
%\label{eq:kalmanApproxOSD}
\underline{\mybold{k}}^{a^{T}}_{i,\frameVar}\approx
\frac{
\frac{R}{M}
\underline{p}^{a}_{i,\frameVar|\frameVar-1}
\left(
\underline{\mybold{\phi}}_{i,\frameVar}
\circ
\underline{\mybold{\hat{w}}}_{i,\frameVar|\frameVar}
\right)^H
}{
\frac{R}{M}
\underline{p}^{a}_{i,\frameVar|\frameVar-1}
\left(
\underline{\mybold{\phi}}_{i,\frameVar}
\circ
\underline{\mybold{\hat{w}}}_{i,\frameVar|\frameVar}
\right)^H
\left(
\underline{\mybold{\phi}}_{i,\frameVar}
\circ
\underline{\mybold{\hat{w}}}_{i,\frameVar|\frameVar}
\right)
+
\underline{\sigma}^{2}_{\frameVar}.
}.
\end{align*}

\section{Results}
\label{sec:results}

In this section, we evaluate the performance of the algorithms derived in this work by comparing them to other approaches in the literature. We refer to~\algref{alg:linearChannel},~\algref{alg:nonlinearCoeffs} from Section~\ref{sec:algorithm} as the \emph{exact} Kalman algorithm in cascade structure. In addition, the \emph{approximated} Kalman algorithm is taken from Section~\ref{sec:approximations}. From the literature, we employ the Kalman algorithm in parallel structure~\cite{MalikEnzner2012b} and both the time-domain nonlinear NLMS and RLS in parallel structure.

The section is organized as follows. First, we define the performance metrics in Section~\ref{sec:metrics}. Then, the computational complexity is analyzed in Section~\ref{sec:complexity}. By performing simulations, we look into time convergence behavior in Section~\ref{sec:convergence}, global performance of the proposed algorithms in Sections~\ref{sec:global} and use cases and~\ref{sec:usecases}.  

\subsection{Performance metrics}
\label{sec:metrics} 
We define the key metrics that serve as performance indicators for the adaptive SI cancellation algorithms. First, we introduce definitions according to the mean-squared error (MSE) principle. The average signal to residual-interference-and-noise ratio (SRINR) is given as follows
\begin{align}
\label{eq:srinrDef}
\text{SRINR}_{\frameVar}=
\frac{
\expv\left[\left| \mybold{d}^{h}_{\frameVar} \right|^2\right]
}{
\expv\left[\left| \mybold{e}_\frameVar-\mybold{d}^{h}_{\frameVar} \right|^2\right]
}
,
\end{align}
where $\mybold{d}^{h}_{\frameVar}$ denotes the desired signal at the receiver, and $\mybold{e}_\frameVar$ is the error signal before decoding~\eqref{eq:errorSignal} in time domain. 
%We also intend to define a ``true'' SRINR that is independent of $\frameVar$ beyond the point of algorithmic convergence. 
The wireless channel~\eqref{eq:wChanDef} is assumed to be random, thus the process~\eqref{eq:srinrDef} is non-ergodic, and therefore we define the ``global'' SRINR to be the average in time for $\frameVar\to\infty$ over many realizations of the wireless channel. 

The capability of system identification can be measured by the system distance, i.e., the power of the estimation error compared to the power of the system variables. In the domain of Kalman filtering, the internal system state is the unknown quantity to be observed. To evaluate the quality of state estimation, we define the system distance for the linear SI channel coefficients as follows:
\begin{align}
\label{eq:sysDistWDef}
\text{SysDist}^{w}_{\frameVar} &=
\frac{
\expv\left[ 
\left( \mybold{w}_{\frameVar}-\hat{\mybold{w}}_{\frameVar|\frameVar} \right)^H  \left( \mybold{w}_{\frameVar}-\hat{\mybold{w}}_{\frameVar|\frameVar} \right) 
\right]
}{
\expv\left[ \mybold{w}^H_{\frameVar} \mybold{w}_{\frameVar}\right]
}
,
\end{align} 
where $\mybold{w}_{\frameVar}$ is the true SI channel vector per frame, as given by~\eqref{eq:siChanVec}, and $\hat{\mybold{w}}_{\frameVar|\frameVar}$ is the corresponding estimate of $\underline{\hat{\mybold{w}}}_{\frameVar|\frameVar}$ from~\algref{alg:linearChannel} in time domain. Similarly, the system distance for the $i$th nonlinear coefficient is defined as follows:
\begin{align}
\label{eq:sysDistADef}
\text{SysDist}^{a}_{i,\frameVar} &=
\frac{
\expv\left[ 
\left| \underline{a}_{i,\frameVar}-\underline{\hat{a}}_{i,\frameVar|\frameVar} \right|^2
\right]
}{
\expv\left[ \left| \underline{a}_{i,\frameVar} \right|^2 \right]
}
,
\end{align} 
where $\underline{a}_{i,\frameVar}$ and $\underline{\hat{a}}_{i,\frameVar|\frameVar}$ are the nonlinear coefficient from~\eqref{eq:nlCoeffsVecEqual} and its updated estimate at frame $\frameVar$, respectively. We need a fair comparison of the algorithms based on either cascaded (see~\figref{fig:systemmodel}) or parallel (see~\figref{fig:parallelmodel}) channel structure. The coefficients $\hat{\mybold{w}}_{0,\frameVar|\frameVar}$ of the parallel structure correspond to the linear SI channel from the cascade model. The nonlinear coefficients $\underline{\hat{a}}_{i,\frameVar}$ in case of a parallel structure are hence obtained by a Least-Squares solution 
\begin{align}
\label{eq:nlLSEst}
\underline{\hat{a}}_{i,\frameVar|\frameVar}=\frac{\hat{\mybold{w}}^H_{0,\frameVar|\frameVar}\hat{\mybold{w}}_{i,\frameVar|\frameVar}}{\hat{\mybold{w}}^H_{0,\frameVar|\frameVar}\hat{\mybold{w}}_{0,\frameVar|\frameVar}}.
\end{align}
%The adaptive algorithm essentially recreates the SI in order to subsequently cancel its impact. Thus, the quality of the SI reconstruction is a helpful indicator of the overall estimation performance. We define a (self-)interference-to-estimation-error ratio (IEER) by
%\begin{align}
%\label{eq:ieerDef}
%\text{IEER}_{\frameVar} &=
%\frac{
%\expv\left[  \mybold{x}^{H}_{\text{si},\frameVar} \mybold{x}_{\text{si},\frameVar} \right]
%}{
%\expv\left[ 
%\left(\mybold{x}_{\text{si},\frameVar} - \hat{\mybold{x}}_{\text{si},\frameVar}\right)^H
%\left(\mybold{x}_{\text{si},\frameVar} - \hat{\mybold{x}}_{\text{si},\frameVar}\right)
%\right]
%}
%,
%\end{align}
%where $\mybold{x}_{\text{si},\frameVar}$ is the actual SI, as given by~\eqref{eq:siTime}, and $\hat{\mybold{x}}_{\text{si},\frameVar}$ is the time-domain equivalent of the reconstructed SI, as provided by~\eqref{eq:reconstructedSI}. The metric~\eqref{eq:ieerDef} is very similar to the ``true'' Echo Return Loss Enhancement (ERLE)~\cite{Haensler2006}, which is commonly used to evaluate acoustic echo cancelers.

Next, we turn to a metric of digital communications. To simplify the analysis, we assume the wireless channel to be constant over time. Furthermore, since after SI cancellation the residual SI can be approximated by white noise~\cite{src:day2012fullduplex},  we assume that the residual SI process is independent and ergodic. Thus, the information capacity with power constraint $P_d$ at the distant node is given by~\cite{src:cover2006elements} 
\begin{align}
C &=\max_{f(d):\expv\left[|d|^2\right]<P_d}\:I\left(d;y\right) \nonumber\\
&\hintedrel[eq:ratelower1]{\geq}\max_{f(d):\expv\left[|d|^2\right]<P_d}\:I\left(d;y_G\right) \nonumber\\
\label{eq:ratelower}
&\hintedrel[eq:ratelower2]{=} \log\left( 1+\frac{P_d\left\lVert \mybold{h} \right\rVert^2_2}{\sigma^2_{\tilde{e}}} \right) =: R 
\end{align}
where in~$(\hintref{eq:ratelower1})$ we approximate the received signal $y$ by a circular-symmetric complex Gaussian counterpart $y_G$, since Gaussian noise serves as the worst-case noise~\cite{src:shomorony2012Is}. We arrive at~$(\hintref{eq:ratelower2})$, since in a Gaussian point-to-point channel, it is known that a Gaussian input is optimal. Here, $\sigma^2_{\tilde{e}}$ denotes the variance of the residual SI plus independent noise.

%we fix the distribution of $d_{\timeVar}$ to .  
%Since the actual probability distribution of the observation $y_{\timeVar}$ is not given in closed form, we evaluate the lower bound~\eqref{eq:ratelower} by estimations, based on a finite number of measured samples. We utilize a $k$-nearest neighbor estimator (NNE) for the mutual information, which is based on the idea and implementation of~\cite{src:kraskov2004}. Mutual information is a function of joint and marginal probability densities. For  a measure of the joint density, the estimator computes the distance between a tuple of samples and its $k$th-next neighbor. A similar approach is provided for the marginal densities. Results in~\cite{src:kraskov2004} indicate that at least for multivariate Gaussian variables, the estimation error is very low if $>10^4$ samples are used for the estimation.

\subsection{Complexity}
\label{sec:complexity} 
The cascade model of the nonlinear SI channel (recall~\figref{fig:systemmodel}) is essentially motivated by reducing computational complexity of the adaptive cancellation algorithm, which is primarily determined by the number of multiplications and divisions. In our analysis, we focus on the principal complexity only and do not evaluate the cost of each calculation step in detail. Thus, we identify those parts of the algorithms that exhibit the most significant impact on the complexity, depending on the (potentially large) frame shift $R$, the FIR filter length $\firLengthSI$ and on the order of the nonlinear expansion $N$. \tabref{tab:complexity} shows the results. In order to compare time-domain and DFT domain SI estimation and cancellation algorithms, we refer to the complexity as per sample. In all cases, DFT domain approaches are normalized to the frame shift $R=M-\firLengthSI$. This is reducing complexity by an order of magnitude if $R$ is growing while $\firLengthSI$ is kept fixed. 

Now, consider the frame shift $R$. The exact Kalman algorithm in cascade structure, derived in Section~\ref{sec:algorithm}, does have at least quadratic complexity, since it requires the inversion and multiplication of $M\times M$ matrices. On the other hand, considering the approximations of Section~\ref{sec:approximations}, the DFT operation becomes dominant, and therefore we have logarithmic complexity for the Kalman algorithm with nonlinear diagonalization. Similar reasoning holds for the Kalman algorithm in parallel structure after submatrix or full diagonalization~\cite{MalikEnzner2012b}. The complexity of the time-domain algorithms is determined by the the FIR filter size $\firLengthSI$. The classic NLMS algorithm in time domain exhibits linear complexity in $\firLengthSI$, while the parallel RLS algorithm in time domain has quadratic complexity due to the update step. However, a direct comparison of time-domain and DFT domain complexities is more involved and depends on the priorities of Kalman filter design. Certain designs might increase $\firLengthSI$ for finer channel estimation (and thus keep $R$ essentially constant), while others might use larger frame shifts $R$ to diminish the impact of the overlap-save approximation.

The number of nonlinear basis functions $N$ affects the computational complexity of the various algorithms in different ways. In cascade structure, the complexity generally scales linearly with respect to $N$. On the other hand, in parallel structures, Kalman or RLS algorithms take the correlation between all sub-channels into account, and therefore require cubic and quadratic complexity in $N$, respectively. Especially the cubic term might be greater than $\log_2M$ even for small $N$ and thus can have a significant impact. Kalman algorithm with full diagonalization~\cite{MalikEnzner2012b} or the NLMS allow to reduce the complexity to linear scale, since the correlation between parallel channels is neglected. 
\begin{table}
\begin{center}
  \begin{tabular}{ | l | l |}
  \hline
%  Algorithm & Complexity (Frame size) & Complexity (Nonlinear coefficients) \\ \hline
%    \hline
    Kalman, Cascade, Exact~\ref{sec:algorithm} & $\mathcal{O}\left(N\frac{(R+L)^3}{R}\right)$ \\ \hline
    Kalman, Cascade, Approx.~\ref{sec:nonlinearDiag} & $\mathcal{O}\left(N\frac{(R+L)\log_2(R+L)}{R}\right)$  \\ \hline
    Kalman, Parallel, Sub. diag.~\cite{MalikEnzner2012b}& $\mathcal{O}\left(N^3\frac{R+L}{R}\right) + \mathcal{O}\left(N\frac{(R+L)\log_2(R+L)}{R}\right)$ \\ \hline
    Kalman, Parallel, Full diag.~\cite{MalikEnzner2012b} & $\mathcal{O}\left(N\frac{(R+L)\log_2(R+L)}{R}\right)$ \\ \hline
    NLMS, Parallel, Time domain & $\mathcal{O}(NL)$  \\ \hline
    RLS, Parallel, Time domain & $\mathcal{O}(N^2L^2)$  \\ \hline
  \end{tabular}
\end{center}
\caption{Comparison of computational complexity with respect to frame shift $R$, FIR filter length $\firLengthSI$ and expansion order $N$ for different SI estimation and cancellation algorithms.}
\label{tab:complexity}
\end{table}

\subsection{Time convergence behavior}
\label{sec:convergence} 
\begin{figure*}
\ifshowTikzPictures
\begin{tikzpicture}
   \begin{groupplot}[group style={
                      group name=framePlots,
                      horizontal sep=3cm,
                      vertical sep=2cm,
                      group size= 2 by 2},
                      height=5cm,
                      width=8cm,
                      xmin=0,
                      xmax=400,
                      grid=major]
        \pgfmathtruncatemacro{\markRepeatFaster}{\markRepeat/2}
        \nextgroupplot[ylabel={SRINR [dB]}, xlabel={$\frameVar$}]
			\addplot [color=red, mark=+, mark options={solid, red}, mark repeat=\markRepeatFaster,mark phase=\markPOne]
			  table[]{\resSrinrVsSinr NonOrtho-Static-1.tsv};\label{plots:plot1a}												
%			\addplot [color=green, mark=o, mark options={solid, green}, mark repeat=\markRepeat,mark phase=\markPTwo]
%			  table[]{\resSrinrVsSinr NonOrtho-Static-2.tsv};\label{plots:plot2b}			
%			\addplot [color=blue, mark=asterisk, mark options={solid, blue}, mark repeat=\markRepeat,mark phase=\markPThree]
%			  table[]{\resSrinrVsSinr NonOrtho-Static-3.tsv};\label{plots:plot3c}			
			\addplot [color=cyan, mark=star, dashdotted, mark repeat=\markRepeat,mark phase=\markPFour]
			  table[]{\resSrinrVsSinr NonOrtho-Static-4.tsv};\label{plots:plot4d}			
			\addplot [color=magenta, mark=x, mark options={solid, magenta}, mark repeat=\markRepeat,mark phase=\markPFive]
			  table[]{\resSrinrVsSinr NonOrtho-Static-5.tsv};\label{plots:plot5e}			
%			\addplot [color=orange, mark=square, mark options={solid, orange}, mark repeat=\markRepeat,mark phase=\markPSix]
%			  table[]{\resSrinrVsSinr NonOrtho-Static-6.tsv};\label{plots:plot6f}			
			\addplot [color=black, mark=diamond, mark options={solid, black}, mark repeat=\markRepeat,mark phase=\markPSeven]
			  table[]{\resSrinrVsSinr NonOrtho-Static-7.tsv};\label{plots:plot7g}			
			\addplot [color=blue, mark=triangle, mark options={solid, blue}, mark repeat=\markRepeat,mark phase=\markPEight]
			  table[]{\resSrinrVsSinr NonOrtho-Static-8.tsv};\label{plots:plot8h}
			  
        \nextgroupplot[ylabel={System distance [dB]}, xlabel={$\frameVar$}, ymin=-45, ymax=20]
%			\addplot [color=red, mark=+, mark options={solid, red}, mark repeat=\markRepeatFaster,mark phase=\markPOne]
%			  table[]{\resSysDistWVsSinr NonOrtho-Static-1.tsv};	
%			\addplot [color=green, mark=o, mark options={solid, green}, mark repeat=\markRepeat,mark phase=\markPTwo]
%			  table[]{\resSysDistWVsSinr NonOrtho-Static-2.tsv};
			\addplot [color=blue, mark=asterisk, mark options={solid, blue}, mark repeat=\markRepeat,mark phase=\markPThree]
			  table[]{\resSysDistWVsSinr NonOrtho-Static-3.tsv};
			\addplot [color=cyan, mark=star, dashdotted, mark repeat=\markRepeat,mark phase=\markPFour]
			  table[]{\resSysDistWVsSinr NonOrtho-Static-4.tsv};
			\addplot [color=magenta, mark=x, mark options={solid, magenta}, mark repeat=\markRepeat,mark phase=\markPFive]
			  table[]{\resSysDistWVsSinr NonOrtho-Static-5.tsv};
%			\addplot [color=orange, mark=square, mark options={solid, orange}, mark repeat=\markRepeat,mark phase=\markPSix]
%			  table[]{\resSysDistWVsSinr NonOrtho-Static-6.tsv};
			\addplot [color=black, mark=diamond, mark options={solid, black}, mark repeat=\markRepeat,mark phase=\markPSeven]
			  table[]{\resSysDistWVsSinr NonOrtho-Static-7.tsv};
			\addplot [color=blue, mark=triangle, mark options={solid, blue}, mark repeat=\markRepeat,mark phase=\markPEight]
			  table[]{\resSysDistWVsSinr NonOrtho-Static-8.tsv};
			  
        \nextgroupplot[ylabel={System distance [dB]}, xlabel={$\frameVar$}, ymin=-45, ymax=20]
			\addplot [color=red, mark=+, mark options={solid, red}, mark repeat=\markRepeatFaster,mark phase=\markPOne]
			  table[]{\resSysDistAOneVsSinr NonOrtho-Static-1.tsv};	
%			\addplot [color=green, mark=o, mark options={solid, green}, mark repeat=\markRepeat,mark phase=\markPTwo]
%			  table[]{\resSysDistAOneVsSinr NonOrtho-Static-2.tsv};
%			\addplot [color=blue, mark=asterisk, mark options={solid, blue}, mark repeat=\markRepeat,mark phase=\markPThree]
%			  table[]{\resSysDistAOneVsSinr NonOrtho-Static-3.tsv};
			\addplot [color=cyan, mark=star, dashdotted, mark repeat=\markRepeat,mark phase=\markPFour]
			  table[]{\resSysDistAOneVsSinr NonOrtho-Static-4.tsv};
			\addplot [color=magenta, mark=x, mark options={solid, magenta}, mark repeat=\markRepeat,mark phase=\markPFive]
			  table[]{\resSysDistAOneVsSinr NonOrtho-Static-5.tsv};
%			\addplot [color=orange, mark=square, mark options={solid, orange}, mark repeat=\markRepeat,mark phase=\markPSix]
%			  table[]{\resSysDistAOneVsSinr NonOrtho-Static-6.tsv};
			\addplot [color=black, mark=diamond, mark options={solid, black}, mark repeat=\markRepeat,mark phase=\markPSeven]
			  table[]{\resSysDistAOneVsSinr NonOrtho-Static-7.tsv};
			\addplot [color=blue, mark=triangle, mark options={solid, blue}, mark repeat=\markRepeat,mark phase=\markPEight]
				  table[]{\resSysDistAOneVsSinr NonOrtho-Static-8.tsv};
			  
        \nextgroupplot[xlabel={Rounds}, ylabel={System distance [dB]}, xlabel={$\frameVar$}, ymin=-45, ymax=20]
			\addplot [color=red, mark=+, mark options={solid, red}, mark repeat=\markRepeatFaster,mark phase=\markPOne]
			  table[]{\resSysDistATwoVsSinr NonOrtho-Static-1.tsv};	
%			\addplot [color=green, mark=o, mark options={solid, green}, mark repeat=\markRepeat,mark phase=\markPTwo]
%			  table[]{\resSysDistATwoVsSinr NonOrtho-Static-2.tsv};
%			\addplot [color=blue, mark=asterisk, mark options={solid, blue}, mark repeat=\markRepeat,mark phase=\markPThree]
%			  table[]{\resSysDistATwoVsSinr NonOrtho-Static-3.tsv};
			\addplot [color=cyan, mark=star, dashdotted, mark repeat=\markRepeat,mark phase=\markPFour]
			  table[]{\resSysDistATwoVsSinr NonOrtho-Static-4.tsv};
			\addplot [color=magenta, mark=x, mark options={solid, magenta}, mark repeat=\markRepeat,mark phase=\markPFive]
			  table[]{\resSysDistATwoVsSinr NonOrtho-Static-5.tsv};
%			\addplot [color=orange, mark=square, mark options={solid, orange}, mark repeat=\markRepeat,mark phase=\markPSix]
%			  table[]{\resSysDistATwoVsSinr NonOrtho-Static-6.tsv};
			\addplot [color=black, mark=diamond, mark options={solid, black}, mark repeat=\markRepeat,mark phase=\markPSeven]
			  table[]{\resSysDistATwoVsSinr NonOrtho-Static-7.tsv};
			\addplot [color=blue, mark=triangle, mark options={solid, blue}, mark repeat=\markRepeat,mark phase=\markPEight]
			  table[]{\resSysDistATwoVsSinr NonOrtho-Static-8.tsv};
    \end{groupplot}
    
\node[text width=8cm,align=center,anchor=north] at ([yshift=-\subcaptionSpacing cm]framePlots c1r1.south) {\subcaption{Signal-to-residual-interference-and-noise ratios~\eqref{eq:srinrDef}.\label{fig:srinrNonOrtho}}};
\node[text width=8cm,align=center,anchor=north] at ([yshift=-\subcaptionSpacing cm]framePlots c2r1.south) {\subcaption{System distance~\eqref{eq:sysDistWDef} to the linear SI channel $\underline{\mybold{w}}_{\frameVar}$\label{fig:sysdistWNonOrtho}}};
\node[text width=8cm,align=center,anchor=north] at ([yshift=-\subcaptionSpacing cm]framePlots c1r2.south) {\subcaption{System distance~\eqref{eq:sysDistADef} to the nonlinear coefficient $\underline{a}_{1,\frameVar}$\label{fig:sysdistA1NonOrtho}}};
\node[text width=8cm,align=center,anchor=north] at ([yshift=-\subcaptionSpacing cm]framePlots c2r2.south) {\subcaption{System distance~\eqref{eq:sysDistADef} to the nonlinear coefficient $\underline{a}_{2,\frameVar}$\label{fig:sysdistA2NonOrtho}}};

\path (framePlots c1r1.north west|-current bounding box.north)--
      coordinate(legendposFramePlots)
      (framePlots c2r1.north east|-current bounding box.north);
\matrix[
    matrix of nodes,
    anchor=south,
    draw,
    inner sep=0.2em,
    nodes={anchor=west}
  ]at([yshift=1ex]legendposFramePlots)
  {
    \ref*{plots:plot1a}& Kalman, Cascade, Exact~\ref{sec:algorithm} &[5pt]
%    \ref*{plots:plot2b}& Kalman, Cascade, Overlap-Save Diag.~\ref{sec:overlapSaveDiag} &\\
%    \ref*{plots:plot3c}& Kalman, Cascade, Intra-channel decorr.~\ref{sec:intraChannelApprox} &[5pt]
    \ref*{plots:plot4d}& Kalman, Cascade, Approximated~\ref{sec:nonlinearDiag} & [5pt]
    \ref*{plots:plot5e}& Kalman, Parallel, Sub. diag.~\cite{MalikEnzner2012b}& \\
%    \ref*{plots:plot6f}& Kalman, Parallel, Full diag.~\cite{MalikEnzner2012b} & \\
    \ref*{plots:plot7g}& NLMS, Parallel, time domain & [5pt]
    \ref*{plots:plot8h}& RLS, Parallel, time domain  \\
    };
%\node at ($(legendpos2)+(0,3)$) {{\Large \storeInfoText}};
\end{tikzpicture}
\fi
\vspace{-0.25cm}
\caption{Time convergence performance of SI cancellation algorithms over the frame index $\frameVar$ with non-orthogonalized inputs with static SI path}
\label{fig:convergence}
\end{figure*}
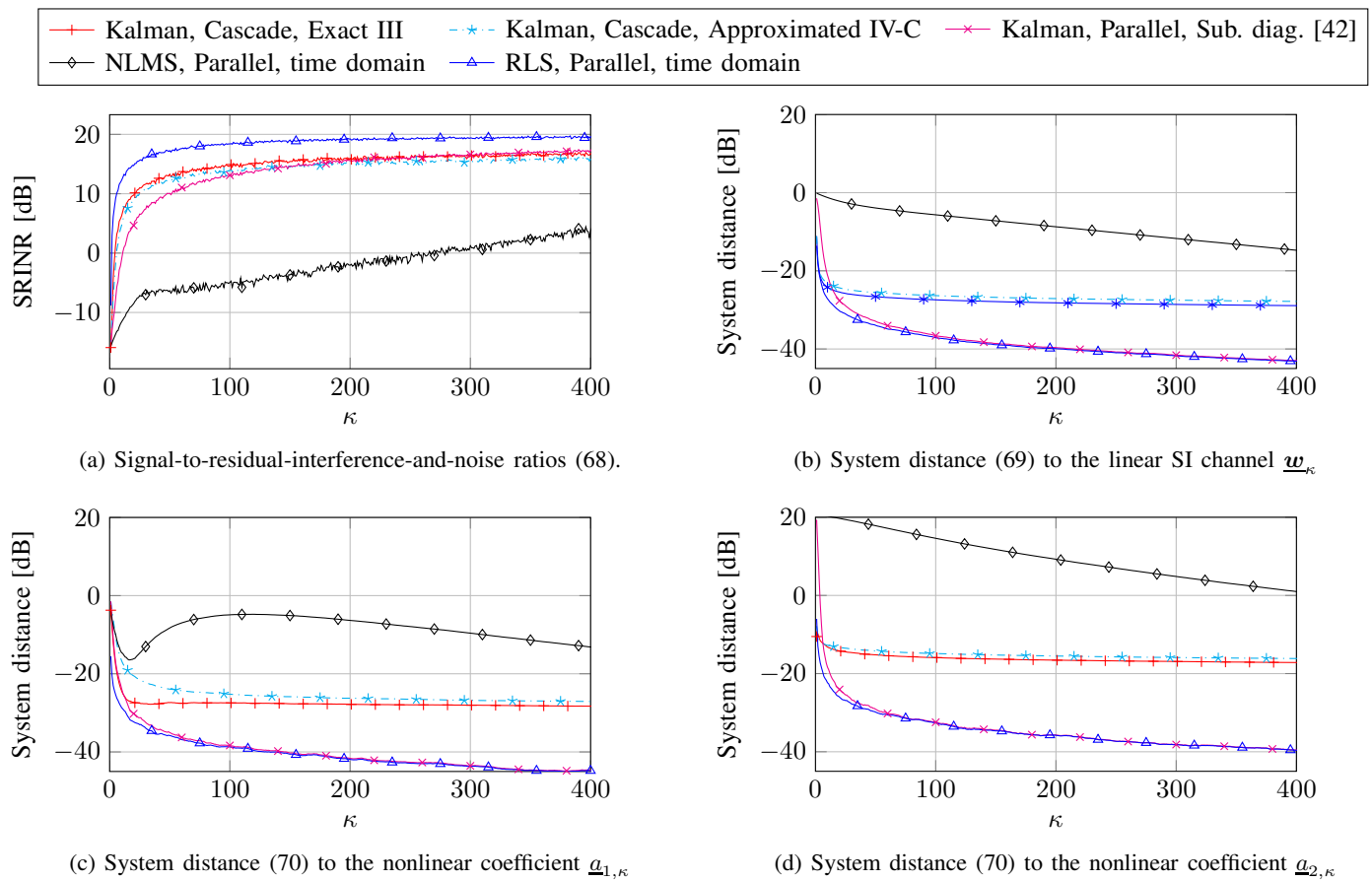

In this Section, we analyze the performance of the adaptive estimations over the course of time. We evaluate the performance of the algorithm by simulations.  We choose the parameters $M=64$, $R=56$ and select $N=3$ basis functions. The basis function are chosen according to the following emphases: (i) the linear part $\phi_0(x_{\timeVar})=x_{\timeVar}$ of the SI, then (ii) the widely-linear component $\phi_1(x_{\timeVar})=x^*_{\timeVar}$~\cite{src:korpi2014widely} and finally, (iii), the $3^{\text{rd}}$ order nonlinear component $\phi_2(x_{\timeVar})=x^2_{\timeVar}x^*_{\timeVar}$ with initial nonlinear coefficients $\underline{a}_{1,0}=\underline{a}_{2,0}=-10$~dB. The additive noise level is set to $-50$~dB. For the SI signal and the SoI, we generate zero-mean, standard normal i.i.d. samples. The Kalman filter parameters are perfectly matched to the simulated time-variant SI channel conditions. The NLMS step size is $10^{-2}$, while the RLS forgetting factor is matched to the Markov model of~\eqref{eq:siChanMarkov}.
%\tcr{For the Kalman approach, all parameters can be systematically chosen, for NLMS and RLS this is more difficult to do}
%\tcr{The approach is working if it is properly matched to the scale of channel state transition.  }
 Let the \emph{input SINR} be the ratio of the SoI power to the SI-and-noise power at the receiver before any signal processing occurs. The input SINR is fixed to $-15$~dB, thus, the SI is much stronger than the SoI. Consider~\figref{fig:convergence}, which shows the time convergence behavior of SI cancellation algorithms over the frame index $\frameVar$.
The inputs are not orthogonalized. Both the linear SI channel and the nonlinear coefficients remain constant over time. The SRINR results of~\figref{fig:srinrNonOrtho} demonstrate that without prior orthogonalization, the algorithms with strong inherent decorrelation feature, such as the Kalman in parallel structure with submatrix diagonalization or the RLS, are advantageous. Conversely, the NLMS is much slower in convergence. The Kalman approaches with cascade structure are bounded from above by the RLS and from below by the NLMS. Interestingly, the performance difference between the exact Kalman approach in cascade structure and its approximated counterpart 
%(intra-channel, overlap-save and nonlinear diagonalization) 
are almost negligible. Sometimes the exact calculation performs slightly worse. This has been identified as a particularity of the iterative design~\algref{alg:linearChannel} and~\algref{alg:nonlinearCoeffs}, which does not jointly estimate the linear SI channel and the nonlinear coefficients. The results of the system distance~(\Cref{fig:sysdistWNonOrtho,fig:sysdistA1NonOrtho,fig:sysdistA2NonOrtho}) essentially support the aforementioned findings. It can be observed that the system distances of the linear SI channel coefficients $\underline{\mybold{w}}_{\frameVar}$ of~\figref{fig:sysdistWNonOrtho} and the first nonlinear coefficient $\underline{a}_{1,\frameVar}$ in~\figref{fig:sysdistA1NonOrtho} are ranked consistently. In contrast to that, the distance of second nonlinear coefficient $\underline{a}_{2,\frameVar}$ in~\figref{fig:sysdistA2NonOrtho} is much larger, especially for the NLMS. This is due to the fact that the input signal and its $3$rd order component are correlated, and thus the performance deteriorates if the algorithms lack a decorrelation feature. The system distances in~\Cref{fig:sysdistWNonOrtho,fig:sysdistA1NonOrtho,fig:sysdistA2NonOrtho} of the Kalman approach in parallel structure with submatrix diagonalization and the RLS are almost identical. But the SRINR result (\figref{fig:srinrNonOrtho}) indicates that the Kalman algorithm does not precisely achieve the same level as the RLS. However, this observation can be explained. Both algorithms are devised in parallel structure. Thus, the nonlinear coefficients are computed by~\eqref{eq:nlLSEst}, which essentially is averaging over the channel coefficients and thus removing potential estimation errors. The parallel Kalman algorithm is slightly more susceptible to errors than the RLS since it is formulated in DFT domain with overlap-save approximation similar to the method described in sub-section~\ref{sec:overlapSaveDiag}. These error phenomena are compensated by reducing the number of estimated variables, such as by using~\eqref{eq:nlLSEst}. 

\subsection{Global performance}
\label{sec:global} 
\begin{figure*}
\ifshowTikzPictures
\begin{tikzpicture}
   \begin{groupplot}[group style={
                      group name=globalPlots,
                      horizontal sep=3cm,
                      vertical sep=2.5cm,
                      group size= 2 by 1},
                      height=5cm,
                      width=8cm,
                      xmin=-20,
                      xmax=20,
                      grid=major]
                      
        \nextgroupplot[ylabel={SRINR [dB]}, xlabel={Input SINR [dB]}, ymin=-5, ymax=25]
			\addplot [color=red, mark=+, mark options={solid, red}, mark repeat=2,mark phase=1]
			  table[]{\resSrinrConvVsSinr NonOrtho-Static-1.tsv};\label{plots:global1}												
%			\addplot [color=green, mark=o, mark options={solid, green}, mark repeat=2,mark phase=2]
%			  table[]{\resSrinrConvVsSinr NonOrtho-Static-2.tsv};\label{plots:global2}			
%			\addplot [color=blue, mark=asterisk, mark options={solid, blue}, mark repeat=2,mark phase=1]
%			  table[]{\resSrinrConvVsSinr NonOrtho-Static-3.tsv};\label{plots:global3}			
			\addplot [color=cyan, mark=star, dashdotted, mark repeat=2,mark phase=2]
			  table[]{\resSrinrConvVsSinr NonOrtho-Static-4.tsv};\label{plots:global4}			
			\addplot [color=magenta, mark=x, mark options={solid, magenta}, mark repeat=2,mark phase=1]
			  table[]{\resSrinrConvVsSinr NonOrtho-Static-5.tsv};\label{plots:global5}			
%			\addplot [color=orange, mark=square, mark options={solid, orange}, mark repeat=2,mark phase=2]
%			  table[]{\resSrinrConvVsSinr NonOrtho-Static-6.tsv};\label{plots:global6}			
			\addplot [color=black, mark=diamond, mark options={solid, black}, mark repeat=2,mark phase=1]
			  table[]{\resSrinrConvVsSinr NonOrtho-Static-7.tsv};\label{plots:global7}			
			\addplot [color=blue, mark=triangle, mark options={solid, blue}, mark repeat=2,mark phase=2]
			  table[]{\resSrinrConvVsSinr NonOrtho-Static-8.tsv};\label{plots:global8}
			\addplot [color=black, mark options={none}, thick] coordinates {(0,0) (20,20)};		
			\node[text width=2cm,align=center] at ($(axis cs:14,2.5)+(0,0)$) {Treating-interference-as-noise threshold};	  
			\addplot [color=black, mark options={none}, thick] coordinates {(-20,20) (20,20)};		
			\node[text width=2cm,align=center] at ($(axis cs:-10,23)+(0,0)$) {SNR};
			\addplot [color=black, mark options={none}, thick] coordinates {(-20,0) (0,0)};		
			\node[text width=2cm,align=center] at ($(axis cs:-10,3)+(0,0)$) {min. SRINR};
			  
        \nextgroupplot[ylabel={System distance [dB]}, xlabel={Input SINR [dB]}, ymin=-50,ymax=20]%Input SINR
			\addplot [color=red, mark=+, mark options={solid, red},mark repeat=2, mark phase=1]
			  table[]{\resSysDistWConvVsSinr NonOrtho-Static-1.tsv};	
%			\addplot [color=green, mark=o, mark options={solid, green}, mark repeat=2,mark phase=2]
%			  table[]{\resSysDistWConvVsSinr NonOrtho-Static-2.tsv};
%			\addplot [color=blue, mark=asterisk, mark options={solid, blue}, mark repeat=2,mark phase=1]
%			  table[]{\resSysDistWConvVsSinr NonOrtho-Static-3.tsv};
			\addplot [color=cyan, mark=star, dashdotted, mark repeat=2, mark phase=2]
			  table[]{\resSysDistWConvVsSinr NonOrtho-Static-4.tsv};
			\addplot [color=magenta, mark=x, mark options={solid, magenta}, mark repeat=2,mark phase=1]
			  table[]{\resSysDistWConvVsSinr NonOrtho-Static-5.tsv};
%			\addplot [color=orange, mark=square, mark options={solid, orange}, mark repeat=2,mark phase=2]
%			  table[]{\resSysDistWConvVsSinr NonOrtho-Static-6.tsv};
			\addplot [color=black, mark=diamond, mark options={solid, black}, mark repeat=2,mark phase=1]
			  table[]{\resSysDistWConvVsSinr NonOrtho-Static-7.tsv};
			\addplot [color=blue, mark=triangle, mark options={solid, blue}, mark repeat=2,mark phase=2]
			  table[]{\resSysDistWConvVsSinr NonOrtho-Static-8.tsv};

    \end{groupplot}
    
\node[text width=8cm,align=center,anchor=north] at ([yshift=-\subcaptionSpacing cm]globalPlots c1r1.south) {\subcaption{Signal-to-residual -interference-and-noise ratios~\eqref{eq:srinrDef}\label{fig:srinrConvNonOrtho}}};
\node[text width=8cm,align=center,anchor=north] at ([yshift=-\subcaptionSpacing cm]globalPlots c2r1.south) {\subcaption{System distance~\eqref{eq:sysDistWDef} to the linear SI channel $\underline{\mybold{w}}_{\frameVar}$\label{fig:sysdistWConvNonOrtho}}};
%\node[text width=8cm,align=center,anchor=north] at ([yshift=-\subcaptionSpacing cm]globalPlots c1r2.south) {\subcaption{System distance~\eqref{eq:sysDistADef} to the nonlinear coefficient $\underline{a}_{1,\frameVar}$\label{fig:sysdistA1ConvNonOrtho}}};
%\node[text width=8cm,align=center,anchor=north] at ([yshift=-\subcaptionSpacing cm]globalPlots c2r2.south) {\subcaption{System distance~\eqref{eq:sysDistADef} to the nonlinear coefficient $\underline{a}_{2,\frameVar}$\label{fig:sysdistA2ConvNonOrtho}}};

\path (globalPlots c1r1.north west|-current bounding box.north)--
      coordinate(legendposglobalPlots)
      (globalPlots c2r1.north east|-current bounding box.north);
\matrix[
    matrix of nodes,
    anchor=south,
    draw,
    inner sep=0.2em,
    nodes={anchor=west}
  ]at([yshift=1ex]legendposglobalPlots)
  {
    \ref*{plots:global1}& Kalman, Cascade, Exact~\ref{sec:algorithm} &[5pt]
%    \ref*{plots:global2}& Kalman, Cascade, Overlap-Save Diag.~\ref{sec:overlapSaveDiag} &\\
%    \ref*{plots:global3}& Kalman, Cascade, Intra-channel decorr.~\ref{sec:intraChannelApprox} &[5pt]
    \ref*{plots:global4}& Kalman, Cascade, Approximated~\ref{sec:nonlinearDiag} & [5pt]
    \ref*{plots:global5}& Kalman, Parallel, Sub. diag.~\cite{MalikEnzner2012b}&\\
%    \ref*{plots:global6}& Kalman, Parallel, Full diag.~\cite{MalikEnzner2012b} & \\
    \ref*{plots:global7}& NLMS, Parallel, time domain &
    \ref*{plots:global8}& RLS, Parallel, time domain  \\
    };
%\node at ($(legendpos2)+(0,3)$) {{\Large \storeInfoText}};
\end{tikzpicture}
\fi
\vspace{-0.25cm}
\caption{Global performance of SI cancellation algorithms over input SINR with non-orthogonalized inputs and static SI path.}
\label{fig:global}
\end{figure*}
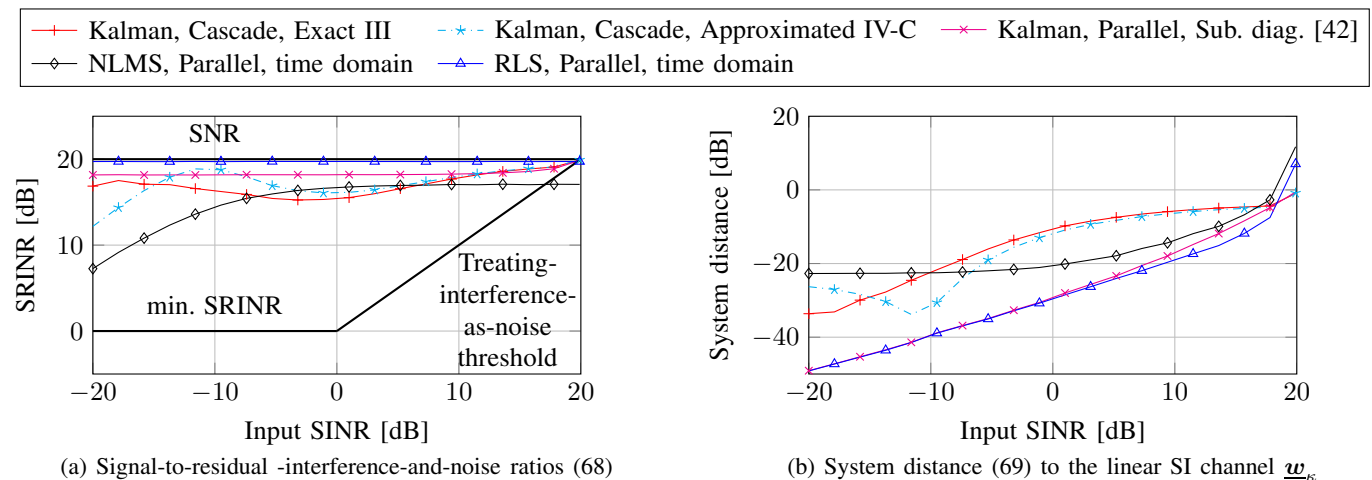

For the global performance, we evaluate SRINR (after cancellation) and system distances with respect to certain input SINR (before cancellation), which is illustrated in~\figref{fig:global}. 
Again, the input signals are not orthogonalized and both the linear SI channel and the nonlinear coefficients remain static. Consider~\figref{fig:srinrConvNonOrtho} first, which depicts the SRINR. The solid black line labeled ``SNR'' designates the maximum possible SRINR, since the noise floor is fixed at $-20$~dB with respect to the SoI. If the SRINR vs. input SINR graph crosses the ``treating-interference-as-noise threshold'', then it is no longer reasonable to use the algorithm and better to rather accept the SI as noise. The ``min. SRINR'' line is simply attained when the algorithm selects $\underline{\hat{\mybold{x}}}_{\text{si},\frameVar}=\underline{\mybold{y}}_{\frameVar}$ as output such that the received signal is completely suppressed. The results show that all algorithms can be used over the whole range of SINR scenarios. However, as already previously indicated, algorithms with already existing decorrelation property (Kalman in parallel structure with submatrix diagonalization or the RLS) have an advantage. This is especially prominent for the low SINR regime, where a great amount of cancellation is required. On the other hand, in the high SINR regime, the performance difference between the algorithms significantly diminishes. The system distance results of the linear SI channel coefficients $\underline{\mybold{w}}_{\frameVar}$ are shown in~\Cref{fig:sysdistWConvNonOrtho}. If we compare the performance of the various algorithms, the graphs are essentially reflecting the expected ``upside-down'' pattern compared to the SRINR figure, since a negative system distance naturally pays off in terms of SRINR. Here, however, the overall system distance is much better in the low SINR regime. This is intuitive, since at low SINR, the SI power is much larger than the SoI power, and thus the SI system components can be identified more accurately. Conversely, in the high SINR regime, the SI is buried in the SoI and the noise floor, and therefore the coefficient estimations are less reliable. While the detailed system identification is more difficult in the high SINR regime, the SRINR performance (denoting the overall SI estimation) is, however, not degraded.

 \subsection{Use cases}
\label{sec:usecases} 
\begin{figure*}
\ifshowTikzPictures
\begin{tikzpicture}[spy using outlines=
                      	{rectangle, magnification=4, connect spies}
                      	%,
]    
   \begin{groupplot}[group style={
                      group name=ratePlots,
                      horizontal sep=3cm,
                      vertical sep=2cm,
                      group size= 2 by 2},
                      height=5cm,
                      width=8cm,
                      xmin=-20,
                      xmax=20,
                      grid=major                      
                     ]
                      
        \nextgroupplot[ylabel={Rate [bits/sample]}, xlabel={Input SINR [dB]}, ymin=0]
			\addplot [color=red, mark=+, mark options={solid, red}, mark repeat=2,mark phase=1]
			  table[]{\resRateVsSinr NonOrtho-Static-1.tsv};\label{plots:rate1}												
%			\addplot [color=green, mark=o, mark options={solid, green}, mark repeat=2,mark phase=2]
%			  table[]{\resRateVsSinr NonOrtho-Static-2.tsv};\label{plots:rate2}			
%			\addplot [color=blue, mark=asterisk, mark options={solid, blue}, mark repeat=2,mark phase=1]
%			  table[]{\resRateVsSinr NonOrtho-Static-3.tsv};\label{plots:rate3}			
			\addplot [color=cyan, mark=star, dashdotted, mark repeat=2,mark phase=2]
			  table[]{\resRateVsSinr NonOrtho-Static-4.tsv};\label{plots:rate4}			
			\addplot [color=magenta, mark=x, mark options={solid, magenta}, mark repeat=2,mark phase=1]
			  table[]{\resRateVsSinr NonOrtho-Static-5.tsv};\label{plots:rate5}			
%			\addplot [color=orange, mark=square, mark options={solid, orange}, mark repeat=2,mark phase=2]
%			  table[]{\resRateVsSinr NonOrtho-Static-6.tsv};\label{plots:rate6}			
			\addplot [color=black, mark=diamond, mark options={solid, black}, mark repeat=2,mark phase=1]
			  table[]{\resRateVsSinr NonOrtho-Static-7.tsv};\label{plots:rate7}			
			\addplot [color=blue, mark=triangle, mark options={solid, blue}, mark repeat=2,mark phase=2]
			  table[]{\resRateVsSinr NonOrtho-Static-8.tsv};\label{plots:rate8}
			\addplot [color=blue, mark=none, thick, mark options={solid, blue}]
			  table[]{\resRateVsSinr NonOrtho-Static-9.tsv};\label{plots:rate9}

        \nextgroupplot[ylabel={Rate [bits/sample]}, xlabel={Input SINR [dB]}, ymin=0]
			\addplot [color=red, mark=+, mark options={solid, red}, mark repeat=2,mark phase=1]
			  table[]{\resRateVsSinr Orthogonal-Static-1.tsv};												
%			\addplot [color=green, mark=o, mark options={solid, green}, mark repeat=2,mark phase=2]
%			  table[]{\resRateVsSinr Orthogonal-Static-2.tsv};
%			\addplot [color=blue, mark=asterisk, mark options={solid, blue}, mark repeat=2,mark phase=1]
%			  table[]{\resRateVsSinr Orthogonal-Static-3.tsv};
			\addplot [color=cyan, mark=star, dashdotted, mark repeat=2,mark phase=2]
			  table[]{\resRateVsSinr Orthogonal-Static-4.tsv};
			\addplot [color=magenta, mark=x, mark options={solid, magenta}, mark repeat=2,mark phase=1]
			  table[]{\resRateVsSinr Orthogonal-Static-5.tsv};
%			\addplot [color=orange, mark=square, mark options={solid, orange}, mark repeat=2,mark phase=2]
%			  table[]{\resRateVsSinr Orthogonal-Static-6.tsv};
			\addplot [color=black, mark=diamond, mark options={solid, black}, mark repeat=2,mark phase=1]
			  table[]{\resRateVsSinr Orthogonal-Static-7.tsv};
			\addplot [color=blue, mark=triangle, mark options={solid, blue}, mark repeat=2,mark phase=2]
			  table[]{\resRateVsSinr Orthogonal-Static-8.tsv};
			\addplot [color=blue, mark=none, thick, mark options={solid, blue}]
			  table[]{\resRateVsSinr Orthogonal-Static-9.tsv};

        \nextgroupplot[ylabel={Rate [bits/sample]}, xlabel={Input SINR [dB]}, ymin=0]
			\addplot [color=red, mark=+, mark options={solid, red}, mark repeat=2,mark phase=1]
			  table[]{\resRateVsSinr NonOrtho-Varying-1.tsv};												
%			\addplot [color=green, mark=o, mark options={solid, green}, mark repeat=2,mark phase=2]
%			  table[]{\resRateVsSinr NonOrtho-Varying-2.tsv};
%			\addplot [color=blue, mark=asterisk, mark options={solid, blue}, mark repeat=2,mark phase=1]
%			  table[]{\resRateVsSinr NonOrtho-Varying-3.tsv};
			\addplot [color=cyan, mark=star, dashdotted, mark repeat=2,mark phase=2]
			  table[]{\resRateVsSinr NonOrtho-Varying-4.tsv};
			\addplot [color=magenta, mark=x, mark options={solid, magenta}, mark repeat=2,mark phase=1]
			  table[]{\resRateVsSinr NonOrtho-Varying-5.tsv};
%			\addplot [color=orange, mark=square, mark options={solid, orange}, mark repeat=2,mark phase=2]
%			  table[]{\resRateVsSinr NonOrtho-Varying-6.tsv};
			\addplot [color=black, mark=diamond, mark options={solid, black}, mark repeat=2,mark phase=1]
			  table[]{\resRateVsSinr NonOrtho-Varying-7.tsv};
			\addplot [color=blue, mark=triangle, mark options={solid, blue}, mark repeat=2,mark phase=2]
			  table[]{\resRateVsSinr NonOrtho-Varying-8.tsv};
			\addplot [color=blue, mark=none, thick, mark options={solid, blue}]
			  table[]{\resRateVsSinr NonOrtho-Varying-9.tsv};
			\coordinate (spypoint1) at (axis cs:-15,2);
			\coordinate (magnifyglass1) at (axis cs:12,2.25);

        \nextgroupplot[ylabel={Rate [bits/sample]}, xlabel={Input SINR [dB]}, ymin=0]
			\addplot [color=red, mark=+, mark options={solid, red}, mark repeat=2,mark phase=1]
			  table[]{\resRateVsSinr Orthogonal-Varying-1.tsv};												
%			\addplot [color=green, mark=o, mark options={solid, green}, mark repeat=2,mark phase=2]
%			  table[]{\resRateVsSinr Orthogonal-Varying-2.tsv};
%			\addplot [color=blue, mark=asterisk, mark options={solid, blue}, mark repeat=2,mark phase=1]
%			  table[]{\resRateVsSinr Orthogonal-Varying-3.tsv};
			\addplot [color=cyan, mark=star, dashdotted, mark repeat=2,mark phase=2]
			  table[]{\resRateVsSinr Orthogonal-Varying-4.tsv};
			\addplot [color=magenta, mark=x, mark options={solid, magenta}, mark repeat=2,mark phase=1]
			  table[]{\resRateVsSinr Orthogonal-Varying-5.tsv};
%			\addplot [color=orange, mark=square, mark options={solid, orange}, mark repeat=2,mark phase=2]
%			  table[]{\resRateVsSinr Orthogonal-Varying-6.tsv};
			\addplot [color=black, mark=diamond, mark options={solid, black}, mark repeat=2,mark phase=1]
			  table[]{\resRateVsSinr Orthogonal-Varying-7.tsv};
			\addplot [color=blue, mark=triangle, mark options={solid, blue}, mark repeat=2,mark phase=2]
			  table[]{\resRateVsSinr Orthogonal-Varying-8.tsv};
			\addplot [color=blue, mark=none, thick, mark options={solid, blue}]
			  table[]{\resRateVsSinr Orthogonal-Varying-9.tsv};
			\coordinate (spypoint2) at (axis cs:-15,2);
			\coordinate (magnifyglass2) at (axis cs:12,2.25);
			
    \end{groupplot}
%\spy [black, width=2.25cm, height=2.25cm] on (spypoint1)
%	in node[fill=white] at (magnifyglass1); 
\spy [black, width=2.25cm, height=2.25cm] on (spypoint2)
	in node at (magnifyglass2); 
	    
%\begin{scope}[spy using outlines=
%                      	{rectangle, magnification=4, connect spies}
%                      	%,
%]    
%\spy [black, width=2.25cm, height=2.25cm] on (spypoint2)
%	in node at (magnifyglass2);    
%\end{scope}	

%[fill=white]
\node[text width=8cm,align=center,anchor=north] at ([yshift=-\subcaptionSpacing cm]ratePlots c1r1.south) {\subcaption{Non-orthogonalized inputs, static SI path.\label{fig:rateNonOrthoStatic}}};
\node[text width=8cm,align=center,anchor=north] at ([yshift=-\subcaptionSpacing cm]ratePlots c2r1.south) {\subcaption{Orthogonalized inputs, static SI path.\label{fig:rateOrthoStatic}}};
\node[text width=8cm,align=center,anchor=north] at ([yshift=-\subcaptionSpacing cm]ratePlots c1r2.south) {\subcaption{Non-orthogonalized inputs, time-variant SI path.\label{fig:rateNonOrthoVarying}}};
\node[text width=8cm,align=center,anchor=north] at ([yshift=-\subcaptionSpacing cm]ratePlots c2r2.south) {\subcaption{Orthogonalized inputs, time-variant SI path.\label{fig:rateOrthoVarying}}};

\path (ratePlots c1r1.north west|-current bounding box.north)--
      coordinate(legendposratePlots)
      (ratePlots c2r1.north east|-current bounding box.north);
\matrix[
    matrix of nodes,
    anchor=south,
    draw,
    inner sep=0.2em,
    nodes={anchor=west}
  ]at([yshift=1ex]legendposratePlots)
  {
    \ref*{plots:rate1}& Kalman, Cascade, Exact~\ref{sec:algorithm} &[5pt]
%    \ref*{plots:rate2}& Kalman, Cascade, Overlap-Save Diag.~\ref{sec:overlapSaveDiag} &\\
%    \ref*{plots:rate3}& Kalman, Cascade, Intra-channel decorr.~\ref{sec:intraChannelApprox} &[5pt]
    \ref*{plots:rate4}& Kalman, Cascade, Approximated~\ref{sec:nonlinearDiag} & [5pt]
    \ref*{plots:rate5}& Kalman, Parallel, Sub. diag.~\cite{MalikEnzner2012b}& \\
%    \ref*{plots:rate6}& Kalman, Parallel, Full diag.~\cite{MalikEnzner2012b} & \\
    \ref*{plots:rate7}& NLMS, Parallel, time domain & [5pt]
    \ref*{plots:rate8}& RLS, Parallel, time domain & [5pt]
    \ref*{plots:rate9}& Capacity &[5pt] \\
%					  & 	&
    };
\end{tikzpicture}
\fi
\vspace{-0.25cm}
\caption{Communication rate~\eqref{eq:ratelower} over the input SINR for different cases of input processing and temporal variations.}
\label{fig:rates}
\end{figure*}
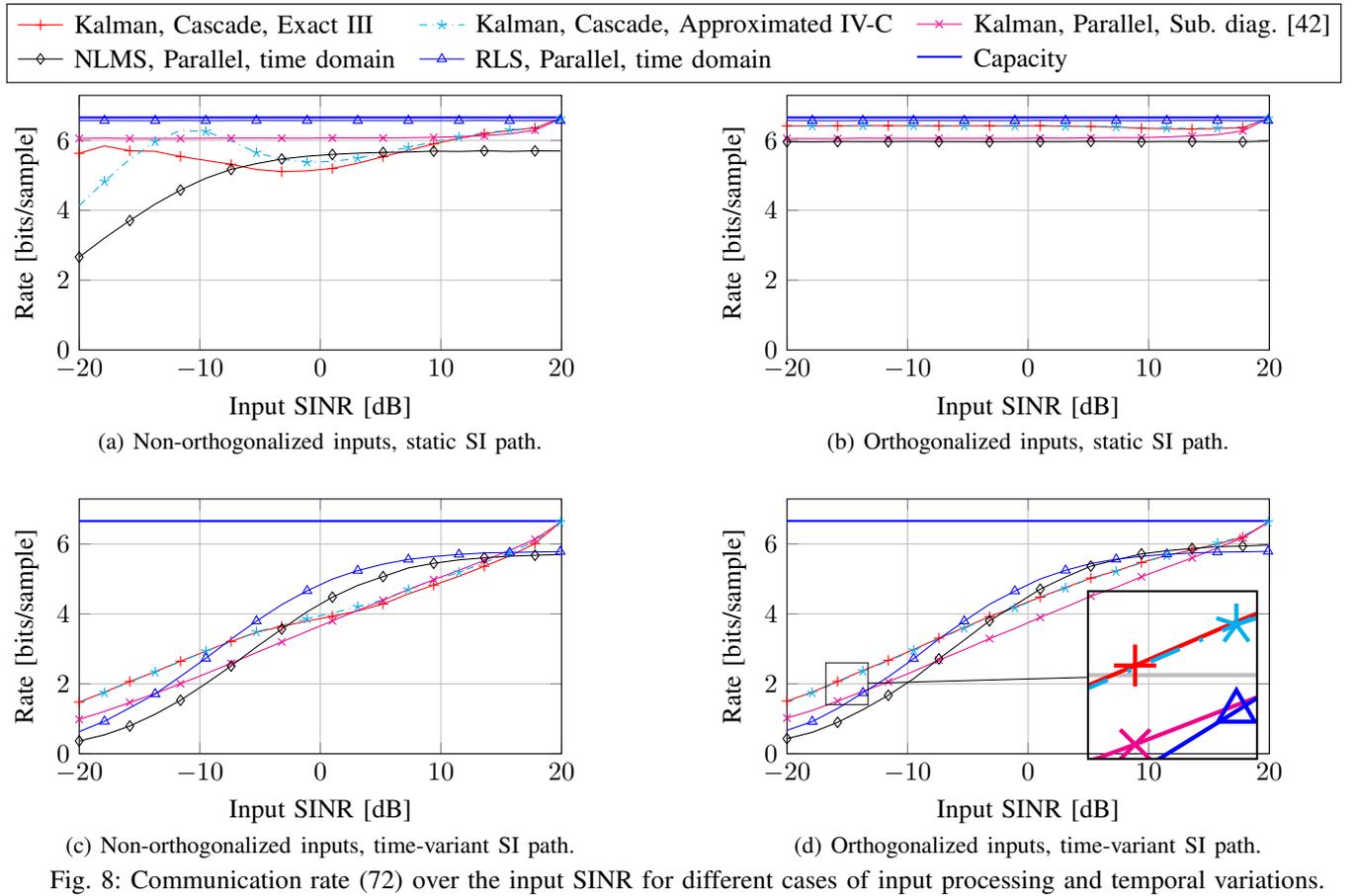

In this Section, we investigate the impact of time-variant channels on the performance of the SI estimation and cancellation algorithms. Static and time-variant channels represent fundamentally different use cases that significantly influence the proper choice of algorithms. As a performance metric, we apply the communication rate~\eqref{eq:ratelower}. The time-variant case is characterized by the (normalized) coherence time $\frameVar^{w}_{\text{coh}}=10^3$ for the linear SI channel and $\frameVar^{a}_{\text{coh}}=10^4$ for the nonlinear coefficients. We assume that the algorithm parameters are perfectly matched to the simulated channel conditions. Now consider~\figref{fig:rates}, where we illustrate the rate over the input SINR for different cases of temporal variations with or without orthogonalization. 
In each figure, we have provided the capacity as the upper bound for all rates. In~\figref{fig:rateNonOrthoStatic}, we have depicted the rate for static conditions of both the linear SI channel and the nonlinear coefficients, while the input signals are not orthogonalized. This is essentially a similar situation shown as in~\figref{fig:srinrConvNonOrtho}, but now the performance gaps in the low SINR regime are even more pronounced. Algorithms with inherent decorrelation like the Kalman in parallel structure with submatrix diagonalization perform the best regardless of the input SINR, while the RLS actually almost achieves the capacity. The cascaded and computationally efficient Kalman algorithms proposed in this work have a middle position, while the simple NMLS exhibits worse performance.
 
However, if orthogonalization of the input signal is applied, the situation fundamentally changes, as shown in~\figref{fig:rateOrthoStatic}. Now all the algorithm performances are completely independent of the input SINR. Furthermore, the performance gaps are almost insignificant, and thus a system designer can switch to a different metric in order to choose the best algorithm, such as the computational complexity. 
%As an intermediate conclusion here, in the static scenario with orthogonalization, it is not beneficial to rely on more sophisticated algorithms. 
In the time-convergence behavior, where we can obverse the same evolution for all the algorithms except the NLMS, but we omit the figure due to space constraints. In addition, the system distance results (not shown here) are very similar for all studied algorithms. 

\figref{fig:rateNonOrthoVarying} illustrates the communication rate in time-variant scenario, but without using input orthogonalization. Compared to the result of~\figref{fig:srinrConvNonOrtho}, it can clearly be seen that now the input SINR has a significant impact on the performance of all algorithms. 
%The Kalman approaches with cascade structure seem to slightly outperform the other algorithms for very low SINR. This is due to the fact that the Kalman approach allows to more accurately cover the temporal variations and the power of the SoI under severe conditions. 
\figref{fig:rateOrthoVarying} shows the results under time-variant conditions for orthogonalized input signals. In this case, the orthogonalization does not seem to help either, since the temporal variations represent the overall performance limitation.
%, and thus the orthogonalization step can be simply omitted.

\subsection{Decoding}
\label{sec:decodingRes} 
The decoding step has been introduced in Section~\ref{sec:decoding}. It takes place after the Kalman prediction, right before the update step is performed. Now, we evaluate the benefits of the information decoding to the SI estimation and cancellation. We essentially regard two relevant cases. First, we consider the situation that the algorithm does not decode the desired information at all. As a consequence, the SoI is considered as unknown noise to the adaptive algorithm, and thus the estimation performance is equal to the results of the previous Sections. Second, in the case of perfect decoding, the desired information is obtained without any errors, and thus the remaining residual error signal contains the residual SI only and some independent noise with $\text{SNR}=20$~dB. 

\begin{figure*}
\ifshowTikzPictures
\begin{tikzpicture}
   \begin{groupplot}[group style={
                      group name=decPlots,
                      horizontal sep=3cm,
                      vertical sep=2.5cm,
                      group size= 2 by 1},
                      height=5cm,
                      width=8cm,
                      xmin=-20,
                      xmax=20,
                      grid=major]
                      
        \nextgroupplot[ylabel={Rate [bits/sample]}, xlabel={Input SINR [dB]}, ymin=0]
%			\addplot [color=red, mark=+, mark options={solid, red}, mark repeat=2,mark phase=1]
%			  table[]{\resRateVsSinr NonOrtho-Static-1.tsv};\label{plots:dec1}														
			\addplot [color=cyan, mark=star, dashdotted, mark repeat=2,mark phase=1]
			  table[]{\resRateVsSinr NonOrtho-Static-4.tsv};\label{plots:dec4}			
%			\addplot [color=magenta, mark=x, mark options={solid, magenta}, mark repeat=2,mark phase=1]
%			  table[]{\resRateVsSinr PerfectDecNonOrtho-Static-1.tsv};\label{plots:dec5}			
			\addplot [color=orange, mark=square, mark options={solid, orange}, mark repeat=2,mark phase=2]
			  table[]{\resRateVsSinr PerfectDecNonOrtho-Static-4.tsv};\label{plots:dec6}
			\addplot [color=blue, mark=triangle, mark options={solid, blue}, mark repeat=2,mark phase=1]
			  table[]{\resRateVsSinr PerfectDecNonOrtho-Static-8.tsv};\label{plots:dec8}
			\addplot [color=green, mark=diamond, mark options={solid, green}, mark repeat=2,mark phase=2]
			  table[]{\resRateVsSinr PerfectDecNonOrtho-Static-10.tsv};\label{plots:dec10}
  			\addplot [color=blue, mark=none, thick, mark options={solid, blue}]
			 table[]{\resRateVsSinr Orthogonal-Static-11.tsv};\label{plots:dec11}			
			  
        \nextgroupplot[ylabel={Rate [bits/sample]}, xlabel={Input SINR [dB]}, ymin=0]
%			\addplot [color=red, mark=+, mark options={solid, red}, mark repeat=2,mark phase=1]
%			  table[]{\resRateVsSinr NonOrtho-Varying-1.tsv};														
			\addplot [color=cyan, mark=star, dashdotted, mark repeat=2,mark phase=1]
			  table[]{\resRateVsSinr NonOrtho-Varying-4.tsv};			
%			\addplot [color=magenta, mark=x, mark options={solid, magenta}, mark repeat=2,mark phase=1]
%			  table[]{\resRateVsSinr PerfectDecNonOrtho-Varying-1.tsv};			
			\addplot [color=orange, mark=square, mark options={solid, orange}, mark repeat=2,mark phase=2]
			  table[]{\resRateVsSinr PerfectDecNonOrtho-Varying-4.tsv};
			\addplot [color=blue, mark=triangle, mark options={solid, blue}, mark repeat=2,mark phase=1]
						  table[]{\resRateVsSinr PerfectDecNonOrtho-Varying-8.tsv};
			\addplot [color=green, mark=diamond, mark options={solid, green}, mark repeat=2,mark phase=2]
						  table[]{\resRateVsSinr PerfectDecNonOrtho-Varying-10.tsv};
  			\addplot [color=blue, mark=none, thick, mark options={solid, blue}]
				 table[]{\resRateVsSinr Orthogonal-Varying-11.tsv};
			  
    \end{groupplot}
    
\node[text width=8cm,align=center,anchor=north] at ([yshift=-\subcaptionSpacing cm]decPlots c1r1.south) {\subcaption{Communication rate~\eqref{eq:ratelower} for static SI path.\label{fig:DecRateNonOrthoStatic}}};
\node[text width=8cm,align=center,anchor=north] at ([yshift=-\subcaptionSpacing cm]decPlots c2r1.south) {\subcaption{Communication rate~\eqref{eq:ratelower} for time-variant SI path.\label{fig:DecRateNonOrthoVarying}}};

\path (decPlots c1r1.north west|-current bounding box.north)--
      coordinate(legendposdecPlots)
      (decPlots c2r1.north east|-current bounding box.north);
\matrix[
    matrix of nodes,
    anchor=south,
    draw,
    inner sep=0.2em,
    nodes={anchor=west}
  ]at([yshift=1ex]legendposdecPlots)
  {
%    \ref*{plots:dec1}& Kalman, Cascade, Exact~\ref{sec:algorithm}, No decoding &[5pt]
    \ref*{plots:dec4}& Kalman, Cascade, Approximated~\ref{sec:nonlinearDiag}, No decoding & 
%    \ref*{plots:dec5}& Kalman, Cascade, Exact~\ref{sec:algorithm}, Perfect decoding &[5pt]
    \ref*{plots:dec6}& Kalman, Cascade, Approximated~\ref{sec:nonlinearDiag}, Perfect decoding & \\
    \ref*{plots:dec8}& RLS, Parallel, time domain, No decoding & [5pt]
    \ref*{plots:dec10}& RLS, Parallel, time domain, Perfect decoding & [5pt] \\
    \ref*{plots:dec11}& Capacity &[5pt] \\
    };
\end{tikzpicture}
\fi
\vspace{-0.25cm}
\caption{Impact of decoding strategies on the SI estimation and cancellation performance.}
\label{fig:decoding}
\end{figure*}
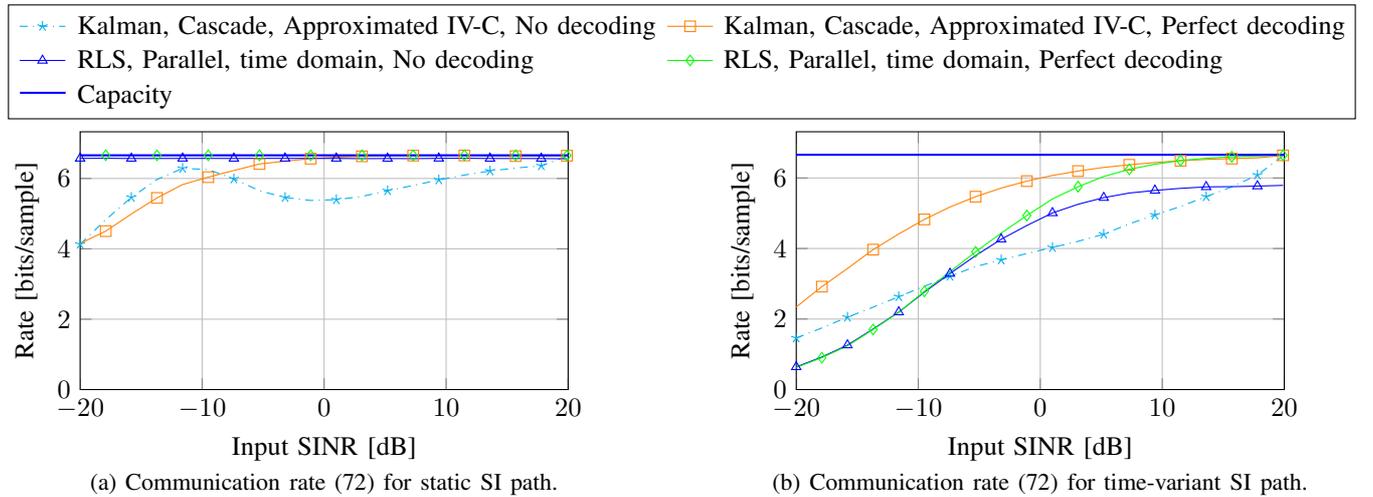

%We focus on these two extreme cases, namely perfect decoding or no decoding at all. The case without any decoding essentially reflects the results from previous Sections and thus serves as a benchmark. 
Consider~\figref{fig:DecRateNonOrthoStatic}.
It shows the communication rate~\eqref{eq:ratelower} for different input SINR in static~(\figref{fig:DecRateNonOrthoStatic}) and time-variant~(\figref{fig:DecRateNonOrthoVarying}) environments without input orthogonalization. 
%We keep our focus on Kalman algorithms in cascade structure with either exact configuration~(Section~\ref{sec:algorithm}) or with all approximations~(Section~\ref{sec:nonlinearDiag}). 
We keep our focus on Kalman algorithms in cascade structure with approximations~(Section~\ref{sec:nonlinearDiag}) and the time-domain RLS in parallel configuration. At first glance, it can be presumed that with perfect decoding, the performance is no longer dependent on the input SINR. However, this is only the case for the RLS algorithm in static environments. In the low SINR regime, the SI power is strong and thus the approximated algorithm is rather limited by estimation errors or favors certain levels of SI power.  Moreover, the probability of decoding errors is much higher in that regime and thus the potential rate gain is in fact even lower. This leads to the conclusion that in low SINR regime for static environments, the decoding step is of less relevance. However, the picture is different if the environment is time-variant, as illustrated in~\figref{fig:DecRateNonOrthoVarying}. The primary limiting factor is the temporal variations. The Kalman algorithm benefits twofold from the perfect decoding: First, the SoI is removed from the received signal before the SI channel update, and second, the signal covariance is adjusted to the noise floor. This improves adaptation to the time-variant context and makes it possible for the Kalman algorithm to be uniformly superior over the RLS in both low and high input SINR regimes. The actual rate performance of decoding with partial errors is expected to lie somewhere in between the two extreme cases of either perfect or no decoding at all. Thus,~\figref{fig:DecRateNonOrthoVarying} provide insights on the lower and upper performance bounds. Decoding increases the rate performance especially in moderate-to-high input SINR regimes.
%In~\figref{fig:DecRateNonOrthoVarying}, it is illustrated that both the exact and the approximated Kalman algorithm exhibit almost the same performance, since the limiting factor are the temporal variations here. 
%Clearly, both algorithms benefit significantly from the perfect decoding. 
%The information rate, as defined in~\eqref{eq:ratelower}, delivers insights into the performance of a FD communication system. The information rate is depicted in~\figref{fig:MIVsSinr} for different input SINR. We show the rate for the cases where the algorithm assumes (i) a time-variant channel with $\frameVar^{w}_{\text{coh}}=10^3$  and (ii) a static channel with $\frameVar^{w}_{\text{coh}}=10^{12}$. Furthermore, as a reference, the capacity of the wireless channel is depicted assuming perfect cancellation of the SI.

%In the scenario discussed in this paper, we assume $\underline{A}^{w}=A^{w}\eye{M}$ and $\underline{A}^{a}=A^{a}\eye{M}$, i.e., the Markov parameters do not depend on the DFT bin or the frame index. 

\section{Conclusion}
\label{sec:conclusion}
In this work, we have proposed an adaptive SI cancellation algorithm for full-duplex communication in the digital domain. It is based on a state-space model of the nonlinear SI channel in cascade structure. The signal model is derived in DFT domain. Linear and nonlinear components, although statistically coupled by the system model, can be linearized and then estimated separately in a two-step iterative approach, each by a Kalman filter. The algorithm comprises input orthogonalization, prediction, decoding of the desired signal, and updating the estimations. 
%After cancellation, a postfilter improves recovery of the desired signal. 
The state-space representation has introduced a more fine-grained handling of the time-variant nature of the SI channel. It underlines the significance of a-priori knowledge on the scale of temporal variations, which is represented by the state-space system model. Our simulation results indicate the following conclusion. While the choice of the underlying algorithmic principle like NLMS, RLS or Kalman do influence the performance of SI cancellation, the structural elements such as orthogonalization, parallel or cascade model, and the decoding have a much more significant impact. Especially the orthogonalization equalizes the performance gain for all algorithms. This work also demonstrates that the temporal variations are determining the fundamental performance limitation.
%If some computational power is available, the Kalman algorithm in parallel implementation represents the excellent trade-off between complexity and performance with monolithic structure.

%\begin{itemize}
%\item Algorithms with inherent orthogonalization feature are delivering better results than other approaches if the nonlinear contributions are correlated 
%\item The cascade modeling is reducing the susceptibility to over-fitting system identification
%\item System identification is more accurate in the low SINR regime, but the overall SRINR performance is worse then 
%\item In the case of static SI channels, the use of input orthogonalization negates the performance gaps between the algorithms for all input SINR
%\item In time-variant SI channels, the orthogonalization step does not help
%\item The proposed state-space algorithm approach is showing best performance in the low SINR regime for time-variant SI channels
%\item Decoding the SoI before the update step is very beneficial for convergence speed and estimation accuracy
%\item 
%\end{itemize}

\balance
\bibliographystyle{IEEEtran}
\bibliography{IEEEabrv,Conf_abrv_new,references}

\begin{thebibliography}{10}
\providecommand{\url}[1]{#1}
\csname url@samestyle\endcsname
\providecommand{\newblock}{\relax}
\providecommand{\bibinfo}[2]{#2}
\providecommand{\BIBentrySTDinterwordspacing}{\spaceskip=0pt\relax}
\providecommand{\BIBentryALTinterwordstretchfactor}{4}
\providecommand{\BIBentryALTinterwordspacing}{\spaceskip=\fontdimen2\font plus
\BIBentryALTinterwordstretchfactor\fontdimen3\font minus
  \fontdimen4\font\relax}
\providecommand{\BIBforeignlanguage}[2]{{%
\expandafter\ifx\csname l@#1\endcsname\relax
\typeout{** WARNING: IEEEtran.bst: No hyphenation pattern has been}%
\typeout{** loaded for the language `#1'. Using the pattern for}%
\typeout{** the default language instead.}%
\else
\language=\csname l@#1\endcsname
\fi
#2}}
\providecommand{\BIBdecl}{\relax}
\BIBdecl

\bibitem{src:sexton20175g}
C.~Sexton, N.~Kaminski, J.~Marquez-Barja, N.~Marchetti, and L.~A. DaSilva,
  ``{5G: Adaptable Networks Enabled by Versatile Radio Access Technologies},''
  \emph{{IEEE} Commun. Surveys Tuts.}, vol.~PP, no.~99, pp. 1--1, 2017.

\bibitem{src:sabharwal2014band}
A.~Sabharwal, P.~Schniter, D.~Guo, D.~W. Bliss, S.~Rangarajan, and R.~Wichman,
  ``{In-Band Full-Duplex Wireless: Challenges and Opportunities},''
  \emph{{IEEE} J. Sel. Areas Commun.}, vol.~32, no.~9, pp. 1637--1652, 2014.

\bibitem{src:aijaz2017simultaneous}
A.~Aijaz and P.~Kulkarni, ``{Simultaneous Transmit and Receive Operation in
  Next Generation IEEE 802.11 WLANs: A MAC Protocol Design Approach},''
  \emph{{IEEE} Wireless Commun.}, vol.~24, no.~6, pp. 128--135, Dec 2017.

\bibitem{src:jain2011practical}
M.~Jain, J.~I. Choi, T.~Kim, D.~Bharadia, S.~Seth, K.~Srinivasan, P.~Levis,
  S.~Katti, and P.~Sinha, ``Practical, real-time, full duplex wireless,'' in
  \emph{Proc. of the 17th annual int. conf. on Mobile computing and
  networking}.\hskip 1em plus 0.5em minus 0.4em\relax ACM, 2011, pp. 301--312.

\bibitem{src:riihonen2011hybrid}
T.~Riihonen, S.~Werner, and R.~Wichman, ``{Hybrid Full-Duplex/Half-Duplex
  Relaying with Transmit Power Adaptation},'' \emph{{IEEE} Trans. Wireless
  Commun.}, vol.~10, no.~9, pp. 3074--3085, September 2011.

\bibitem{src:duarte2012experiment}
M.~Duarte, C.~Dick, and A.~Sabharwal, ``{Experiment-Driven Characterization of
  Full-Duplex Wireless Systems},'' \emph{{IEEE} Trans. Wireless Commun.},
  vol.~11, no.~12, pp. 4296--4307, December 2012.

\bibitem{src:khandani2013two}
A.~K. Khandani, ``{Two-Way (True Full-duplex) Wireless},'' in \emph{13th
  Canadian Workshop on Information Theory}, Toronto, Ontario, Canada, June
  2013, pp. 33--38.

\bibitem{src:duarte2014design}
M.~Duarte, A.~Sabharwal, V.~Aggarwal, R.~Jana, K.~K. Ramakrishnan, C.~W. Rice,
  and N.~K. Shankaranarayanan, ``{Design and Characterization of a Full-Duplex
  Multiantenna System for WiFi Networks},'' \emph{{IEEE} Trans. Veh. Technol.},
  vol.~63, no.~3, pp. 1160--1177, March 2014.

\bibitem{src:debaillie2014analog}
B.~Debaillie, D.~J. van~den Broek, C.~Lavín, B.~van Liempd, E.~A.~M.
  Klumperink, C.~Palacios, J.~Craninckx, B.~Nauta, and A.~Pärssinen,
  ``{Analog/RF Solutions Enabling Compact Full-Duplex Radios},'' \emph{{IEEE}
  J. Sel. Areas Commun.}, vol.~32, no.~9, pp. 1662--1673, Sept 2014.

\bibitem{src:ahmed2015all}
E.~Ahmed and A.~M. Eltawil, ``{All-Digital Self-Interference Cancellation
  Technique for Full-Duplex Systems},'' \emph{{IEEE} Trans. Wireless Commun.},
  vol.~14, no.~7, pp. 3519--3532, July 2015.

\bibitem{src:askar2016agile}
R.~Askar, B.~Schubert, W.~Keusgen, and T.~Haustein, ``Agile full-duplex
  transceiver: The concept and self-interference channel characteristics,'' in
  \emph{22th European Wireless Conf.}, Oulu, Finland, May 2016, pp. 1--7.

\bibitem{src:avestimehr2008approximate}
A.~S. Avestimehr, A.~Sezgin, and D.~N.~C. Tse, ``Approximate capacity of the
  two-way relay channel: A deterministic approach,'' in \emph{46th Annual
  Allerton Conf. on Communication, Control, and Computing}, Sept 2008, pp.
  1582--1589.

\bibitem{src:nunn2017antenna}
D.~G. Wilson-Nunn, A.~Chaaban, A.~Sezgin, and M.~S. Alouini, ``{Antenna
  Selection for Full-Duplex MIMO Two-Way Communication Systems},'' \emph{{IEEE}
  Commun. Lett.}, vol.~21, no.~6, pp. 1373--1376, June 2017.

\bibitem{src:kariminezhad2017fullduplex}
A.~Kariminezhad, S.~Gherekhloo, and A.~Sezgin, ``{Full-Duplex vs. Half-Duplex:
  Delivery-Time Optimization in Cellular Downlink},'' in \emph{23th European
  Wireless Conf.}, May 2017, pp. 1--6.

\bibitem{src:zheng2013improving}
G.~Zheng, I.~Krikidis, J.~Li, A.~P. Petropulu, and B.~Ottersten, ``{Improving
  Physical Layer Secrecy Using Full-Duplex Jamming Receivers},'' \emph{{IEEE}
  Trans. Signal Process.}, vol.~61, no.~20, pp. 4962--4974, Oct 2013.

\bibitem{src:vogt2016practical}
H.~Vogt, K.~Ramm, and A.~Sezgin, ``{Practical Secret-Key Generation by
  Full-Duplex Nodes with Residual Self-Interference},'' in \emph{20th Int. ITG
  Workshop on Smart Antennas (WSA)}, March 2016, pp. 1--5.

\bibitem{src:zeng2015full}
Y.~Zeng and R.~Zhang, ``{Full-Duplex Wireless-Powered Relay With Self-Energy
  Recycling},'' \emph{{IEEE} Commun. Lett.}, vol.~4, no.~2, pp. 201--204, April
  2015.

\bibitem{src:bi2016accumulate}
Y.~Bi and H.~Chen, ``{Accumulate and Jam: Towards Secure Communication via A
  Wireless-Powered Full-Duplex Jammer},'' \emph{{IEEE} J. Sel. Areas Commun.},
  vol.~10, no.~8, pp. 1538--1550, Dec 2016.

\bibitem{src:bharadia2013full}
D.~Bharadia, E.~McMilin, and S.~Katti, ``Full duplex radios,'' in \emph{ACM
  SIGCOMM Computer Communication Review}, vol.~43, no.~4.\hskip 1em plus 0.5em
  minus 0.4em\relax ACM, 2013, pp. 375--386.

\bibitem{src:day2012fullduplex}
B.~P. Day, A.~R. Margetts, D.~W. Bliss, and P.~Schniter, ``{Full-Duplex
  Bidirectional MIMO: Achievable Rates Under Limited Dynamic Range},''
  \emph{{IEEE} Trans. Signal Process.}, vol.~60, no.~7, pp. 3702--3713, July
  2012.

\bibitem{src:masmoudi2016maximum}
A.~Masmoudi and T.~Le-Ngoc, ``{A Maximum-Likelihood Channel Estimator for
  Self-Interference Cancelation in Full-Duplex Systems},'' \emph{{IEEE} Trans.
  Veh. Technol.}, vol.~65, no.~7, pp. 5122--5132, July 2016.

\bibitem{src:bernhardt2018self}
M.~Bernhardt, F.~H. Gregorio, J.~Cousseau, and T.~Riihonen,
  ``{Self-Interference Cancellation through Advanced Sampling},'' \emph{{IEEE}
  Trans. Signal Process.}, vol.~PP, no.~99, pp. 1--1, 2018.

\bibitem{src:heino2015recent}
M.~Heino, D.~Korpi, T.~Huusari, E.~Antonio-Rodriguez, S.~Venkatasubramanian,
  T.~Riihonen, L.~Anttila, C.~Icheln, K.~Haneda, R.~Wichman, and M.~Valkama,
  ``Recent advances in antenna design and interference cancellation algorithms
  for in-band full duplex relays,'' \emph{{IEEE} Commun. Mag.}, vol.~53, no.~5,
  pp. 91--101, May 2015.

\bibitem{src:bharadia2014full}
D.~Bharadia and S.~Katti, ``{Full duplex MIMO radios},'' in \emph{USENIX
  Symposium on Networked Systems Design and Implementation (NSDI)}, Seattle,
  WA, USA, April 2014, pp. 1--15.

\bibitem{src:stimming2018nonlinear}
A.~{Balatsoukas-Stimming}, ``{Non-Linear Digital Self-Interference Cancellation
  for In-Band Full-Duplex Radios Using Neural Networks},'' \emph{ArXiv
  e-prints}, Nov. 2017.

\bibitem{src:korpi2015adaptive}
D.~Korpi, Y.~S. Choi, T.~Huusari, L.~Anttila, S.~Talwar, and M.~Valkama,
  ``{Adaptive Nonlinear Digital Self-Interference Cancellation for Mobile
  Inband Full-Duplex Radio: Algorithms and RF Measurements},'' in \emph{IEEE
  Global Communications Conf.}, San Diego, CA, USA, Dec. 2015, pp. 1--7.

\bibitem{src:ferrand2017multi}
P.~Ferrand and M.~Duarte, ``Multi-tap digital canceller for full-duplex
  applications,'' in \emph{IEEE 18th Int. Workshop on Signal Processing
  Advances in Wireless Communications (SPAWC)}, July 2017, pp. 1--5.

\bibitem{src:lemos2015fullduplex}
J.~S. Lemos, F.~A. Monteiro, I.~Sousa, and A.~Rodrigues, ``{Full-duplex
  relaying in MIMO-OFDM frequency-selective channels with optimal adaptive
  filtering},'' in \emph{IEEE Global Conf. on Signal and Information Processing
  (GlobalSIP)}, Dec 2015, pp. 1081--1085.

\bibitem{src:emara2017nonlinear}
M.~Emara, K.~Roth, L.~G. Baltar, M.~Faerber, and J.~A. Nossek, ``{Nonlinear
  Digital Self-Interference Cancellation with Reduced Complexity for Full
  Duplex Systems},'' in \emph{Int. ITG Workshop on Smart Antennas}, Berlin,
  Germany, March 2017.

\bibitem{src:kiayani2018adaptive}
A.~Kiayani, M.~Z. Waheed, L.~Anttila, M.~Abdelaziz, D.~Korpi, V.~Syrjälä,
  M.~Kosunen, K.~Stadius, J.~Ryynänen, and M.~Valkama, ``{Adaptive Nonlinear
  RF Cancellation for Improved Isolation in Simultaneous Transmit-Receive
  Systems},'' \emph{{IEEE} Trans. Microw. Theory Techn.}, vol.~PP, no.~99, pp.
  1--14, 2018.

\bibitem{src:everett2014passive}
E.~Everett, A.~Sahai, and A.~Sabharwal, ``{Passive Self-Interference
  Suppression for Full-Duplex Infrastructure Nodes},'' \emph{{IEEE} Trans.
  Wireless Commun.}, vol.~13, no.~2, pp. 680--694, February 2014.

\bibitem{src:benesty2001advances}
J.~Benesty, T.~G\"ansler, D.~R. Morgan, M.~M. Sondhi, and S.~L. Gay,
  \emph{Advances in Network and Acoustic Echo Cancellation}, 1st~ed.\hskip 1em
  plus 0.5em minus 0.4em\relax Springer Publishing Company, Incorporated, 2001.

\bibitem{Haensler97}
E.~H{\"a}nsler, ``From algorithms to systems - it's a rocky road,'' in
  \emph{Proc.\ Intl.\ Workshop on Acoustic Echo and Noise Control (IWAENC)},
  London, UK, September 1997.

\bibitem{Haensler2006}
E.~H{\"a}nsler and G.~Schmidt, Eds., \emph{Topics in Acoustic Echo and Noise
  Control}.\hskip 1em plus 0.5em minus 0.4em\relax Berlin: Springer, 2006.

\bibitem{Haykin2002}
S.~Haykin, \emph{Adaptive Filter Theory}, 4th~ed.\hskip 1em plus 0.5em minus
  0.4em\relax Upper Saddle River, NJ: Prentice-Hall, 2002.

\bibitem{Enzner2014}
G.~Enzner, H.~Buchner, A.~Favrot, and F.~Kuech, ``{Acoustic Echo Control},'' in
  \emph{Academic Press Library in Signal Processing}, R.~Chellappa and
  S.~Theodoridis, Eds.\hskip 1em plus 0.5em minus 0.4em\relax Elsevier, 2014,
  vol.~4, pp. 807--878.

\bibitem{src:gay1995the}
S.~L. Gay and S.~Tavathia, ``The fast affine projection algorithm,'' in
  \emph{Int. Conf. on Acoustics, Speech, and Signal Processing}, vol.~5, May
  1995, pp. 3023--3026 vol.5.

\bibitem{src:glicacho2012nonlinear}
J.~M. Gil-Cacho, T.~van Waterschoot, M.~Moonen, and S.~H. Jensen, ``Nonlinear
  acoustic echo cancellation based on a parallel-cascade kernel affine
  projection algorithm,'' in \emph{IEEE Int. Conf. on Acoustics, Speech and
  Signal Processing (ICASSP)}, March 2012, pp. 33--36.

\bibitem{Enzner06}
G.~Enzner and P.~Vary, ``Frequency-domain adaptive {Kalman} filter for acoustic
  echo control in hands-free telephones,'' \emph{Signal Processing, Elsevier},
  vol.~86, no.~6, pp. 1140--1156, June 2006.

\bibitem{Enzner2010}
G.~Enzner, ``Bayesian inference model for applications of time-varying acoustic
  system identification,'' in \emph{Proc.\ Europ. Signal Proc. Conf.\
  (EUSIPCO)}, Aalborg, DN, August 2010.

\bibitem{Malik2011a}
S.~Malik and G.~Enzner, ``Recursive {Bayesian} control of multichannel acoustic
  echo cancellation,'' \emph{IEEE Signal Proc. Lett.}, vol.~18, no.~11, pp.
  619--622, November 2011.

\bibitem{MalikEnzner2012b}
------, ``{State-Space Frequency-Domain Adaptive Filtering for Nonlinear
  Acoustic Echo Cancellation},'' \emph{{IEEE/ACM} Trans. Audio, Speech,
  Language Process.}, vol.~20, no.~7, pp. 2065--2079, September 2012.

\bibitem{src:malik2013variational}
------, ``{A Variational Bayesian Learning Approach for Nonlinear Acoustic Echo
  Control},'' \emph{{IEEE} Trans. Signal Process.}, vol.~61, no.~23, pp.
  5853--5867, Dec 2013.

\bibitem{src:laakso96splitting}
T.~I. Laakso, V.~Valimaki, M.~Karjalainen, and U.~K. Laine, ``{Splitting the
  unit delay [FIR/all pass filters design]},'' \emph{{IEEE} Signal Process.
  Mag.}, vol.~13, no.~1, pp. 30--60, Jan 1996.

\bibitem{src:sohaib2017alow}
M.~Sohaib, H.~Nawaz, K.~Ozsoy, O.~Gurbuz, and I.~Tekin, ``{A Low Complexity
  Full-Duplex Radio Implementation with a Single Antenna},'' \emph{{IEEE}
  Trans. Veh. Technol.}, vol.~PP, no.~99, pp. 1--1, 2017.

\bibitem{src:komatsu2017freq}
K.~Komatsu, Y.~Miyaji, and H.~Uehara, ``{Frequency-Domain Hammerstein
  Self-Interference Canceller for In-Band Full-Duplex OFDM Systems},'' in
  \emph{IEEE Wireless Communications and Networking Conf. (WCNC)}, March 2017,
  pp. 1--6.

\bibitem{src:korpi2014full}
D.~Korpi, T.~Riihonen, V.~Syrjälä, L.~Anttila, M.~Valkama, and R.~Wichman,
  ``{Full-Duplex Transceiver System Calculations: Analysis of ADC and Linearity
  Challenges},'' \emph{{IEEE} Trans. Wireless Commun.}, vol.~13, no.~7, pp.
  3821--3836, July 2014.

\bibitem{src:nadh2017ataylor}
A.~Nadh, J.~Samuel, A.~Sharma, S.~Aniruddhan, and R.~K. Ganti, ``{A Taylor
  Series Approximation of Self-Interference Channel in Full-Duplex Radios},''
  \emph{{IEEE} Trans. Wireless Commun.}, vol.~16, no.~7, pp. 4304--4316, July
  2017.

\bibitem{src:bishop2006pattern}
C.~M. Bishop, \emph{{Pattern Recognition and Machine Learning (Information
  Science and Statistics)}}.\hskip 1em plus 0.5em minus 0.4em\relax Secaucus,
  NJ, USA: Springer-Verlag New York, Inc., 2006.

\bibitem{src:proakis2007digital}
J.~G. Proakis, \emph{Digital Communications 5th Edition}.\hskip 1em plus 0.5em
  minus 0.4em\relax McGraw Hill, 2007.

\bibitem{src:scharf1991statistical}
L.~L. Scharf, \emph{Statistical signal processing}.\hskip 1em plus 0.5em minus
  0.4em\relax Addison-Wesley Reading, MA, 1991, vol.~98.

\bibitem{src:cover2006elements}
T.~M. Cover and J.~A. Thomas, \emph{Elements of Information Theory}.\hskip 1em
  plus 0.5em minus 0.4em\relax Wiley-Interscience, 2006.

\bibitem{src:shomorony2012Is}
I.~Shomorony and A.~S. Avestimehr, ``{Is Gaussian noise the worst-case additive
  noise in wireless networks?}'' in \emph{IEEE Int. Symposium on Information
  Theory}, Cambridge, MA, USA, July 2012, pp. 214--218.

\bibitem{src:korpi2014widely}
D.~Korpi, L.~Anttila, V.~Syrjälä, and M.~Valkama, ``{Widely Linear Digital
  Self-Interference Cancellation in Direct-Conversion Full-Duplex
  Transceiver},'' \emph{{IEEE} J. Sel. Areas Commun.}, vol.~32, no.~9, pp.
  1674--1687, Sept 2014.

\end{thebibliography}

\end{document}